\documentclass[12pt]{article}
\usepackage{setspace}
\usepackage[utf8]{inputenc}
\usepackage[english]{babel}  
\usepackage[colorlinks=true, linkcolor=blue, citecolor = blue]{hyperref} \usepackage[ruled,vlined]{algorithm2e}    
\usepackage{url}  
\usepackage{booktabs}       
\usepackage{amsfonts}       
\usepackage{nicefrac}       
\usepackage{arxiv}
\usepackage{microtype}
\usepackage[top=1 in,bottom=1in, left=1 in, right=1 in]{geometry}
\usepackage{graphicx}
\usepackage{subfig}
\usepackage{rotating}
\usepackage{caption} 

\usepackage{float}
\usepackage{wrapfig}
\usepackage{amsmath}
\usepackage{enumitem}
\usepackage{amssymb}
\usepackage{amsthm}
\usepackage[round]{natbib}
\usepackage{tabularx}
\usepackage[export]{adjustbox}
\usepackage{multicol}
\usepackage{color}
\usepackage{multirow}
\usepackage[thinlines]{easytable}
\usepackage{multirow}
\usepackage[table,xcdraw]{xcolor}
\hypersetup{
    colorlinks=true,
    linkcolor=blue,
    filecolor=blue,      
    urlcolor=blue,
}
\usepackage{tcolorbox}
\usepackage{fancyhdr}
\newcommand\independent{\protect\mathpalette{\protect\independenT}{\perp}}
\def\independenT#1#2{\mathrel{\rlap{$#1#2$}\mkern2mu{#1#2}}}
\newtheorem{theorem}{Theorem}
\newtheorem{lemma}{Lemma}
\newtheorem{corollary}{Corollary}

\newtheorem{assumption}{Assumption}
\newtheorem{condition}{Condition}
\newtheorem{definition}{Definition}

\title{\textbf{Network Structural Equation Models for Causal Mediation and Spillover Effects}}
\author
{Ritoban Kundu$^1$ and 
Peter X.K. Song$^{1*}$\\
Department of Biostatistics, University of Michigan, Ann Arbor, Michigan, U.S.A. \\
$^{*}$Corresponding Author}

\begin{document}
\maketitle

\centerline{\Large \bf Abstract}
\vspace{0.1in}
\noindent
Social network interference induces complex dependencies where a unit’s outcome is influenced not only by its own exposure and mediator but also by those of connected neighbors. In such settings, a significant challenge lies in distinguishing direct exposure effects from interference-driven spillover effects, and further separating these from indirect effects mediated by intermediate variables. To address this, we propose a theoretical framework utilizing structural graphical models. Central to our approach is the Random Effects Network Structural Equation Model (REN-SEM), which extends the exposure mapping paradigm to capture these multifaceted spillover and mediation mechanisms while accounting for latent dependencies within mediators and outcomes. We establish general identification conditions and derive decomposition formulas for six distinct mechanistic estimands. Furthermore, for the class of Linear REN-SEMs, we develop a maximum likelihood estimation framework and establish a rigorous asymptotic theory tailored to non-i.i.d. network data, proving the consistency of our estimators and the validity of the variance estimates. The robustness and practical utility of our methodology are demonstrated through simulation experiments and an analysis of the Twitch Gamers Network, underscoring its effectiveness in quantifying intricate network-mediated exposure effects.

\vspace{0.2in}


\section{Introduction}\label{sec-intro}
\begin{figure}[ht]
\centering
 \includegraphics[width=0.7\linewidth]{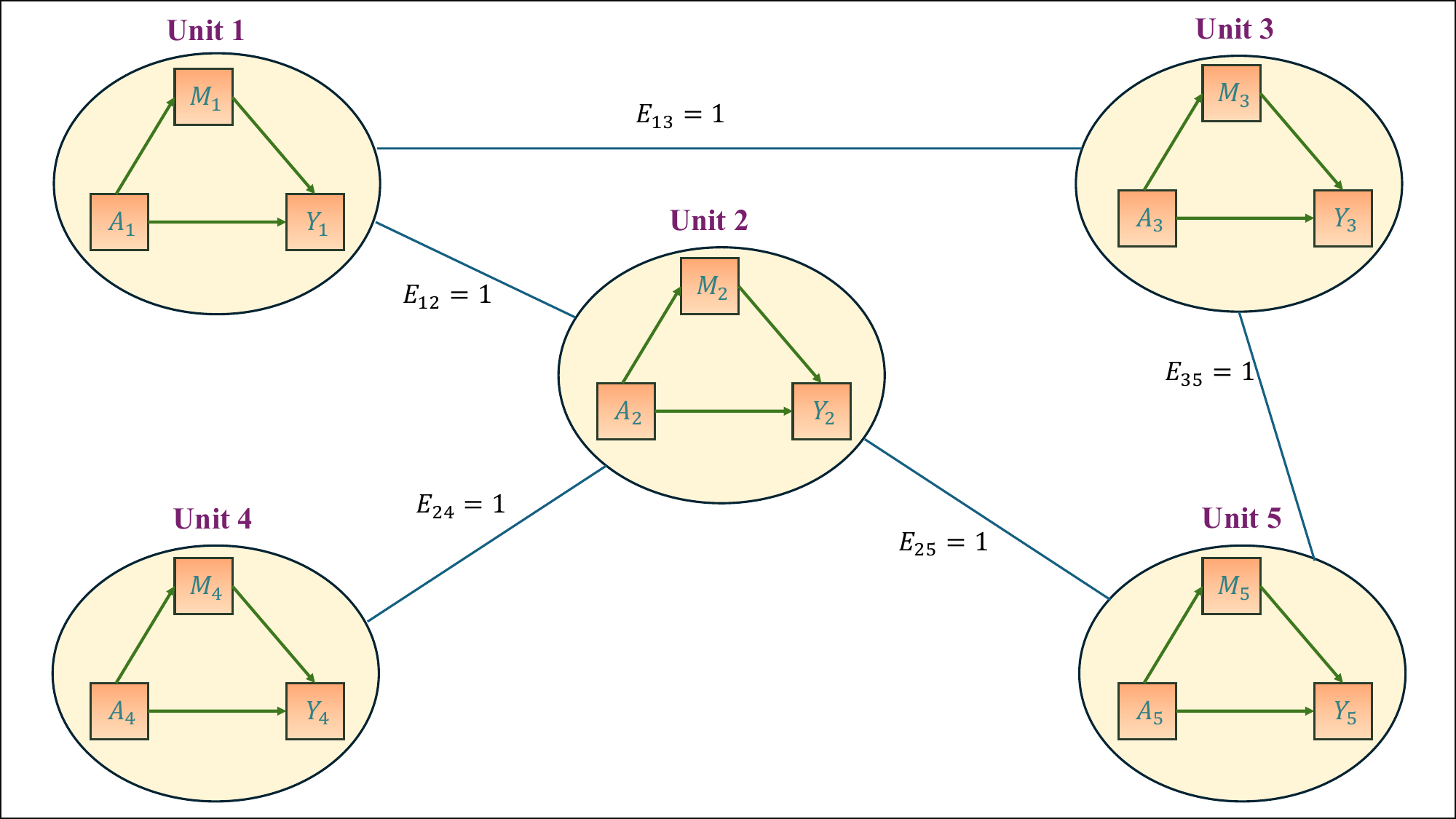}
 \caption{We consider a setup where $N=5$. Each node or DAG-clique consists of a triplet $(A,M,Y)$. The relationship between a DAG-clique $i$ and $j$, where $i\neq j$ and $i,j\in\mathcal{V}=\{1,2,3,4,5\}$ is characterized by the elements $E_{ij}$ of the adjacency matrix $E$ of network connectivity.}
   \label{fig:DAG1}
\end{figure}

In observational settings, causal mediation analysis becomes substantially more intricate when units are embedded within a social network. Let $\mathcal{V} = \{1, \dots, N\}$ denote a set of $N$ units connected by edges, with the network structure represented by a binary adjacency matrix $E \in \{0,1\}^{N \times N}$. Here, $E_{ij} = 1$ indicates a direct connection between nodes $i$ and $j$, and we adopt the convention that $E_{ii} = 1$. For each unit $i$, we define the set of first-degree neighbors as $\mathcal{N}^\dagger_i = \{j \neq i : E_{ij} = 1\}$ and the local community (or closed neighborhood) as $\mathcal{N}_i = \mathcal{N}^\dagger_i \cup \{i\}$. The set of second-degree neighbors, denoted $\mathcal{N}^\ddagger_i$, comprises units connected to $i$ strictly through $\mathcal{N}^\dagger_i$: $\mathcal{N}^\ddagger_i = \{k \notin \mathcal{N}_i : \exists j \in \mathcal{N}^\dagger_i \text{ such that } E_{jk} = 1\}$.  We denote the cardinality of these sets as $n_i = |\mathcal{N}_i^{\dagger}|$ (the degree of unit $i$) and $n^{(2)}_i = |\mathcal{N}_i^{\ddagger}|$. Figure~\ref{fig:DAG1} illustrates a representative network with $N = 5$; for instance, unit 1 has first-degree neighbors $\mathcal{N}^\dagger_1 = \{2,3\}$ and second-degree neighbors $\mathcal{N}^\ddagger_1 = \{4,5\}$.
\par
As illustrated in Figure~\ref{fig:DAG1}, the causal structure for each unit $i$ consists of a directed acyclic graph (DAG) defined by three variables: a binary exposure $A_i \in \{0,1\}$, a univariate continuous mediator $M_i \in \mathbb{R}$, and a continuous outcome $Y_i \in \mathbb{R}$. Additionally, let $\boldsymbol{C}_i$ denote a vector of $p$ observed confounders—which may be continuous or categorical—that influence the variables $(A_i, M_i, Y_i)$, thereby confounding the exposure-mediator, exposure-outcome, and mediator-outcome relationships. Throughout this work, we restrict our attention to binary exposures and univariate continuous mediators.

\noindent
In classical mediation analysis, the primary objective is to estimate the natural direct effect (NDE) and natural indirect effect (NIE) of an exposure \( A \) on an outcome \( Y \), where the NIE operates through a mediator \( M \). The NDE captures the direct pathway \( A \rightarrow Y \), while the NIE quantifies the indirect pathway \( A \rightarrow M \rightarrow Y \). Foundational work by \citet{baron1986moderator}, followed by formal causal developments in \citet{robins1992identifiability}, laid the groundwork for an extensive literature, including influential contributions by \citet{10.5555/2074022.2074073, robins2003semantics, rubin2004direct, vanderweele2008simple, vanderweele2009marginal, imai2010identification, ding2024first}. These frameworks generally assume independently sampled units, with no interference between them (i.e., \( E_{ij} = 0 \) for all \( i \neq j \)).\\

\noindent
In social network settings, interference introduces substantial complexity: the exposure or treatment assigned to one unit can influence the outcomes of neighboring units through social or physical connections, giving rise to a phenomenon known as the \emph{spillover effect}, represented as \(\boldsymbol{A}_{\mathcal{N}^{\dagger}_i} \rightarrow Y_i\). Classical causal inference relies on the Stable Unit Treatment Value Assumption (SUTVA) \citep{rubin1980randomization}, which precludes spillover effects by assuming that each unit’s outcome depends solely on its own exposure. However, recent work has increasingly relaxed this assumption to accommodate interference. Key contributions include methodological developments for causal inference in network-based randomized trials \citep{hudgens2008toward, tchetgen2012causal, manski2013identification, aronow2017estimating, leung2020treatment, savje2021average, hu2022average}, as well as extensions to observational settings \citep{tchetgen2021auto, ogburn2022causal}. Several of these studies adopt exposure mapping techniques to construct lower-dimensional representations of neighborhood-level exposures, facilitating the identification of causal effects in the presence of interference \citep{manski2013identification, aronow2017estimating, leung2022unconfoundedness}.\\

\noindent
However, the aforementioned studies were restricted to settings involving only exposures and outcomes. The complexity of causal inference under interference is substantially magnified when mediators are incorporated into the analysis. For instance, in Figure~\ref{fig:DAG1}, altering the neighboring exposure \(A_2\) impacts \(Y_1\) both directly and indirectly through both self mediator \(M_1\) and neighboring mediator \(M_2\) via their network ties—pathways that are not captured by classical mediation models. Properly addressing such network-induced interdependencies necessitates the development of new estimands, identification conditions, and inferential strategies. Although recent extensions of mediation analysis to networked settings—such as treating networks as mediators \citep{sweet2019modeling, liu2021social} or embedding them within exposure decomposition frameworks \citep{shpitser2017modeling, vanderweele2012components, vanderweele2013mediation}—represent important progress, critical gaps remain. Specifically, while the framework of \citet{vanderweele2013mediation} addresses spillover effects mediated through neighboring mediators in group-randomized trials, and \citet{ohnishi2025bayesian} provide a Bayesian nonparametric approach to mediation and spillover in cluster randomized trials, these methods are restricted to experimental settings. \\

\noindent
Consequently, there remains a lack of statistical theory for observational studies that simultaneously accounts for mediators and network interference. Developing a unified framework for this setting is non-trivial, primarily due to the complex dependence structures and identification challenges inherent in observational network data. To bridge this gap, we introduce a rigorous theoretical framework designed to identify and analyze mechanistic pathways involving both mediation and network interference. This approach enables a comprehensive decomposition of joint effects, thereby expanding the applicability of causal inference methods to complex network data.\\

\noindent
The remainder of this article is organized as follows. In Section~\ref{sec:ren-sem}, we propose the Random Effects Network Structural Equation Model (REN-SEM), a novel extension of the classical mediation SEM framework designed to capture the direct and indirect influences of neighboring exposures and mediators. Section~\ref{sec:potout} generalizes the exposure mapping framework to explicitly define spillover effects involving both neighborhood exposures and mediators. Building on this, Section~\ref{sec:effects} introduces a set of new mediation and interference estimands that extend beyond the classical natural direct and indirect effects. We establish the identifiability conditions and derive the general identification formula in Section~\ref{sec:iden}. In Section~\ref{sec:lren-sem}, we specialize our framework to the class of Linear REN-SEMs (LREN-SEM), deriving closed-form expressions for the causal estimands. Section~\ref{sec:est} develops the corresponding maximum likelihood estimation procedures, establishing asymptotic consistency and variance estimation under non-i.i.d. network dependence. Section~\ref{sec:simu} presents simulation results assessing finite-sample performance, followed by an application to the Twitch Gamers Social Network in Section~\ref{sec:twitch}. Concluding remarks are provided in Section~\ref{sec:con}, with technical proofs and supplementary details available in the Supplementary Material.

\section{Framework}

\subsection{Random Effects Network Structural Equation Model (REN-SEM)}\label{sec:ren-sem}
\begin{figure}
\centering
 \includegraphics[width=0.7\linewidth]{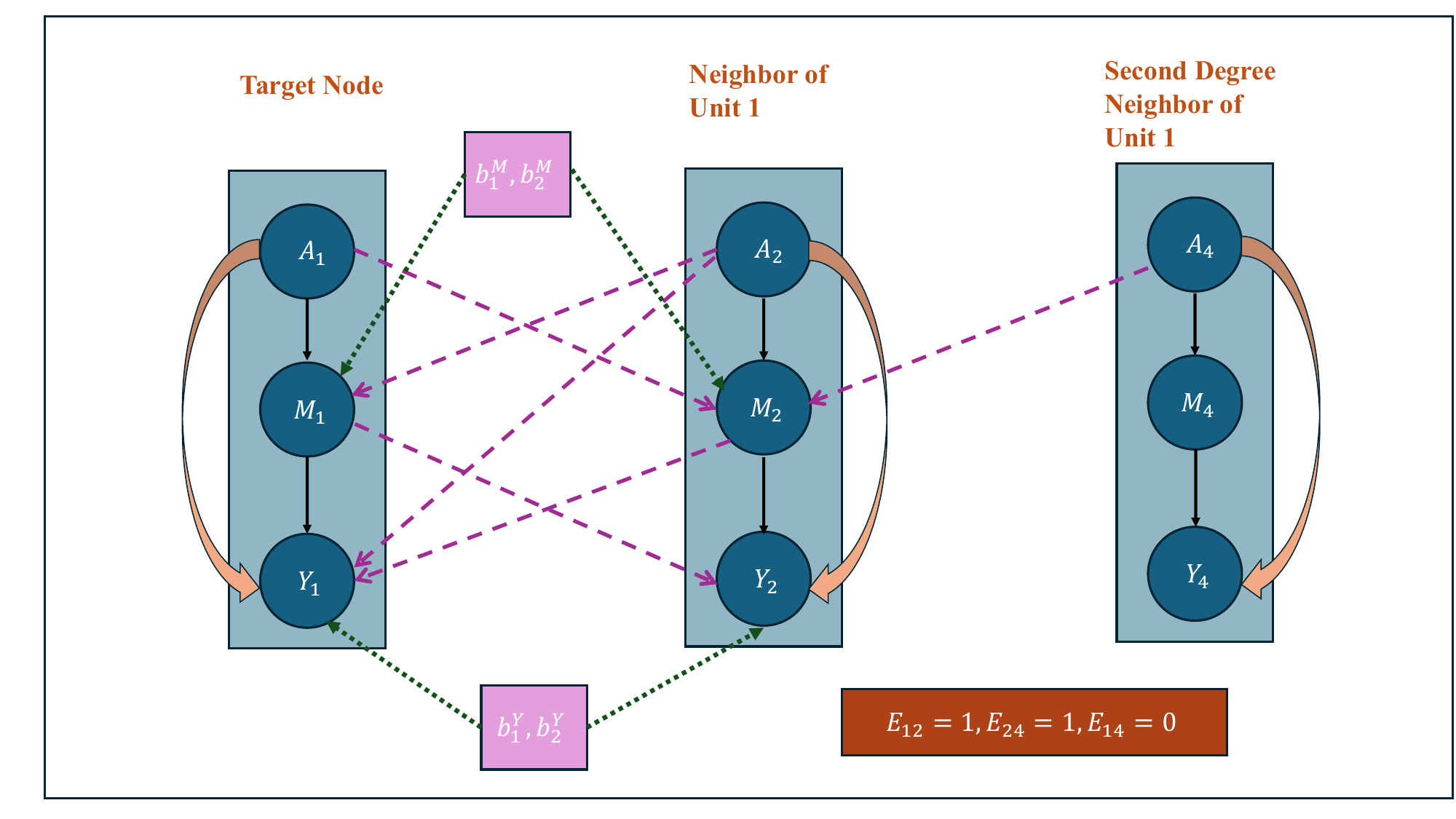}
 \caption{Different mechanistic pathways arising from the exposures of unit 1, its first degree and second degree neighbors, to the outcome of unit 1, where units 2 and 4 are regarded as representatives of first and second degree neighbors of unit 1, respectively, according to the network topology given in Figure~\ref{fig:DAG1}.}
   \label{fig:DAG2}
\end{figure}

To formally characterize the mechanistic pathways through which the exposure vector $\boldsymbol{A} = (A_1, \dots, A_N)$ influences individual outcomes in the presence of both mediation and network interference, we extend the classical Structural Equation Modeling (SEM) framework. We term this generalized approach the Random Effects Network Structural Equation Model (REN-SEM). The proposed REN-SEM framework defines a system of structural equations where the relationships within each unit-specific DAG $(A_i, M_i, Y_i)$ are modeled as functions of self-confounders $\boldsymbol{C}_i$, as well as the variables and confounders of neighboring DAGs, denoted by $(\boldsymbol{A}_{\mathcal{N}^{\dagger}_i}, \boldsymbol{M}_{\mathcal{N}^{\dagger}_i}, \boldsymbol{Y}_{\mathcal{N}^{\dagger}_i})$ and $\boldsymbol{C}_{\mathcal{N}^{\dagger}_i}$, respectively, for all units $i = 1, \dots, N$. Additionally to explicitly account for the network dependencies within the mediator vector $\boldsymbol{M} = (M_1, \dots, M_N)$ and the outcome vector $\boldsymbol{Y} = (Y_1, \dots, Y_N)$, we incorporate unit-specific random effects $b^m_i \in \mathbb{R}$ and $b^y_i \in \mathbb{R}$. This approach leverages latent variables to capture unobserved heterogeneity and correlation structures, similar to the strategy employed by \citet{bind2016causal} for longitudinal mediation. Under a fixed network structure, all distributional assumptions, as well as mediation and interference estimands, are specified conditional on a given network adjacency matrix \(E\). This approach aligns with the popular analytical framework in the current literature considered by, for example, \citet{ogburn2022causal}. The REN-SEMs is defined by the system of equations presented in (\ref{eq:SEM}). This formulation encodes a first-order Markov interference assumption, restricting the scope of network dependence to a unit's immediate neighbors. For each unit $i=1, \ldots, N$ within a network characterized by the adjacency matrix $E$, the model is given by:
\begin{equation}
  \begin{cases}
    \! 
    \begin{alignedat}{2}
      Y_i & =f_y(A_i,\boldsymbol A_{\mathcal{N}^{\dagger}_i}, M_i,
   \boldsymbol  M_{\mathcal{N}^{\dagger}_i},\boldsymbol C_i,\boldsymbol C_{\mathcal{N}^{\dagger}_i}, b^y_i,\boldsymbol b^y_{\mathcal{N}^{\dagger}_i},\epsilon^y_i,E), 
      \\
      M_i & = f_m(A_i,\boldsymbol A_{\mathcal{N}^{\dagger}_i},C_i,\boldsymbol C_{\mathcal{N}^{\dagger}_i},b^m_i,\boldsymbol b^m_{\mathcal{N}^{\dagger}_i},\epsilon^m_i,E),
      \\
       A_i &= f_a(\boldsymbol C_i,\boldsymbol C_{\mathcal{N}^{\dagger}_i},\epsilon^a_i,E),
    \end{alignedat}
  \end{cases}\label{eq:SEM}
\end{equation} 
where $\epsilon^y_i,\epsilon^m_i,\epsilon^a_i$ are the corresponding error (or noise) terms for the distributions of the triplets  $(Y_i,M_i,A_i)$ in a DAG of target node $i$, respectively; $f_y(.),f_m(.), f_a(.)$ are generic notations for the respective machinery to generate the three outputs given the inputs and random noises on the right-hand side of the equations. Regarding the specification of the exposure model in \eqref{eq:SEM}, we assume the absence of exposure-specific random effects. This is formally characterized by the following conditional independence structure on the exposure assignment mechanism:
\begin{assumption}\label{ass:assexp}
    For any $i \neq j$ in $\mathcal{V}$,
    $A_i\independent A_j|\boldsymbol C_i,\boldsymbol C_{\mathcal{N}^{\dagger}_i},\boldsymbol C_j,\boldsymbol C_{\mathcal{N}^{\dagger}_j}.$
\end{assumption}
\noindent
Assumption \ref{ass:assexp} implies that conditional on self and neighboring confounders, individual exposures are independent. 
Consequently, the REN-SEMs in (\ref{eq:SEM}) imply that the exposure vector $\boldsymbol{A}=(A_1,\cdots,A_N)$ affects the outcome $Y$ of a focal unit via seven distinct causal pathways. 
Such effects are present through seven pathways given as follows: $A\rightarrow Y$;  $A\rightarrow M \rightarrow Y$; $A\rightarrow \boldsymbol M_{\mathcal{N}^{\dagger}}\rightarrow Y$; $\boldsymbol A_{\mathcal{N}^{\dagger}}\rightarrow Y$; $\boldsymbol A_{\mathcal{N}^{\dagger}}\rightarrow M\rightarrow Y$; $\boldsymbol A_{\mathcal{N}^{\dagger}}\rightarrow \boldsymbol M_{\mathcal{N}^{\dagger}}\rightarrow Y$ and $\boldsymbol A_{\mathcal{N}^\ddagger}\rightarrow \boldsymbol M_{\mathcal{N}^{\dagger}}\rightarrow Y$. Note that the exposure vector of second-degree neighbors, $\boldsymbol A_{\mathcal{N}^\ddagger_i}$, influences the outcome $Y_i$ exclusively through the mediator vector of the first-degree neighbors, $\boldsymbol M_{\mathcal{N}^{\dagger}_i}$.
To pictorially elucidate these seven paths, Figure \ref{fig:DAG2} depicts the causal structure imposed by the 5-node network topology of Figure \ref{fig:DAG1}. This visualization highlights the distinct mechanistic pathways influencing the outcome $Y_1$ of the focal unit 1, originating from its own exposure $A_1$, as well as the exposures of its first-degree neighbor (unit 2 in this example) and second-degree neighbor (unit 4 in this example).  Specifically, the seven pathways are identified as: $A_1\rightarrow Y_1$; $A_1\rightarrow M_1 \rightarrow Y_1$; $A_1\rightarrow M_2 \rightarrow Y_1$; $A_2\rightarrow Y_1$; $A_2 \rightarrow M_1\rightarrow Y_1$; $A_2 \rightarrow M_2\rightarrow Y_1$; and finally, $A_4 \rightarrow M_2\rightarrow Y_1$. Next we provide an example for REN-SEM.

\subsection*{Example : Linear Random Effects Network SEM (LREN-SEM)}
In this section, we introduce a special class of REN-SEMs. For any vector $\boldsymbol V$ of $\text{dim}(\boldsymbol V)=n_i$, where $n_i=|\mathcal{N}_i|=\sum_{i=1}^NE_{ij}>0$,
define an operator $S_{1i}(.):\mathbb{R}^{n_i} \rightarrow \mathbb{R}$ by $S_{1i}(\boldsymbol V)=\frac{1}{n_i} \sum_{j\neq i}E_{ij}V_j$, a weighted average of neighboring elements in $\boldsymbol V$ that connect to unit $i$. Moreover, for any $n_i\times p$ matrix $\boldsymbol V=(\boldsymbol V_1,\boldsymbol V_2,\cdots,\boldsymbol V_p)$, we apply the operator $ S_{1i}$ on each column vector, yielding a $p$-dimensional operator 
$\boldsymbol S_{1i}(\boldsymbol V)= [S_{1i}(\boldsymbol V_1), \cdots, S_{1i}(\boldsymbol V_p)]$. In this example, we specify the outcome generative machinery $f_y(\cdot)$ by a linear random-effects model of the form: 
\begin{align}
    & Y_i=\beta_0+\beta_1A_i+\beta_2S_{1i}(\boldsymbol A_{\mathcal{N}^{\dagger}_i})+\beta_3M_i +\beta_4S_{1i}(\boldsymbol M_{\mathcal{N}^{\dagger}_i})+\boldsymbol\beta_5^T \boldsymbol C_i+\boldsymbol\beta_6^T \boldsymbol S_{1i}(\boldsymbol C_{\mathcal{N}^{\dagger}_i}) + E_{i.}^{T} \boldsymbol b^y+\epsilon_i^y \label{eq:eqY}
\end{align}
where $\boldsymbol \epsilon^y =(\epsilon_1^y,\ldots, \epsilon_N^y)^T\sim \text{MVN}(\boldsymbol 0,\sigma_y^2\mathcal{I}_{N\times N})$ and $\boldsymbol b^y = (b_1^y, \ldots, b_N^y)^T \sim \text{MVN}(\boldsymbol 0,\sigma_{b^y}^2\mathcal{I}_{N\times N})$ and $E_{i.}$ denotes the $i^{\text{th}}$ row of the adjacency matrix $E$. Next the mediator generative machinery $f_m(\cdot)$ also takes a linear random-effects model:
\begin{align}
        & M_i=\gamma_0+\gamma_1A_i+\gamma_2 S_{1i}(\boldsymbol A_{\mathcal{N}^{\dagger}_i})+ \boldsymbol\gamma_3^T \boldsymbol C_i +\boldsymbol\gamma_4^T  \boldsymbol S_{1i}(\boldsymbol C_{\mathcal{N}^{\dagger}_i}) + E_{i.}^{T}\boldsymbol b^m+\epsilon_i^m, \label{eq:eqM}
\end{align}
where $\boldsymbol \epsilon^m = (\epsilon_1^m,\ldots, \epsilon_N^m)^T \sim \text{MVN}(\boldsymbol 0,\sigma_m^2\mathcal{I}_{N\times N})$ and $\boldsymbol b^m = (b_1^m, \ldots, b_N^m)^T \sim \text{MVN}(\boldsymbol 0,\sigma_{b^m}^2\mathcal{I}_{N\times N})$. In the above models $\boldsymbol \epsilon^y,\boldsymbol \epsilon^m,\boldsymbol b^y,\boldsymbol b^m$ are assumed to be mutually independent. Finally, machinery $f_a(\cdot)$ is specified as:  $A_i|\boldsymbol C_i,\boldsymbol S_{1i}(\boldsymbol C_{\mathcal{N}^{\dagger}_i})$ follows a Bernoulli distribution with
\begin{align}
       P\left[A_i=1|\boldsymbol C_i,\boldsymbol S_{1i}(\boldsymbol C_{\mathcal{N}^{\dagger}_i})\right]=f\left(\boldsymbol C_i,\boldsymbol S_{1i}(\boldsymbol C_{\mathcal{N}^{\dagger}_i}),\boldsymbol \alpha \right), \label{eq:eqA}     
\end{align}
where $f(.)$ is a certain function parameterized by $\boldsymbol \alpha$. The resulting system of models (\ref{eq:eqY}), (\ref{eq:eqM}) and (\ref{eq:eqA}) is called \emph{Linear REN-SEM (LREN-SEM)} in this paper. 

\subsection{Potential Outcomes and Extended Exposure Mappings}\label{sec:potout}
Let $M_i(\boldsymbol{a})$ denote the potential value of the mediator for unit $i$ given the exposure vector $\boldsymbol{A}=\boldsymbol{a} \in \{0,1\}^N$, mapping $\{0,1\}^N \to \mathbb{R}$. Similarly, let $Y_i(\boldsymbol{a}, \boldsymbol{m})$ denote the potential outcome for unit $i$ when $\boldsymbol{A}$ is set to $\boldsymbol{a}$ and the mediator vector $\boldsymbol{M}$ is set to $\boldsymbol{m}$, where $Y_i(\cdot, \cdot): \{0,1\}^N \times \mathbb{R}^N \to \mathbb{R}$. Furthermore, $Y_i(\boldsymbol{a}, \boldsymbol{M}(\boldsymbol{a'}))$ represents the potential outcome if $\boldsymbol{A}$ were set to $\boldsymbol{a}$ and $\boldsymbol{M}$ took the value it would naturally assume under $\boldsymbol{A}=\boldsymbol{a'}$. However, causal inference based on the simultaneous manipulation of the full, high-dimensional vector $\boldsymbol{A}$ is often intractable due to severe positivity violations and the practical impossibility of such global interventions. Instead, realistic inquiries typically focus on lower-dimensional summaries of the network exposure. To rigorously disentangle individual from peer effects in a tractable manner, we extend the exposure mapping framework—originally established by \citet{manski2013identification}—to incorporate interference mechanisms operating through both exposures and mediators.\\

\noindent
\textbf{Exposure Mapping Functions for Mediators:} Based on the REN-SEMs system in \eqref{eq:SEM} we observe that the mediator $M_i$ depends on the global exposure vector $\boldsymbol{A}$ solely through the unit's own exposure $A_i$ and the exposures of its immediate neighbors, $\boldsymbol{A}_{\mathcal{N}^{\dagger}_i}$.  To operationalize this local dependence parsimoniously, we assume that the influence of the high-dimensional neighbor vector $\boldsymbol{A}_{\mathcal{N}^{\dagger}_i}$ is transmitted entirely through a lower-dimensional summary function, or exposure mapping, denoted by $\boldsymbol T^a_{1i}(\cdot): \{0,1\}^{n_i} \rightarrow \mathcal{A}_{1i}$, where $\text{dim}(\mathcal{A}_{1i}) \leq n_i$. Mathematically, this dimension reduction implies that for any realization $\boldsymbol a = (a_i,\boldsymbol a_{\mathcal{N}^{\dagger}_i}, \boldsymbol a_{-\mathcal{N}^{\dagger}_i})$, the mediator mapping can be expressed as,
\begin{equation}
    M_i(\boldsymbol a)=\widetilde{M_i}(a_i,\boldsymbol T^a_{1i}(\boldsymbol a_{\mathcal{N}^{\dagger}_i})), ~ a_i \in \{0,1\}, \boldsymbol a_{\mathcal{N}^{\dagger}_i} \in \{0,1\}^{n_i}, \label{eq:eq1} 
\end{equation}
where $\boldsymbol T^a_{1i}(.)$ may be regarded as of a sufficient dimension reduction on the full exposures involving only the community of first-degree neighbors, and the resulting mapping $\widetilde{M}_i(\cdot,\cdot):\{0,1\}\times \mathcal{A}_{1i}\rightarrow \mathbb{R}$. In the context of the LREN-SEM in (\ref{eq:eqY})-(\ref{eq:eqA}), this mapping $\boldsymbol T^a_{1i}(.)$ in model \eqref{eq:eqM} takes a simple form: the average number of first-degree neighbors receiving the exposure. Specifically,  for node $i$, we have mapping $\boldsymbol T^a_{1i}(\boldsymbol a_{\mathcal N^{\dagger}_i}) = S_{1i}(\boldsymbol a_{\mathcal N^{\dagger}_i})\cdot$\\

\noindent
\textbf{Extended Exposure Mapping Functions for Outcomes.} The REN-SEM system \eqref{eq:SEM} implies that the outcome $Y_i$ is dependent on the unit's own exposure and mediator, the vectors of first-degree neighbor exposures and mediators, and the vector of second-degree neighbor exposures (only through mediators of first-degree neighbors). To render this dependence tractable, we assume that $Y_i$ is influenced by these neighborhood vectors solely through lower-dimensional aggregate summaries. Specifically, we employ the summary functions $\boldsymbol T^a_{1i}(\cdot)$ defined in \eqref{eq:eq1}, along with $\boldsymbol T^a_{2i}: \{0,1\}^{n_i} \to \mathcal{A}_2$, $\boldsymbol T^a_{3i}: \{0,1\}^{n^{(2)}_i} \to \mathcal{A}_3$, and $\boldsymbol T_{1i}^m: \mathbb{R}^{n_i} \to \mathcal{M}_1$, where the dimensions of the co-domains satisfy $\text{dim}(\mathcal{A}_2), \text{dim}(\mathcal{M}_1) \leq n_i$ and $\text{dim}(\mathcal{A}_3) \leq n^{(2)}_i$. Mathematically, once again we utilize the strategy of sufficient dimension reduction by imposing the following simplification: With the decomposition $\boldsymbol a = (a_i , \boldsymbol a_{\mathcal N^{\dagger}_i},\boldsymbol a_{\mathcal N^{\ddagger}_i}, \boldsymbol a_{-(\mathcal N_i\cup \mathcal N^{\ddagger}_i)})$,
\begin{align}
    Y_i(\boldsymbol a)=Y_i(\boldsymbol a,\boldsymbol M(\boldsymbol a))& =\widetilde{Y}_i(a_i,\boldsymbol T^a_{1i}(\boldsymbol a_{\mathcal{N}^{\dagger}_i}),M_i(\boldsymbol a),\boldsymbol T^m_{1i}\circ \boldsymbol M_{\mathcal{N}^{\dagger}_i}(\boldsymbol a)) \label{eq:eq2} \\
     & =\widetilde{Y}_i(a_i,\boldsymbol T^a_{1i}(\boldsymbol a_{\mathcal{N}^{\dagger}_i}),\widetilde{M}_i(a_i,\boldsymbol T^a_{1i}(\boldsymbol a_{\mathcal{N}^{\dagger}_i})),\widetilde{\boldsymbol  T}^m_{1i}(a_i,\boldsymbol T^a_{2i}(\boldsymbol a_{\mathcal{N}^{\dagger}_i}),\boldsymbol T^a_{3i}(\boldsymbol a_{\mathcal{N}^{\ddagger}_i}))) \cdot\nonumber
\end{align}
where the resulting mappings, $\widetilde{Y_i}(\cdot):\{0,1\}\times \mathcal{A}_1\times \mathbb{R}\times \mathcal{M}_1\rightarrow \mathbb{R}$, $\widetilde{\boldsymbol  T}^m_{1i}(\cdot):\{0,1\}\times \mathcal{A}_2\times \mathcal{A}_3 \rightarrow \mathcal{M}_1$ and $``\circ"$ denotes the operator for the compound of two functions. Thus, the neighborhood summary $\boldsymbol T^a_{1i}(\boldsymbol a_{\mathcal{N}^{\dagger}_i})$, defined in \eqref{eq:eq1}, influences the outcome both directly and indirectly through the individual mediator $M_i$. In contrast, $\boldsymbol T^a_{2i}(\boldsymbol a_{\mathcal{N}^{\dagger}_i})$ and the second-degree summary $\boldsymbol T^a_{3i}(\boldsymbol a_{\mathcal{N}^{\ddagger}_i})$ affect the outcome exclusively through the neighborhood mediator summary function $\widetilde{\boldsymbol T}^m_{1i}(\cdot)$. \\

\noindent
\textit{Example:} In the LREN-SEM specified by \eqref{eq:eqY} and \eqref{eq:eqM}, it is easy to show that for unit $i$, the summary functions take the following forms:
\begin{align}
    & \boldsymbol T^a_{1i}(\boldsymbol a_{\mathcal{N}^{\dagger}_i})=S_{1i}(\boldsymbol a_{\mathcal{N}^{\dagger}_i}), \quad \boldsymbol T^a_{2i}(\boldsymbol a_{\mathcal{N}^{\dagger}_i})=\left(S_{1i}(\boldsymbol a_{\mathcal{N}^{\dagger}_i}),S_{2i}(\boldsymbol a_{\mathcal{N}^{\dagger}_i}):=\sum_{j\neq i}\frac{1}{n_i n_j} \sum_{k\neq j,k\in \mathcal{N}^{\dagger}_i} E_{ij}E_{jk}a_k\right),\nonumber\\
    & \boldsymbol T^a_{3i}(\boldsymbol a_{\mathcal{N}^{\ddagger}_i})=S_{3i}(\boldsymbol a_{\mathcal{N}^{\dagger}_i}):=\sum_{j\neq i}\frac{1}{n_i n_j} \sum_{k\neq j,k\in \mathcal{N}^\ddagger_i} E_{ij}E_{jk}a_k,\nonumber\\
    & \boldsymbol T^m_{1i}\circ\boldsymbol M_{\mathcal{N}^{\dagger}_i}(\boldsymbol a)=S^{m}_{1i}(a_i,S_{1i}^a(\boldsymbol a_{\mathcal{N}^{\dagger}_i}),S_{2i}^a(\boldsymbol a_{\mathcal{N}^{\dagger}_i}),S_{3i}^a(\boldsymbol a_{\mathcal{N}^{\ddagger}_i})):=S_{1i}\circ\boldsymbol M_{\mathcal{N}^{\dagger}_i}(\boldsymbol a),\label{eq:lsum}
\end{align}
We distinguish between the primary summary functions, $\boldsymbol T^a_{1i}$ and $\widetilde{\boldsymbol T}^m_{1i}$, which are investigator-specified to operationalize the causal hypotheses of substantive interest, and the secondary functions $\boldsymbol T^a_{2i}$ and $\boldsymbol T^a_{3i}$. The latter typically emerge as structural byproducts of the network topology and the primary specifications, serving as necessary adjustment controls rather than independent targets of intervention. We further emphasize that while the preceding example utilizes linear aggregations, the proposed framework is flexible: depending on the scientific context, these mappings may be readily specified as multi-dimensional or non-linear functions.

\subsection{Consistency Assumption}
Next we state the consistency assumption that is one of the fundamental identifiability conditions in the potential outcomes paradigm. For notational convenience, we denote the observed exposure summary measures as $\boldsymbol T^a_{ki} := \boldsymbol T^a_{ki}(\boldsymbol A_{\mathcal{N}^{\dagger}_i})$ for $k \in \{1, 2\}$, and $\boldsymbol T^a_{3i} := \boldsymbol T^a_{3i}(\boldsymbol A_{\mathcal{N}^{\ddagger}_i})$.

\begin{assumption}\label{ass:asscons}
    (Consistency Assumption) In a network of $N$ units connected by a given adjacency matrix $E =(E_{ij})$, the potential outcomes satisfy: 
\begin{itemize}
    \item[(i)]$\widetilde{M}_i(a,\boldsymbol t^a_{1})$ in equation \eqref{eq:eq1} equals to observed $M_i$ when observed $(A_i,\boldsymbol T^a_{1i})=(a,\boldsymbol t^a_{1})$ for any $a \in \{0,1\}$ and $\boldsymbol t^a_{1}\in \text{Supp}(\boldsymbol T^a_{1i})$;
    \item[(ii)] $\widetilde{\boldsymbol  T}^m_{1i}(a,\boldsymbol t^a_{2},\boldsymbol t^a_{3})$ equals to observed $\widetilde{\boldsymbol  T}^m_{1i}$ when observed $(A_i,\boldsymbol T^a_{2i},\boldsymbol T^a_{3i})=(a,\boldsymbol t^a_{2},\boldsymbol t^a_{3})$ for any $a \in \{0,1\}$, $\boldsymbol t^a_{2} \in \text{Supp}(\boldsymbol T^a_{2i})$ and $\boldsymbol t^a_{3} \in \text{Supp}(\boldsymbol T^a_{3i})$.
    \item[(iii)] $\widetilde{Y}_i(a,\boldsymbol t_1^a,m,\boldsymbol  t^m_{1})$ in equation \eqref{eq:eq2} equals to observed $Y_i$ when observed 
    $(A_i,\boldsymbol T_{1i}^a,M_i,\widetilde{\boldsymbol T}^m_{1i})=(a,\boldsymbol t_1^a,m,\boldsymbol t^m_{1})$ for any $a \in \{0,1\}$, $\boldsymbol t_1^a \in \text{Supp}(\boldsymbol T^a_{1i})$, $m \in \mathbb{R}$ and $\boldsymbol  t^m_{1} \in \text{Supp}(\widetilde{\boldsymbol T}^m_{1i})$;
\end{itemize}
\end{assumption} 

\section{Definitions of Mediation and Interference Pathways}\label{sec:effects}

\subsection{Meaningful Interventions}
As formalized in Equation \eqref{eq:eq2}, interventions on the global exposure vector $\boldsymbol{A}$ influence the potential outcome $\widetilde{Y}_i(\cdot)$ via the sufficient summary tuple $(A_i, \boldsymbol{T}_{1i}^a, \boldsymbol{T}_{2i}^a, \boldsymbol{T}_{3i}^a)$. Here, $\boldsymbol{T}_{1i}^a$ and $\boldsymbol{T}_{2i}^a$ aggregate first-degree neighbor exposures, while $\boldsymbol{T}_{3i}^a$ summarizes second-degree exposures. In practice, however, enforcing simultaneous interventions across broad network communities is often precluded by administrative and financial constraints. Furthermore, structural components like $\boldsymbol{T}^a_{3i}$ (or secondary dimensions of $\boldsymbol{T}^a_{2i}$ in the LREN-SEM) often lack direct policy relevance, primarily because they cannot be manipulated independently of the primary summary $\boldsymbol{T}^a_{1i}$ given the network topology. Consequently, our objective is to evaluate the effect of $\boldsymbol{A}$ transmitted solely through the self-exposure $A_i$ and the investigator-specified summary $\boldsymbol{T}^a_{1i}$. To isolate these effects, we marginalize over the joint distribution of the nuisance components $(\boldsymbol{T}_{2i}^a, \boldsymbol{T}_{3i}^a)$ conditional on fixed values of $\{\boldsymbol{T}^a_{1i}, \boldsymbol X_i\}$, where $\boldsymbol X_i:=(\boldsymbol{C}_i, \boldsymbol{C}_{\mathcal{N}^{\dagger}_i}, \boldsymbol{C}_{\mathcal{N}^{\ddagger}_i})$. Under Assumption \ref{ass:assexp}, these neighbor summaries are independent of self-exposure $A_i$ conditional on $\boldsymbol X_i$. Moreover, the assumption implies that $\boldsymbol{T}^a_{3i} \independent (\boldsymbol{T}_{1i}^a, \boldsymbol{T}_{2i}^a) \mid \boldsymbol X_i$. This independence structure allows the marginalization to proceed in two tractable steps: first, integrating out $\boldsymbol{T}^a_{3i}$ conditional on $\boldsymbol X_i$, and second, integrating out $\boldsymbol{T}^a_{2i}$ conditional on the specific realization of $\boldsymbol{T}_{1i}^a$ and $\boldsymbol X_i$. We now proceed to define the causal estimands of interest, beginning with the total effect.

\subsection{Key Estimands of Interest}
\begin{figure}[ht]
\centering
  \includegraphics[width=0.7\linewidth]{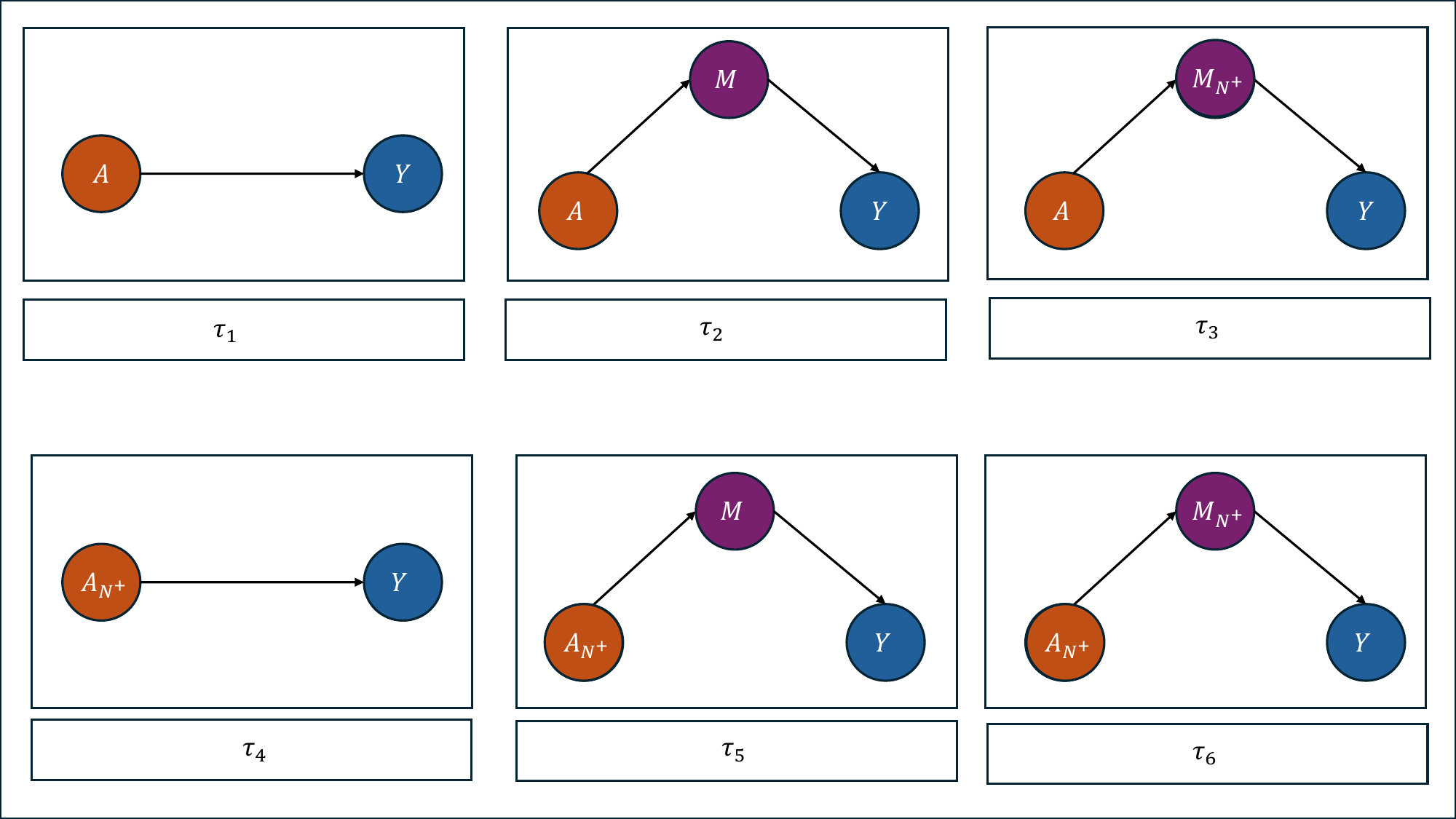}
\caption{Six DAGs represent the mediation and interference pathways for the corresponding six estimands, $\tau_1,\tau_2,\tau_3,\tau_4,\tau_5,\tau_6$, respectively with individual's self exposure ($A$), self mediator ($M$), self response ($Y$), neighborhood exposure ($\boldsymbol A_{\mathcal{N}^{\dagger}}$), neighbourhood mediator ($\boldsymbol M_{\mathcal{N}^{\dagger}}$). These DAGs do not contain the confounders for simplicity.}
\label{fig:indi_dags}
\end{figure}

We define the total effect corresponding to a change in the aggregate exposure $\boldsymbol D_i := (A_i, \boldsymbol{T}^a_{1i})$ from a reference level $\boldsymbol d' = (a', \boldsymbol t^{a'}_{1})$ to a target level $\boldsymbol d = (a, \boldsymbol t^{a}_{1})$, for any $a, a' \in \{0,1\}$ and $\boldsymbol t^{a}_{1}, \boldsymbol t^{a'}_{1} \in \text{Supp}(\boldsymbol{T}^a_{1i})$ for all $i \in \{1,2,\cdots,N\}$. To formalize the marginalization procedure described in the preceding section, we introduce the following notation. For any random vector $\boldsymbol V$, we define the operators
\begin{align}
    &\vec{\mathbb{E}}^a_{i}(\boldsymbol V)=\mathbb{E}_{\boldsymbol X_i}\left[\mathbb{E}_{\boldsymbol T^a_{3i}|\boldsymbol X_i}\left\{\mathbb{E}_{\boldsymbol T_{2i}^a|\boldsymbol T_{1i}^a=\boldsymbol t_{1}^a,\boldsymbol X_i}\mathbb{E}(\boldsymbol V|\boldsymbol X_i, \boldsymbol T_{3i}^a,\boldsymbol T_{2i}^a)\right\}\right],\label{eq:marg1}\\
    & \vec{\mathbb{E}}^{a'}_{i}(\boldsymbol V)=\mathbb{E}_{\boldsymbol X_i}\left[\mathbb{E}_{\boldsymbol T^a_{3i}|\boldsymbol X_i}\left\{\mathbb{E}_{\boldsymbol T_{2i}^a|\boldsymbol T_{1i}^a=\boldsymbol t_{1}^{a'},\boldsymbol X_i}\mathbb{E}(\boldsymbol V|\boldsymbol X_i, \boldsymbol T_{3i}^a,\boldsymbol T_{2i}^a)\right\}\right]\cdot\label{eq:marg2}
\end{align}
\begin{definition}\label{def:tot}
The \emph{first-order marginal total effect}, $\tau_{\text{tot}}(\boldsymbol d',\boldsymbol d)$ is given as follows:  
\begin{align*}
    &\frac{1}{N}\sum_{i=1}^N \left[\vec{\mathbb{E}}_i^a\left\{\widetilde{Y}_i(a,\boldsymbol t^{a}_{1},\widetilde{M}_i(a,\boldsymbol t^a_{1}),\widetilde{\boldsymbol  T}^m_{1i}(a,\boldsymbol T^a_{2i},\boldsymbol T^a_{3i}) \right\}-\vec{\mathbb{E}}^{a'}_{i}\left\{\widetilde{Y}_i(a',\boldsymbol t^{a'}_{1},\widetilde{M}_i(a',\boldsymbol t^{a'}_{1}),\widetilde{\boldsymbol  T}^m_{1i}(a',\boldsymbol T^a_{2i},\boldsymbol T^a_{3i}) \right\}\right],
\end{align*}
where the marginalization operation of $\vec{\mathbb{E}}$ is carried as described in equations \eqref{eq:marg1}-\eqref{eq:marg2}. This operation will also be repeated in the other definitions below that correspond to various mechanistic routes through which an individual's outcome may be influenced by self and neighbors' exposures within the network.
\end{definition}\label{sec:toteff}
\noindent
By decomposing the total effect across the pathways illustrated in Figure \ref{fig:indi_dags}, we derive six key estimands of interest. These quantities capture direct and mediated effects—both within-unit and via network interference—obtained after marginalizing over the secondary summary functions $\boldsymbol T^a_{2i}$ and $\boldsymbol T^a_{3i}$. We classify these estimands into five distinct categories, defined as follows. 
\begin{definition}\label{def:dir} 
The direct effect of $A$ on $Y$ $(A\rightarrow Y)$, denoted by $\tau_{1}(\boldsymbol d', \boldsymbol d)$, is  
\begin{align*}
    &\frac{1}{N}\sum_{i=1}^N \left[\vec{\mathbb{E}}_i^{a'}\left\{\widetilde{Y}_i(a,\boldsymbol t^{a'}_{1},\widetilde{M}_i(a',\boldsymbol t^{a'}_{1}),\widetilde{\boldsymbol  T}^m_{1i}(a',\boldsymbol T^a_{2i},\boldsymbol T^a_{3i}) \right\}-\vec{\mathbb{E}}^{a'}_{i}\left\{\widetilde{Y}_i(a',\boldsymbol t^{a'}_{1},\widetilde{M}_i(a',\boldsymbol t^{a'}_{1}),\widetilde{\boldsymbol  T}^m_{1i}(a',\boldsymbol T^a_{2i},\boldsymbol T^a_{3i}) \right\}\right],
\end{align*}
The estimand $\tau_1$ quantifies the average direct effect resulting from a change in the unit's own exposure, holding all external network conditions and mediators constant. This corresponds to the natural direct effect (NDE) in the classical mediation literature.
\end{definition}

\begin{definition}\label{def:med} In the network setting, two types of mediation effects from own exposure $A$ are defined.\\
\noindent
(a) The effect of $A$ on $Y$ via $M$ $(A\rightarrow M\rightarrow Y)$ denoted by $\tau_{2}(\boldsymbol d',  \boldsymbol d)$, is
\begin{align*}
    &\frac{1}{N}\sum_{i=1}^N \left[\vec{\mathbb{E}}_i^a\left\{\widetilde{Y}_i(a,\boldsymbol t^{a}_{1},\widetilde{M}_i(a,\boldsymbol t^a_{1}),\widetilde{\boldsymbol  T}^m_{1i}(a,\boldsymbol T^a_{2i},\boldsymbol T^a_{3i}) \right\}-\vec{\mathbb{E}}^{a}_{i}\left\{\widetilde{Y}_i(a,\boldsymbol t^{a}_{1},\widetilde{M}_i(a',\boldsymbol t^{a}_{1}),\widetilde{\boldsymbol  T}^m_{1i}(a,\boldsymbol T^a_{2i},\boldsymbol T^a_{3i}) \right\}\right],
\end{align*}
The estimand $\tau_2$ corresponds to the natural indirect effect (NIE) in classical mediation analysis with independent units. It captures the average mediation effect internal to an individual unit, isolated from any neighborhood influence.\\
\noindent
(b) The effect of $A$ on $Y$ via first-degree neighbors' mediators $\boldsymbol M_{\mathcal{N}^{\dagger}}$ $(A\rightarrow \boldsymbol M_{\mathcal{N}^{\dagger}}\rightarrow Y)$, denoted by $\tau_{3}(\boldsymbol d',\boldsymbol d)$, is given by 
\begin{align*}
   &\frac{1}{N}\sum_{i=1}^N \left[\vec{\mathbb{E}}_i^a\left\{\widetilde{Y}_i(a,\boldsymbol t^{a}_{1},\widetilde{M}_i(a',\boldsymbol t^a_{1}),\widetilde{\boldsymbol  T}^m_{1i}(a,\boldsymbol T^a_{2i},\boldsymbol T^a_{3i}) \right\}-\vec{\mathbb{E}}^{a}_{i}\left\{\widetilde{Y}_i(a,\boldsymbol t^{a}_{1},\widetilde{M}_i(a',\boldsymbol t^{a}_{1}),\widetilde{\boldsymbol T}^m_{1i}(a',\boldsymbol T^a_{2i},\boldsymbol T^a_{3i}) \right\}\right],
\end{align*}
The estimand $\tau_3$ constitutes a novel component that arises solely from network connectivity. It quantifies the effect of the focal unit's exposure $A_i$ mediated specifically through the mediators of the first-degree neighborhood.
\end{definition}
\begin{definition}\label{def:spill} The effect of $\boldsymbol A_{\mathcal{N}^{\dagger}}$ on $Y$ $(\boldsymbol A_{\mathcal{N}^{\dagger}}\rightarrow Y)$, denoted by $\tau_{4}(\boldsymbol d',\boldsymbol d)$ is:
\begin{align*}
    &\frac{1}{N}\sum_{i=1}^N \left[\vec{\mathbb{E}}_i^a\left\{\widetilde{Y}_i(a,\boldsymbol t^{a}_{1},\widetilde{M}_i(a',\boldsymbol t^{a}_{1}),\widetilde{\boldsymbol  T}^m_{1i}(a',\boldsymbol T^a_{2i},\boldsymbol T^a_{3i}) \right\}-\vec{\mathbb{E}}^{a}_{i}\left\{\widetilde{Y}_i(a,\boldsymbol t^{a'}_{1},\widetilde{M}_i(a',\boldsymbol t^{a}_{1}),\widetilde{\boldsymbol T}^m_{1i}(a',\boldsymbol T^a_{2i},\boldsymbol T^a_{3i}) \right\}\right],
\end{align*}
The estimand $\tau_4$, commonly referred to as the spillover effect (or peer effect) in the literature on interference, quantifies the average change in the outcome attributable to shifts in the aggregated exposure of the first-degree neighbors, holding all other factors fixed.
\end{definition}  
\begin{definition}\label{def:spillmed} 
(a) The effect of $\boldsymbol A_{\mathcal{N}^{\dagger}}$ on $Y$ via self mediator $M$ $(\boldsymbol A_{\mathcal{N}^{\dagger}}\rightarrow M \rightarrow Y)$, denoted by $\tau_{5}(\boldsymbol d',\boldsymbol d)$, is given by 
\begin{align*}
    &\frac{1}{N}\sum_{i=1}^N \left[\vec{\mathbb{E}}_i^a\left\{\widetilde{Y}_i(a,\boldsymbol t^{a'}_{1},\widetilde{M}_i(a',\boldsymbol t^{a}_{1}),\widetilde{\boldsymbol  T}^m_{1i}(a',\boldsymbol T^a_{2i},\boldsymbol T^a_{3i}) \right\}-\vec{\mathbb{E}}^{a}_{i}\left\{\widetilde{Y}_i(a,\boldsymbol t^{a'}_{1},\widetilde{M}_i(a',\boldsymbol t^{a'}_{1}),\widetilde{\boldsymbol T}^m_{1i}(a',\boldsymbol T^a_{2i},\boldsymbol T^a_{3i}) \right\}\right],
\end{align*}
The novel estimand $\tau_5$ captures a distinct mediated spillover effect. It quantifies the average change in the outcome resulting from a shift in the unit's own mediator ($M_i$), which is itself induced by changes in the aggregated exposure of the first-degree neighborhood.\\
\noindent
(b) The effect from neighborhood exposure $\boldsymbol A_{\mathcal{N}^{\dagger}}$ on $Y$ through neighborhood mediator $\boldsymbol M_{\mathcal{N}^{\dagger}}$ $(\boldsymbol A_{\mathcal{N}^{\dagger}}\rightarrow \boldsymbol M_{\mathcal{N}^{\dagger}} \rightarrow Y)$, denoted by $\tau_{6}(\boldsymbol d',\boldsymbol d)$, takes the form: 
\begin{align*}
    &\frac{1}{N}\sum_{i=1}^N \left[\vec{\mathbb{E}}_i^a\left\{\widetilde{Y}_i(a,\boldsymbol t^{a'}_{1},\widetilde{M}_i(a',\boldsymbol t^{a'}_{1}),\widetilde{\boldsymbol  T}^m_{1i}(a',\boldsymbol T^a_{2i},\boldsymbol T^a_{3i}) \right\}-\vec{\mathbb{E}}^{a'}_{i}\left\{\widetilde{Y}_i(a,\boldsymbol t^{a'}_{1},\widetilde{M}_i(a',\boldsymbol t^{a'}_{1}),\widetilde{\boldsymbol T}^m_{1i}(a',\boldsymbol T^a_{2i},\boldsymbol T^a_{3i}) \right\}\right],
\end{align*}
Finally, the novel estimand $\tau_6$ quantifies the neighborhood-mediated spillover effect. It measures the average change in the outcome driven by shifts in the first-degree neighbors' mediators, which are in turn induced by changes in those neighbors' own exposures. This effect is particularly complex as it encapsulates a causal process occurring entirely within the neighborhood community, involving both peer exposures and peer mediators.
\end{definition}

\noindent
The six estimands $\tau_1$ through $\tau_6$ are defined conditional on the sample size $N$ and the fixed adjacency matrix $E$, a formulation consistent with \citet{ogburn2022causal}. Consequently, these quantities are network-specific (or data-dependent), meaning their true values are intrinsically tied to the particular topology and size of the observed network. \\

\noindent
\textbf{Remark 1:} Another fundamental challenge in causal inference under network interference stems from the inherent structural constraints on the support of neighborhood summary measures derived from binary exposures. Specifically, for a fixed adjacency matrix, the discrete nature of the summary statistic $\boldsymbol T_1^a$ implies that not all theoretical intervention values are feasible for every unit.  Consider the LREN-SEM, where the component $S_{1i}^a=\sum_{j\neq i}\frac{E_{ij}A_{ij}}{n_i}$, represents the proportion of treated neighbors. If we define a causal contrast based on a shift from $\boldsymbol D' = (0, 0)$ to $\boldsymbol D = (1, 0.5)$, the target value $S_{1i}^a = 0.5$ is achievable only for units with even degrees (i.e., where $n_i$ is a multiple of two). Consequently, enforcing such an intervention strictly renders the counterfactual undefined for odd-degree nodes, implicitly subsetting the analysis and potentially yielding unrepresentative inferences. While one could restrict the set of interventions to those universally feasible (e.g., $S_{1i}^a \in \{0, 1\}$), this severely limits the scope of inquiry. A more robust alternative lies in the broad class of approximation strategies. Among the many possibilities in this category, one common approach is to define a valid range $S_{1i}^a \in [0.5 - \epsilon, 0.5 + \epsilon]$ for a given $\epsilon$ of choice.

\section{Identifiability}\label{sec:iden}
\subsection{Identifiability Conditions} \label{sec:identify}
Technically, estimating the six estimands in Definitions \ref{def:dir}-\ref{def:spillmed} requires generically estimating the following quantity under marginalization with various configurations of self and neighbors' exposures. Specifically, for any $a, a^*, a'' \in \{0,1\}$ and $\boldsymbol t_{1}^{a'}, \boldsymbol t_{1}^{a\dagger}, \boldsymbol t_{1}^{a \ddagger} \in \text{Supp}(\boldsymbol T_{1i}^a)$, 
\begin{align}
    \Psi_i&=\vec{\mathbb{E}}_i^{a \ddagger}\left\{\widetilde{Y_i}\left(a,\boldsymbol t_{1}^{a'},\widetilde{M_i}\left(a^*,\boldsymbol t_{1}^{a\dagger}\right),\widetilde{\boldsymbol T}^m_{1i}\left(a'',\boldsymbol T_{2i}^a,\boldsymbol T_{3i}^a\right)\right) \right\}. \label{eq:eq5}
\end{align}
Moreover, let $\boldsymbol t^a_{2} \in \text{Supp}\left(\boldsymbol  T_{2i}^a|\boldsymbol  T_{1i}^a=\boldsymbol t_{1}^{a \ddagger}\right)$,  $\boldsymbol t^a_{3} \in\text{Supp}(\boldsymbol T_{3i}^a)$ and  $\boldsymbol t_{1}^m \in \text{Supp}(\widetilde{\boldsymbol T}^m_{1i})$ respectively. We define the composite random effects vector for unit $i$ as $\boldsymbol u_i = (\boldsymbol b^y_{\mathcal{N}_i}, \boldsymbol b^m_{\mathcal{N}_i}, \boldsymbol b^m_{\mathcal{N}^\ddagger_i})$. To identify and estimate the quantity $\Psi_i$ in (\ref{eq:eq5}) in the class of REN-SEMs models, we assume the following identifiability conditions.

\begin{condition}\label{cond:c1}
$\widetilde{Y_i}\left(a,\boldsymbol t_{1}^{a'},m,\boldsymbol t_{1}^m\right)\independent (A_i,\boldsymbol T_{1i}^a,\boldsymbol T_{2i}^a,\boldsymbol T_{3i}^a)|\boldsymbol X_i,\boldsymbol u_i$.
\end{condition} 
\noindent
Condition \ref{cond:c1} holds if there are no unmeasured factors confounding the relationship between the outcome and the set of exposures vector—comprising self ($A_i$), first-degree ($\boldsymbol T_{1i}^a,\boldsymbol T_{2i}^a$), and second-degree components ($\boldsymbol T_{3i}^a$), given the observed covariates $\boldsymbol X_i$ and the random effects $\boldsymbol u_i$.

\begin{condition}\label{cond:c2}
$\widetilde{Y_i}\left(a,\boldsymbol t_{1}^{a'},m,\boldsymbol t_{1}^m\right)\independent (M_i, \widetilde{\boldsymbol T}_{1i}^m)|A_i,\boldsymbol T_{1i}^a,\boldsymbol T_{2i}^a,\boldsymbol T_{3i}^a,\boldsymbol X_i,\boldsymbol u_i\cdot$
\end{condition}
\noindent
Condition \ref{cond:c2} holds if there are no unmeasured factors confounding the relationship between the outcome and the set of mediators, comprising both self ($M_i$) and first-degree neighbor mediator summary ($\widetilde{\boldsymbol T}_{1i}^m$), conditional on the set of exposures vector, comprising self ($A_i$), first-degree ($\boldsymbol T_{1i}^a,\boldsymbol T_{2i}^a$), and second-degree components ($\boldsymbol T_{3i}^a$), the observed covariates $\boldsymbol X_i$, and the random effects $\boldsymbol u_i$.
\begin{condition}
$\left[\widetilde{M_i}\left(a^*,\boldsymbol t_{1}^{a \dagger}\right),\widetilde{\boldsymbol  T}^m_{1i}\left(a'',\boldsymbol t^a_{2},\boldsymbol t^a_{3}\right)\right]\independent (A_i,\boldsymbol T_{1i}^a,\boldsymbol T_{2i}^a,\boldsymbol T_{3i}^a)|\boldsymbol X_i,\boldsymbol u_i\cdot$\label{cond:c3}
\end{condition}
\noindent
Condition \ref{cond:c3} holds if there are no unmeasured factors confounding the relationship between the set of exposures vector—comprising self ($A_i$), first-degree ($\boldsymbol T_{1i}^a,\boldsymbol T_{2i}^a$), and second-degree components ($\boldsymbol T_{3i}^a$) and the set of mediators, comprising both self ($M_i$) and first-degree neighbor mediator summary ($\widetilde{\boldsymbol T}_{1i}^m$), given the observed covariates $\boldsymbol X_i$ and the random effects $\boldsymbol u_i$.

\begin{condition}\label{cond:c4}
    $\widetilde{M_i}\left(a^*,\boldsymbol t_{1}^{a \dagger}\right)\independent \widetilde{\boldsymbol T}^m_{1i}\left(a'',\boldsymbol t^a_{2},\boldsymbol t^a_{3}\right)|\boldsymbol X_i,\boldsymbol u_i\cdot$
\end{condition}
\noindent
Condition \ref{cond:c4} holds if there are no factors affected by the exposure vector, comprising self ($A_i$), first-degree ($\boldsymbol T_{1i}^a, \boldsymbol T_{2i}^a$), and second-degree ($\boldsymbol T_{3i}^a$) components—that confound the relationship between the self-mediator $M_i$ and the first-degree neighbor mediator summary $\widetilde{\boldsymbol T}_{1i}^m$, given the observed covariates $\boldsymbol X_i$ and the random effects $\boldsymbol u_i$.
\begin{condition}
$\widetilde{Y_i}\left(a,\boldsymbol t_{1}^{a'},m,\boldsymbol t_{1}^m\right)\independent \left[\widetilde{M_i}\left(a^*,\boldsymbol t_{1}^{a \dagger}\right),\widetilde{\boldsymbol T}^m_{1i}\left(a'',\boldsymbol t_{1}^{a \ddagger},\boldsymbol t^a_{2},\boldsymbol t^a_{3}\right)\right]|\boldsymbol X_i,\boldsymbol u_i\cdot$\label{cond:c5}
\end{condition}
\noindent
Condition \ref{cond:c5} holds if there are no factors influenced by the exposure vector—comprising self ($A_i$), first-degree ($\boldsymbol T_{1i}^a, \boldsymbol T_{2i}^a$), and second-degree ($\boldsymbol T_{3i}^a$) components—that confound the relationship between the outcome and the set of mediators, comprising self $M_i$ and the first-degree neighbor summary $\widetilde{\boldsymbol T}_{1i}^m$, given the observed covariates $\boldsymbol X_i$ and the random effects $\boldsymbol u_i$.\\

\noindent
\textbf{Remark 2:} The identifiability conditions outlined above are formulated conditional on the outcome- and mediator-specific random effects. While we acknowledge that verifying assumptions involving latent variables presents non-trivial challenges, this formulation aligns with established precedents in complex settings. For instance, \citet{bind2016causal} similarly posit identification assumptions conditional on random effects to address unobserved heterogeneity in longitudinal mediation analysis. In the context of network interference, standard frameworks often rely on strict conditional independence assumptions, such as the absence of contagion \citep{ogburn2022causal}. Our framework subsumes these settings, where conditioning on random effects would be redundant; however, we explicitly retain these terms in the conditioning set to maintain theoretical generality and to accommodate the potential for contagion within mediators and outcomes. Furthermore, the inability to empirically validate Conditions \ref{cond:c4} and \ref{cond:c5} is not unique to the network setting. Even in classical mediation analysis without interference, the analogous assumption of no exposure-induced mediator-outcome confounding is inherently unverifiable and must be postulated based on substantive knowledge.

\subsection{General Identification Formula}\label{sec:identhm} 
We next derive the key identification formula for the generic quantity $\Psi_i$ in terms of observed data. Notably, this result is established without imposing distributional assumptions on the REN-SEM, relying exclusively on the mapping restrictions defined in equations \eqref{eq:eq1}--\eqref{eq:eq2}.
\begin{theorem}\label{thm:thm1}
    Under the assumptions \ref{ass:assexp}-\ref{ass:asscons} and identifiability conditions \ref{cond:c1}-\ref{cond:c5}, the generic quantity $\Psi_i$ in (\ref{eq:eq5}) under the REN-SEMs (\ref{eq:eqY})-(\ref{eq:eqM}) takes the following form:
\begin{align*}
&\int_{\boldsymbol b^y_{\mathcal{N}_i}}\int_{\boldsymbol b^m_{\mathcal{N}_i}}\int_{\boldsymbol b^m_{\mathcal{N}^\ddagger_i}}\int_{\boldsymbol X_i}\int_{\boldsymbol t^a_{3}}\int_{\boldsymbol t_{2}^a}\int_{\boldsymbol t_{1}^m}\int_{m}\mathbb{E}\left(\left. Y_i \right|A_i=a,\boldsymbol T_{1i}^a=\boldsymbol t_{1}^{a'},M_i=m_i,\widetilde{\boldsymbol T}_{1i}^m=\boldsymbol t_{1}^m,\boldsymbol X_i,\boldsymbol b^y_{\mathcal{N}_i}\right)\\
& \hspace{1cm}  dF(m)_{\left.M_i\right|A_i=a^*,\boldsymbol T_{1i}^a=\boldsymbol t_{1}^{a \dagger},\boldsymbol X_i,\boldsymbol b^m_{\mathcal{N}_i}}\cdot dF(\boldsymbol t_{1}^m)_{\left.\widetilde{\boldsymbol  T}^m_{1i}\right|A_i=a'',\boldsymbol T_{1i}^a=\boldsymbol t_{1}^{a \ddagger},\boldsymbol T_{2i}^a=\boldsymbol t_{2}^{a},\boldsymbol T_{3i}^a=\boldsymbol t_{3}^{a},\boldsymbol X_i,\boldsymbol b^m_{\mathcal{N}_i},\boldsymbol b^m_{\mathcal{N}^\ddagger_i}}\\
& \hspace{1cm}dF(\boldsymbol t_{2}^a)_{\boldsymbol T_{2i}^a|\boldsymbol T_{1i}^a=\boldsymbol t_{1}^{a \ddagger},\boldsymbol X_i}\cdot dF(\boldsymbol t^a_{3})_{\boldsymbol T_{3i}^a|\boldsymbol X_i}\cdot dF(\boldsymbol b^m_{\mathcal{N}^\ddagger_i})\cdot dF(\boldsymbol b^m_{\mathcal{N}_i})\cdot dF(\boldsymbol b^y_{\mathcal{N}_i})\cdot
\end{align*}
\end{theorem}
\noindent
The proof of Theorem \ref{thm:thm1} is provided in Supplementary Materials Section S1.1. This concludes the exposition of the general REN-SEM framework defined in \eqref{eq:SEM}. In the subsequent section, we particularize this methodology to the Linear REN-SEM (LREN-SEM) specified by \eqref{eq:eqY}--\eqref{eq:eqA}, deriving explicit functional forms for the causal estimands, detailing the necessary assumptions, and establishing the associated asymptotic distributions. We emphasize that while the LREN-SEM serves as a tractable demonstration, the general framework accommodates more complex specifications, including non-linear dependencies, higher-order interaction terms, and alternative neighborhood summary functions.

\section{Causal Mediation under LREN-SEM} \label{sec:lren-sem}
In the context of the LREN-SEM defined by \eqref{eq:eqY}--\eqref{eq:eqA}, we specifically focus on the spillover effect of first-degree neighborhood exposures on the outcome solely through \(\boldsymbol T^a_{1i}=S_{1i}(\boldsymbol A_{\mathcal{N}^{\dagger}_i})\), the average number of first-degree neighboring units receiving the exposure ($A=1$).  Accordingly, our objective is to estimate the causal effect of an intervention that shifts the aggregate exposure vector $\boldsymbol D_i$ from $\boldsymbol d'=(a',s_1^{a'})$ to $\boldsymbol d=(a,s_1^a)$. Following the notation in \eqref{eq:lsum}, we define the observed specific summary components as $S^a_{1i}:=S_{1i}(\boldsymbol A_{\mathcal{N}^{\dagger}_i})$, $S^a_{2i}:=S_{2i}(\boldsymbol A_{\mathcal{N}^{\dagger}_i})$, $S^a_{3i}:=S_{3i}(\boldsymbol A_{\mathcal{N}^{\ddagger}_i})$, and $S^m_{1i}:=S^{m}_{1i}(A_i,S_{1i}^a,S_{2i}^a,S_{3i}^a)$. To circumvent the feasibility issue highlighted in Remark 1—whereby exact values of the summary statistic may be unattainable for certain network configurations—and to ensure that the intervention is well-defined for all units, we introduce the following approximation assumption regarding the intervention on $S^a_{1i}$.
\begin{assumption}\label{ass:a2} 
    A change in $S_{1i}^a$ from $s_1^{a'}$ to $s_1^a$ for unit $i$ is equivalent to a change in $S_{1i}^a$ from $\widetilde{s_{1i}}^{a'}:=\frac{[s_1^{a'}n_i]}{n_i}$ to  $\widetilde{s_{1i}}^{a}:=\frac{[s_1^an_i]}{n_i}$, where $\sum_{j\neq i}E_{ij}A_j$ changes plausibly from $[s_1^{a'}n_i]$ to $[s_1^an_i]$. Here $[x]$ denotes the integer nearest to $x$.
\end{assumption}
\noindent
Assumption~\ref{ass:a2} allows us to include all the units of a network in the analysis under the consideration of plausible changes in the average number of first-degree neighbors receiving exposure. Next we derive the explicit identification formula for LREN-SEM as a corollary to Theorem \ref{thm:thm1}. 
\begin{corollary}\label{cor:cor1}
      For any $a,a^*,a''\in \{0,1\},  \widetilde{s_{1i}}^{a'},\widetilde{s_{1i}}^{a\dagger},\widetilde{s_{1i}}^{a\ddagger}\in [0,1]$, under the assumptions \ref{ass:assexp}-\ref{ass:a2} and identifiability conditions \ref{cond:c1}-\ref{cond:c5}, the generic quantity $\Psi_i$ in (\ref{eq:eq5}) under the LREN-SEMs (\ref{eq:eqY})-(\ref{eq:eqM}) takes the following form:
   \begingroup
\small
\begin{align*}
     &\Psi_i=\vec{\mathbb{E}}_i^{a \ddagger}\left\{\widetilde{Y_i}\left(a,\widetilde{s_{1i}}^{a'},\widetilde{M_i}\left(a^*,\widetilde{s_{1i}}^{a\dagger}\right),S^{m}_{1i}\left(a'',\widetilde{s_{1i}}^{a\ddagger},S_{2i}^a,S_{3i}^a\right)\right) \right\}\\
     &=\beta_0+\beta_1a+\beta_2 \widetilde{s_{1i}}^{a'}+\beta_3 \left(\gamma_0+\gamma_1a^*+\gamma_2 \widetilde{s_{1i}}^{a\dagger}\right)+\beta_4\left\{\gamma_0+\gamma_1\widetilde{s_{1i}}^{a\ddagger}+\gamma_2 a''\left( \sum_{j\neq i}\frac{E_{ij}}{n_i n_j}\right)\right.\\
& \left.+\gamma_2  \mathbb{E}_{\boldsymbol X_i}\mathbb{E}\left(S_{2i}^a|S_{1i}^a=\widetilde{s_{1i}}^{a\ddagger},\boldsymbol X_i\right)\right\}+\beta_4\gamma_2 \mathbb{E}_{\boldsymbol X_i} \mathbb{E}(S_{3i}^a|\boldsymbol X_i) + \mathbb{E}_{\boldsymbol X_i}\{(\boldsymbol\beta_5^T+\beta_3 \boldsymbol\gamma_3^T)\boldsymbol C_i+(\boldsymbol \beta_6^T+\beta_3\boldsymbol\gamma_4^T) \boldsymbol S_{5i}+\beta_4\boldsymbol\gamma_4^T \boldsymbol  S_{6i}\}.
\end{align*}
\endgroup
where $\boldsymbol S_{5i}=\boldsymbol S_{1i}(\boldsymbol C_{\mathcal{N}^{\dagger}_i})$ and $S_{6i}=S_{1i}(\boldsymbol S_{5i})$.
\end{corollary}
   
\noindent
The proof of Corollary~\ref{cor:cor1} is provided in Supplementary Materials Section S1.2.
\subsection{Estimands in the LREN-SEM}
Applying the identification formula in Corollary \ref{cor:cor1} yields the explicit parametric forms for the six estimands $\tau_1$ through $\tau_6$, as detailed in Corollary~\ref{cor:cor2}. Notably, the conditional expectation $\mathbb{E}(S_{3i}^a \mid \boldsymbol X_i)$, which appears in the expansion of $\Psi_i$ (Corollary \ref{cor:cor1}), is invariant across the contrasts defining the causal effects; consequently, it cancels out and requires no estimation for any of the six estimands. Similarly, the conditional expectation $\mathbb{E}\left(S_{2i}^a \mid S_{1i}^a=\widetilde{s_{1i}}^{a \ddagger}, \boldsymbol X_i\right)$ cancels out for estimands $\tau_1$ through $\tau_5$. However, for the spillover mediation effect $\tau_6$, this term persists. To rigorously evaluate this requisite conditional expectation without imposing structural constraints on the exposure mechanism, we first present the general summation formula:
\begin{align}\mathbb{E}\left(S_{2i}^a \mid S_{1i}^a=\widetilde{s_{1i}}^{a \ddagger}, \boldsymbol X_i\right)&= \frac{\sum_{\boldsymbol a_{\mathcal{N}i} : S_{1i}^a=\widetilde{s_{1i}}^{a \ddagger}} S_{2i}^a(\boldsymbol a_{\mathcal{N}i}) P(\boldsymbol A_{\mathcal{N}i} = \boldsymbol a_{\mathcal{N}i} \mid \boldsymbol X_i)}{\sum_{\boldsymbol a_{\mathcal{N}i} : S_{1i}^a=\widetilde{s_{1i}}^{a \ddagger}} P(\boldsymbol A_{\mathcal{N}i} = \boldsymbol a_{\mathcal{N}i} \mid \boldsymbol X_i)}\cdot \label{eq:cond_exp_exact}
\end{align}
Implementation of \eqref{eq:cond_exp_exact} requires estimating the joint exposure probabilities $P(\boldsymbol A_{\mathcal{N}i} = \boldsymbol a_{\mathcal{N}i} \mid \boldsymbol X_i)$; typically achieved via a parametric specification for the propensity score in \eqref{eq:eqA}, such as a logistic or probit regression. While this approach is fully general, the summation over neighborhood configurations can be computationally intensive and may obscure the intuitive structure of the estimator. Therefore, to facilitate analytical tractability and provide a transparent demonstration of the spillover mediation mechanism ($\tau_6$), we adopt the following simplifying assumption for our illustrative analysis:
\begin{assumption}\label{ass:a3}
The exposure variables $A_1, \dots, A_N$ are independent and identically distributed (i.i.d.), following a Bernoulli distribution with parameter $p$.
\end{assumption}
\noindent
Under this assumption, the complex combinatorial weights in \eqref{eq:cond_exp_exact} reduce to standard binomial probabilities, allowing us to derive a compact, easily interpretable formula for $\tau_6$ as demonstrated in the following corollary. This assumption can be interpreted as the implementation of a specific hypothetical conditional allocation strategy, as formalized in the potential outcomes framework for interference \citep{hudgens2008toward}.

\begin{corollary} \label{cor:cor2} Let $\boldsymbol d'=(a',s_1^{a'})$ and $\boldsymbol d=(a,s_1^a)$ where $a,a' \in \{0,1\}$ and $s_1^{a'},s_1^a\in [0,1]$. under the assumptions \ref{ass:assexp}-\ref{ass:a3} and identifiability conditions \ref{cond:c1}-\ref{cond:c5}, the closed-form expressions of the six estimands under the LREN-SEMs (\ref{eq:eqY})-(\ref{eq:eqM}) takes the following forms:
\begin{itemize}
    \item[(a)] \textbf{Path 1:} $\tau_1(A\rightarrow Y) = \tau_1(\boldsymbol d',  \boldsymbol d)=\beta_1(a-a')$; 
    \item[(b)] \textbf{Path 2:} $\tau_2(A\rightarrow M\rightarrow Y) = \tau_2(\boldsymbol d',  \boldsymbol d)=\beta_3\gamma_1(a-a')$;
    \item[(c)] \textbf{Path 3:} $\tau_3(A\rightarrow \boldsymbol M_{\mathcal{N}^{\dagger}}\rightarrow Y) = \tau_3(\boldsymbol d',  \boldsymbol d)=\beta_4\gamma_2(a-a') \bar\Delta_{1N}$ with $\bar\Delta_{1N}=\frac{1}{N}\sum_{i=1}^N\sum_{j\neq i}\frac{E_{ij}}{n_in_j}$;
    \item[(d)] \textbf{Path 4:} $\tau_4(\boldsymbol A_{\mathcal{N}^{\dagger}}\rightarrow Y) = \tau_4(\boldsymbol d',  \boldsymbol d)=\beta_2\bar\Delta_{2N}$ with $\bar\Delta_{2N}=\frac{1}{N}\sum_{i=1}^N (\widetilde{s_{1i}}^a-\widetilde{s_{1i}}^{a'})$ ;
    \item[(e)] \textbf{Path 5:} $\tau_5(\boldsymbol A_{\mathcal{N}^{\dagger}}\rightarrow M \rightarrow Y) = \tau_5(\boldsymbol d',  \boldsymbol d)=\beta_3\gamma_2\bar\Delta_{2N}$;
    \item[(f)] \textbf{Path 6:} $\tau_6(\boldsymbol A_{\mathcal{N}^{\dagger}}\rightarrow \boldsymbol M_{\mathcal{N}^{\dagger}} \rightarrow Y) = \tau_6(\boldsymbol d',  \boldsymbol d)= \beta_4\gamma_1\bar\Delta_{2N} + \beta_4 \gamma_2 \bar\Delta_{3N}$\\ with $\bar\Delta_{3N} = N^{-1}\sum_{i=1}^N \sum_{j\neq i} (n_in_j)^{-1} \sum_{k\neq j,k\in \mathcal{N}^{\dagger}_i} E_{ij}E_{jk} \left(\widetilde{s_{1i}}^a-\widetilde{s_{1i}}^{a'}\right)$.
\end{itemize}
\end{corollary}
\noindent
Refer to Definitions~\ref{def:dir} to \ref{def:spillmed} for the meanings of individual effects in (a)-(f) above.
In fact, deriving the expressions of $\tau_1(\boldsymbol d',  \boldsymbol d)$ to $\tau_5(\boldsymbol d',  \boldsymbol d)$ is straightforward and thus is omitted. In the Supplementary Material S1.3 we include a proof for Part (f) of Corollary~\ref{cor:cor2}.\\

\noindent
\textbf{Remark 3}: Notably, the linearity of the LREN-SEM formulations in \eqref{eq:eqY}--\eqref{eq:eqM} ensures that the random intercepts cancel out in the difference of expectations, rendering the point estimates of the causal contrasts invariant to the latent heterogeneity. While this invariance would not hold in the presence of random slopes, explicitly modeling the random effects remains essential; they govern the variance-covariance structure of $\boldsymbol{M}$ and $\boldsymbol{Y}$, and are thus requisite for valid statistical inference regarding the fixed effects.

\subsection{Statistical Theory with Network-Dependent Data under LREN-SEM}\label{sec:est}
As mentioned above, the six estimands $\tau_1$ through $\tau_6$ of interest are inherently functions of the parameters of the LREN-SEMs model in \eqref{eq:eqY}-\eqref{eq:eqA} and are random effects free. Thus, with the availability of model parameter estimates, the corresponding estimates for such estimands are obtained naturally by the plug-in method as well as their inferences. In this paper, we invoke the Maximum Likelihood Estimation (MLE) for the model parameters, in which, however, the large-sample theory is technically nontrivial due to the fact that the network data are not i.i.d samples.

\subsection{Maximum Likelihood Estimation}
We begin with the marginal log-likelihood by integrating out the random effects. This marginalization operation enables to obtain a log-likelihood function of the fixed effects in the REN-SEM, which is deemed sufficient to estimate the six etimands. Let $\boldsymbol Y=\{Y_i\}_{i=1}^N, \boldsymbol M=\{M_i\}_{i=1}^N,\boldsymbol A=\{A_i\}_{i=1}^N$ and $\boldsymbol C=\{\boldsymbol C_i\}_{i=1}^N$. Utilizing the the LREN-SEMs specification, we obtain the joint density of the data as follows: 
\begin{align*}
    & f(\boldsymbol Y,\boldsymbol M,\boldsymbol A|\boldsymbol C)=\int_{\text{Supp}(\boldsymbol b^y,\boldsymbol b^m)}f(\boldsymbol Y,\boldsymbol M,\boldsymbol A|\boldsymbol C,\boldsymbol b^y,\boldsymbol b^m)f(\boldsymbol b^y,\boldsymbol b^m)d\boldsymbol b^yd\boldsymbol b^m\\
    &=\left\{\int_{\text{Supp}(\boldsymbol b^y)}f(\boldsymbol Y|\boldsymbol A,\boldsymbol M,\boldsymbol C,\boldsymbol b^y) d\boldsymbol b^y\int_{\text{Supp}(\boldsymbol b^m)}f(\boldsymbol M|\boldsymbol A,\boldsymbol C,\boldsymbol b^m) d\boldsymbol b^m\right\} f(\boldsymbol A|\boldsymbol C),
\end{align*}
where $\text{Supp}(\boldsymbol b^m)$ and $\text{Supp}(\boldsymbol b^y)$ denote the supports of the random effects $\boldsymbol b^m$ and $\boldsymbol b^y$ respectively. Denote a new matrix $\widetilde{E}$ where the $(i,j)^{\text{th}}$ element is $\widetilde{E}_{ij}=\widetilde{w}_i E_{ij}$, where $\widetilde{w}_i=\frac{1}{n_i}$, if $i\neq j$ and 0 otherwise. It follows that the LREN-SEMs (\ref{eq:eqY})-(\ref{eq:eqM}) may be rewritten in the following matrix form: 
 \begin{align*}
        &\boldsymbol Y=\beta_0\boldsymbol 1+\beta_1 \boldsymbol A+\beta_2\widetilde{E}\boldsymbol A  +\beta_3\boldsymbol M +\beta_4 \widetilde{E}\boldsymbol M + \beta_5 \boldsymbol C+\beta_6 \widetilde{E}\boldsymbol C_{\mathcal{N}} +  E \boldsymbol b^y+ \boldsymbol \epsilon^y;\\
        & \boldsymbol M=\gamma_0\boldsymbol 1+\gamma_1\boldsymbol A+\gamma_2 \widetilde{E}\boldsymbol A + \gamma_3 \boldsymbol C +\gamma_4  \widetilde{E}\boldsymbol C_{\mathcal{N}} + E \boldsymbol b^m+ \boldsymbol \epsilon^m,
    \end{align*}
where random effects $\boldsymbol b^y,\boldsymbol b^m$ and error terms $\boldsymbol \epsilon^y,\boldsymbol \epsilon^m$ are mutually independent and follow the multivariate normal distributions specified in (\ref{eq:eqY})-(\ref{eq:eqM}). Under the normality, we have $\boldsymbol Y|\boldsymbol A,\boldsymbol M,\boldsymbol X\sim \text{MVN}(\boldsymbol \mu_y,\Sigma_y)$ with the mean vector and covariance given by 
\begin{align*}
    & \boldsymbol \mu_{\boldsymbol y}=\mathbb{E}(\boldsymbol Y|\boldsymbol A,\boldsymbol M,\boldsymbol X)=\mathbb{E}_{\boldsymbol b^y}\{\mathbb{E}(\boldsymbol Y|\boldsymbol A,\boldsymbol M,\boldsymbol X,\boldsymbol b^y)\}=\mathbb{E}_{\boldsymbol b^y}(\boldsymbol X_{\boldsymbol y}\boldsymbol \beta+E \boldsymbol b^y)=\boldsymbol X_{\boldsymbol y}\boldsymbol \beta,\\
    & \Sigma_{\boldsymbol y}=\text{Var}(\boldsymbol Y|\boldsymbol A,\boldsymbol M,\boldsymbol X)=\sigma^2_{\boldsymbol y}\mathcal{I}_{N\times N}+\sigma^2_{\boldsymbol b^y}EE^{T},
\end{align*}
where $\boldsymbol X_{\boldsymbol y}=(\boldsymbol 1,\boldsymbol A,\widetilde{E}\boldsymbol A,\boldsymbol M, \widetilde{E}\boldsymbol M,\boldsymbol C,\widetilde{E}\boldsymbol C)$.  Likewise, we have $\boldsymbol M|\boldsymbol A,\boldsymbol X\sim \text{MVN}(\boldsymbol \mu_m,\Sigma_m)$ with the mean vector and covariance given by $ \boldsymbol \mu_{\boldsymbol m}=\boldsymbol X_{\boldsymbol m}\boldsymbol \gamma \hspace{0.4cm}\text{and} \hspace{0.4cm} \Sigma_{\boldsymbol m}=\sigma^2_{\boldsymbol m}\mathcal{I}_{N\times N}+\sigma^2_{\boldsymbol b^m}EE^T,$ where $\boldsymbol X_{\boldsymbol m}=(\boldsymbol 1,\boldsymbol A,\widetilde{E}\boldsymbol A,\boldsymbol C,\widetilde{E}\boldsymbol C_{\mathcal{N}})$. In addition, we allow a covariate-dependent exposure allocation via the logit link function, so the model (\ref{eq:eqA}) may be specified as follows: $P(A_i=1|\boldsymbol C_i)=\pi(\boldsymbol C_i,\boldsymbol \alpha)=\frac{e^{\boldsymbol\alpha^T\boldsymbol C_i}}{1+e^{\boldsymbol\alpha^T\boldsymbol C_i}}$.
Let \(\Sigma_a=\text{Var}(\boldsymbol A|\boldsymbol C)\) be the variance-covariance matrix of the exposure variables.   Denote the grand vector of all model parameters by $\boldsymbol \varphi=(\boldsymbol \beta, \boldsymbol \gamma, \boldsymbol \alpha, \sigma^2_{\boldsymbol y},\sigma^2_{\boldsymbol b^y},\sigma^2_{\boldsymbol m},\sigma^2_{\boldsymbol b^m})$ with a dimension denoted by \(\text{dim}(\boldsymbol \varphi) = K\). Let the parameter space $\Theta$ be a compact subspace of $\mathbb{R}^K$, which contains the true parameter $\boldsymbol \varphi_0$ at which data are generated from the true underlying distribution  $f(\boldsymbol Y,\boldsymbol M,\boldsymbol A|\boldsymbol X, \boldsymbol \varphi_0)$.  
Then the log-likelihood of the LREN-SEMs model parameters, $\boldsymbol \varphi$, is
\begin{align*}
    l(\boldsymbol\varphi)= &l_y(\boldsymbol\varphi)+l_m(\boldsymbol\varphi)+l_a(\boldsymbol\varphi)\\
    =&-\frac{N}{2} \text{log}(2\pi)-\frac{1}{2} \text{log}|\sigma^2_{\boldsymbol y}\mathcal{I}_{N\times N}+\sigma^2_{\boldsymbol b^y} EE^{T}|-\frac{1}{2}(\boldsymbol Y-\boldsymbol X_{\boldsymbol y}\boldsymbol \beta)^T\Sigma_{\boldsymbol y}^{-1}(\boldsymbol Y-\boldsymbol X_{\boldsymbol y}\boldsymbol \beta)\\
    & -\frac{N}{2} \text{log}(2\pi)-\frac{1}{2}\text{log}|\sigma^2_{\boldsymbol m}\mathcal{I}_{N\times N}+\sigma^2_{\boldsymbol b^m} EE^{T}|-\frac{1}{2}(\boldsymbol M-\boldsymbol X_{\boldsymbol m}\boldsymbol \gamma)^T\Sigma_{\boldsymbol m}^{-1}(\boldsymbol M-\boldsymbol X_{\boldsymbol m}\boldsymbol \gamma)\\
    & +\sum_{i=1}^N A_i\text{log}\left(\frac{\pi(\boldsymbol C_i,\boldsymbol \alpha)}{1-\pi(\boldsymbol C_i,\boldsymbol \alpha)}\right)+\sum_{i=1}^N \text{log}(1-\pi(\boldsymbol C_i,\boldsymbol \alpha))\cdot
\end{align*}
As usual, the maximum likelihood estimate $\widehat{\boldsymbol \varphi_N}$ can be obtained through maximizing the above log-likelihood $l(\boldsymbol\varphi)$; that is, $\widehat{\boldsymbol\varphi}_N = \arg\max_{\boldsymbol\varphi} l(\boldsymbol\varphi)$.  Numerically, we perform iterative weighted least square estimations to handle the optimization, most part of which can be done by using existing software packages such as \texttt{optim}.

\subsection{Large Sample Properties}
Under a network data-generating mechanism governed by the LREN-SEMs model, the conventional large-sample analytic tools, including weak consistency and asymptotic normality under i.i.d. data, are no longer applicable. To establish the relevant large-sample theory, we build our work by extending the analytic framework proposed by \citet{sweeting1980uniform} and \citet{mardia1984maximum} in the setting of spatial regression models. Given a substantial extension of their analytics to the LREN-SEMs model MLE in the given network $E$, whose size $N\rightarrow\infty$, we present the relevant theoretical details in this section, including the technical proofs that require accommodating the complex hierarchical dependence structures involving both unit-specific random effects and network connectivity $E$.\\

\noindent
Let \(\mathcal{L}_N(\boldsymbol \varphi) = \nabla^2_{\boldsymbol \varphi\boldsymbol \varphi}\) be the Hessian matrix of the log-likelihood function \(l(\boldsymbol\varphi)\), and let \(J_N(\boldsymbol \varphi) = -\mathcal{L}_N(\boldsymbol \varphi)\) be the observed information matrix. Let \(B_N(\boldsymbol \varphi) =\mathbb{E}\{J_N(\boldsymbol \varphi)\}\) be the Fisher Information Matrix. For a matrix \(\boldsymbol \Phi = (\boldsymbol \varphi_1, \boldsymbol \varphi_2, \cdots, \boldsymbol \varphi_K)\), where \(\boldsymbol \varphi_k \in \Theta\), \(\forall k \in \{1,2,\cdots,K\}\), the \((k,l)\) element of the matrix \(\mathcal{L}_N(\boldsymbol\Phi)\) is defined to be \(\left.\frac{\partial^2 l(\boldsymbol \varphi)}{\partial \boldsymbol\varphi_k \partial \boldsymbol\varphi_l}\right|_{\boldsymbol \varphi = \boldsymbol \varphi_k}\). Essentially $k^{\text{th}}$ row of this ``\textit{mosaic}" matrix $\mathcal{L}_N(\boldsymbol\Phi)$ takes the $k^{\text{th}}$ row of $\mathcal{L}_N(\boldsymbol\varphi_k)$, $\forall k \in \{1,2,\cdots,K\}$.
In all the subsequent assumptions and theorems, we adopt the Frobenius norm \(||H||_F = \sqrt{\text{tr}(H^T H)}\) for a matrix \(H\). Moreover, for a sequence of non-random matrices $\{H_N\}$,  it is said that \(\lim_{N\rightarrow \infty}\hspace{0.1cm} H_N=H\) if  \(\lim_{N\rightarrow \infty}||H_N - H||_F = 0\); for a sequence of random matrices $\{H_N\}$, $H_N\xrightarrow{p}H$ denotes an element-wise weak convergence, namely for any $(i,j)$, $H_N^{(ij)}\xrightarrow{p}H^{(ij)}$, as $N\rightarrow \infty$ where the dimensions of the all matrices are the same, or  $\text{dim}(H_N)=\text{dim}(H)=K$, $\forall N\in \mathbb{N}$.
Below are the regularity conditions on the Hessian matrix required to study the large sample properties of the MLE for \(\boldsymbol \varphi\). 
\begin{assumption}\label{ass:ass9}
    The positive definite Fisher information matrix $B_N(\boldsymbol \varphi)$ is continuous in $\boldsymbol \varphi \in \Theta \subseteq \mathbb{R}^K$. 
    Let $\lambda_{N,1}(\boldsymbol \varphi),\lambda_{N,2}(\boldsymbol \varphi),\cdots \lambda_{N,K}(\boldsymbol \varphi)$ denote the $K$ eigenvalues of $B_N^{-1}(\boldsymbol \varphi)$, where $\lambda_{N,1}(\boldsymbol \varphi)\geq\lambda_{N,2}(\boldsymbol \varphi)\geq\cdots\geq\lambda_{N,K}(\boldsymbol \varphi)>0$.
    $$\lim_{N\rightarrow \infty}\sup_{\boldsymbol \varphi \in \Theta}\lambda_{N,1}(\boldsymbol \varphi)=0 \hspace{0.5cm}\text{or equivalently}\hspace{0.5cm}\lim_{N\rightarrow \infty} \sup_{\boldsymbol \varphi \in \Theta}\hspace{0.1cm}||B_N^{-1}(\boldsymbol \varphi)-\mathcal{O}_{K\times K}||_F= 0$$
   where $\mathcal{O}_{K\times K}$ denotes the $K\times K$ dimensional null matrix.
\end{assumption}
\noindent
Assumption \ref{ass:ass9} implies that a large network provides more information about model parameter $\boldsymbol \varphi$ so to obtain more accurate estimates of $\boldsymbol \varphi$ by MLE.
\begin{assumption}\label{ass:ass10}
$\lim_{N\rightarrow \infty} \sup_{\boldsymbol \varphi \in \Theta}\hspace{0.1cm}\mathbb{E}\{||B_N^{-\frac{1}{2}}(\boldsymbol \varphi) J_N(\boldsymbol \varphi)  B_N^{-\frac{1}{2}}(\boldsymbol \varphi)-\mathcal{I}_{K\times K}||_F^2\}= 0$, where $\mathcal{I}_{K\times K}$ is the $K\times K$ dimensional identity matrix. An equivalent assumption in terms of the eigenvalues of the matrix, $B_N^{-\frac{1}{2}}(\boldsymbol \varphi) J_N(\boldsymbol \varphi)  B_N^{-\frac{1}{2}}(\boldsymbol \varphi)$ is : Let $\mu_{N,1}(\boldsymbol \varphi),\mu_{N,2}(\boldsymbol \varphi),\cdots \mu_{N,K}(\boldsymbol \varphi)$ denote the $K$ eigenvalues of $B_N^{-\frac{1}{2}}(\boldsymbol \varphi) J_N(\boldsymbol \varphi)  B_N^{-\frac{1}{2}}(\boldsymbol \varphi)$, where $\mu_{N,1}(\boldsymbol \varphi)\geq \mu_{N,2}(\boldsymbol \varphi)\geq \cdots \geq \mu_{N,K}(\boldsymbol \varphi)$. As $N \rightarrow \infty$, $\forall k \in \{1,2,\cdots,K\}$, $\sup_{\boldsymbol \varphi}\mu_{N,k}(\boldsymbol \varphi)\xrightarrow{
p}1$. This is equivalent to the condition w.r.t matrices because 
\begin{align*}
    & \sum_{k=1}
^K(\mu_{N,k}(\boldsymbol\varphi)-1)^2=\text{tr}(B_N^{-\frac{1}{2}}(\boldsymbol \varphi) J_N(\boldsymbol \varphi)  B_N^{-\frac{1}{2}}(\boldsymbol \varphi)-\mathcal{I}_{K\times K})
=\mathbb{E}\{||B_N^{-\frac{1}{2}}(\boldsymbol \varphi) J_N(\boldsymbol \varphi)  B_N^{-\frac{1}{2}}(\boldsymbol \varphi)-\mathcal{I}_{K\times K}||_F^2\}
\end{align*}
\end{assumption}
\begin{assumption}\label{ass:ass11}
   For all $\epsilon>0,\eta>0$ and $\forall \boldsymbol \varphi,\boldsymbol \varphi_1,\boldsymbol \varphi_2,\cdots,\boldsymbol \varphi_K \in \Theta$, $\boldsymbol \Phi=(\boldsymbol \varphi_1,\boldsymbol \varphi_2,\cdots,\boldsymbol \varphi_K)$ as $N \rightarrow \infty$ the following conditions are satisfied:\\
\noindent
(i) $\text{sup}\{||B_N^{-\frac{1}{2}}(\boldsymbol \varphi)B_N^{\frac{1}{2}}(\boldsymbol \varphi')-\mathcal{I}_{K\times K}||_F:||B_N^{\frac{1}{2}}(\boldsymbol \varphi)^T(\boldsymbol \varphi-\boldsymbol \varphi')||_F\leq \epsilon;\forall\boldsymbol \varphi\in \Theta,\forall\boldsymbol \varphi' \in \Theta\}\rightarrow 0$ and \\

\noindent
(ii) $P[\text{sup}\{||B_N(\boldsymbol \varphi)^{-\frac{1}{2}}\{J_N(\boldsymbol\Phi)-J_N(\boldsymbol \varphi)\} B_N(\boldsymbol \varphi)^{-\frac{1}{2}}||_F:||B_N(\boldsymbol \varphi)(\boldsymbol \varphi-\boldsymbol \varphi_k)||_F\leq \epsilon;,1\leq k\leq K\}>\eta]\rightarrow 0.$
\end{assumption}
\noindent
Assumptions \ref{ass:ass9} and \ref{ass:ass10} regulate the growth and limiting behaviors of the matrices \(J_N(\boldsymbol \varphi)\) and \(B_N(\boldsymbol \varphi)\) respectively. Assumption \ref{ass:ass11}(i) and \ref{ass:ass11}(ii) require the continuity of \(J_N(\boldsymbol \varphi)\). Note that in the LREN-SEM, \(\Sigma_y(\boldsymbol\varphi)\), \(\Sigma_m(\boldsymbol\varphi)\), and \(\Sigma_a(\boldsymbol\varphi)\) are twice continuously differentiable in \(\boldsymbol \varphi\) because \(\Sigma_y = \sigma^2_{\boldsymbol y}\mathcal{I}_{N\times N} + \sigma^2_{\boldsymbol b^y} EE^{T}\) and \(\Sigma_m = \sigma^2_{\boldsymbol m}\mathcal{I}_{N\times N} + \sigma^2_{\boldsymbol b^m} EE^{T}\). The twice continuously differentiability for \(\Sigma_a(\boldsymbol\varphi)\) is guaranteed by the fact that logit link function is twice continuously differentiable with respect to $\boldsymbol\alpha$.

\begin{theorem}\label{thm:thm2}
    If the LREN-SEMs model satisfies Assumptions \ref{ass:ass9}, \ref{ass:ass10} and \ref{ass:ass11} then the MLE $\widehat{\boldsymbol \varphi_N}$ is weakly consistent for $\boldsymbol \varphi_0$ where $\boldsymbol \varphi_0$ is the true value.  and asymptotically normally distributed, namely 
    $$B_N^{\frac{1}{2}}(\boldsymbol \varphi_0)(\widehat{\boldsymbol \varphi_N}-\boldsymbol \varphi_0)\xrightarrow{d}\mathcal{N}(\boldsymbol 0,\mathcal{I_{K\times K}}), ~ \mbox{as $N\rightarrow \infty$,}
    $$
\end{theorem}
\noindent
The proof of Theorem \ref{thm:thm2} follows from \citet{sweeting1980uniform}, and thus is omitted. We derive the exact form of Hessian $\mathcal{L}_N(\boldsymbol \varphi)$ in the Supplementary Materials Section S1.4, which forms the basis to establish consistent variance estimators for $\hat\tau_1$ through $\hat\tau_6$ of the six estimands $\tau_1$ through $\tau_6$, respectively. We now turn to obtain the asymptotic distributions of the plug-in estimators, $\hat\tau_1$ through $\hat\tau_6$ of the six estimands, as well as their corresponding consistent variance estimators useful for the calculation of the respective standard errors. Since the six estimators take functional forms of the LREN-SEMs model parameter $\boldsymbol \varphi$, it suffices to establish a general consistency result and asymptotic distribution  of a smooth function $g(.)$ of $\boldsymbol \varphi\in \mathbb{R}^K$. Here $g(.)$ may depend on the network size $N$ and satisfy the following assumptions.

\begin{assumption}\label{ass:ass12}
Let $g:\Theta\rightarrow \mathcal{R}$ be a real valued analytic function of $\boldsymbol \varphi$. Let $\nabla_{\boldsymbol\varphi_0} g(\boldsymbol\varphi)$ and $\nabla^2_{\boldsymbol\varphi_0} g(\boldsymbol\varphi)$ denote the gradient and Hessian matrix of $g(.)$ evaluated at $\boldsymbol\varphi_0$ respectively and $Q_N=\{\nabla_{\boldsymbol\varphi_0} g(\boldsymbol\varphi)\}^T B_N^{-1}(\boldsymbol\varphi_0)\{\nabla_{\boldsymbol\varphi_0} g(\boldsymbol\varphi)\}$ be the sandwich variance.\\

\noindent
(i) $\nabla_{\boldsymbol\varphi_0} g(\boldsymbol\varphi)\neq 0$, $\sup_{N \in \mathbb{N}}||\nabla_{\boldsymbol\varphi_0} g(\boldsymbol\varphi)||_F=C_1<\infty$, $\sup_{N\in \mathbb{N},\boldsymbol\varphi'}||\nabla^2_{\boldsymbol\varphi'} g(\boldsymbol\varphi)||_F=C_2<\infty$, where $\boldsymbol\varphi'$ lies in a $\epsilon$ neighborhood of $\boldsymbol\varphi_0$.\\

\noindent
(ii) Let $\boldsymbol v_{N,1}:=\boldsymbol v_{N,1}(\boldsymbol\varphi_0),\boldsymbol v_{N,2}:=\boldsymbol v_{N,2}(\boldsymbol\varphi_0),\cdots, \boldsymbol v_{N,K}:= \boldsymbol v_{N,K}(\boldsymbol\varphi_0)$ be the corresponding orthonormal eigenvectors to the eigenvalues of $B_N^{-1}(\boldsymbol\varphi_0)$, namely $\lambda_{N,1}:=\lambda_{N,1}(\boldsymbol\varphi_0)\geq\lambda_{N,2}:=\lambda_{N,2}(\boldsymbol\varphi_0)\geq\cdots \geq \lambda_{N,K}:=\lambda_{N,K}(\boldsymbol\varphi_0)$. Since $\boldsymbol v_{N,1}, \boldsymbol v_{N,2},\cdots,\boldsymbol v_{N,K}$ are orthonormal vectors, $\nabla_{\boldsymbol\varphi_0} g(\boldsymbol\varphi)\in \text{Span}\{\boldsymbol v_{N,1}, \boldsymbol v_{N,2},\cdots,\boldsymbol v_{N,K}\}$. It follows that $\nabla_{\boldsymbol\varphi_0} g(\boldsymbol\varphi)=\sum_{k=1}^K z_{N,k}\boldsymbol v_{N,k}$ for any $N$. We assume that $z_{N,k}=O(1)$, $\forall k \in \{1,2,\cdots,K\}$.\\

\noindent
(iii) Let $\forall k \in \{1,2,\cdots,K\}$, $\lambda_{N,k}=O(N^{-\delta_k})$, where $\delta_1\leq\delta_2\leq\cdots\leq \delta_K$. Moreover let $\delta_1=\delta_2=\cdots=\delta_r<\delta_{r+1}$, where $1\leq r\leq K$. Let $\boldsymbol u_{N,k}^T=\frac{\sqrt{\lambda_{N,k}}z_{N,k}\boldsymbol v_{N,k}^T}{\sqrt{\sum_{k=1}^K\lambda_{N,k}z_{N,k}^2}}$ satisfy
$|| \boldsymbol u_{N,k} - \boldsymbol u_{N',k}||_2 = o(1)$ for any sufficiently large $N$ and $N'$. 
\end{assumption}
\noindent
Since $||\boldsymbol u_{N,k}^T||_2 \leq 1$, there exists a sub-sequence of $\{\boldsymbol u_{N_t,k}^T\}_{t=1}^{\infty}$, of $\{\boldsymbol u_{N,k}^T\}$, such that $\lim_{N_t\rightarrow \infty}\boldsymbol u_{N_t,k}^T=\boldsymbol w_k^T$, $\forall k\in\{1,2,\cdots,r\}$, where $\boldsymbol w_1^T,\boldsymbol w_2^T,\cdots,\boldsymbol w_r^T\in \mathbb{R}^K \setminus \{0\}$ are the nonzero limits. The condition~\ref{ass:ass12}(iii) implies that $ \lim_{N\rightarrow \infty}\boldsymbol u_{N,k}^T=\boldsymbol w_k^T$ by taking $N' = N_t$. 

\begin{theorem}\label{thm:thm3}
 Let $g:\Theta\rightarrow \mathcal{R}$ be a real valued analytic function of $\boldsymbol \varphi\in \Theta$, which might depend on the network size $N$. Under Assumptions  \ref{ass:ass9}, \ref{ass:ass10}, \ref{ass:ass11} and \ref{ass:ass12} as $N\rightarrow \infty$, we have
 $$(i)\hspace{0.2cm}g(\widehat{\boldsymbol \varphi_N})-g(\boldsymbol \varphi_0) \xrightarrow{p} 0;\hspace{0.2cm} \text{and} \hspace{0.2cm}(ii)\hspace{0.2cm}Q_N^{-\frac{1}{2}}\{g(\widehat{\boldsymbol \varphi_N})-g(\boldsymbol \varphi_0)\}\xrightarrow{d}\mathcal{N}(0,1)\cdot$$
\end{theorem}
\noindent
The proof of Theorem \ref{thm:thm3} is provided in Supplementary Material Section S1.5. Next we derive a consistent estimator of the asymptotic variance of $g(\widehat{\boldsymbol \varphi_N})$, namely $\widehat{Q}_N$. 

\begin{assumption}\label{ass:ass13}
Let $g:\Theta\rightarrow \mathcal{R}$ be a real valued analytic function satisfying Assumption \ref{ass:ass12}. Moreover we assume that\\

\noindent
   (i) $J_N({\boldsymbol\varphi_0})$ and $J_N(\widehat{\boldsymbol\varphi_N})$ are positive definite matrices, $\forall N\in \mathbb{N}$\\
   
   \noindent
   (ii) $\sup_{N,\boldsymbol\varphi'}||\nabla^2_{\boldsymbol\varphi'} \{\nabla_{\boldsymbol\varphi}g(\boldsymbol \varphi)^T\nabla_{\boldsymbol\varphi}g(\boldsymbol \varphi)\}||_F<\infty$, where $\boldsymbol\varphi'$ lies in a $\epsilon-$ neighborhood of $\boldsymbol\varphi_0$.
\end{assumption}

\begin{theorem}\label{thm:thm4}
Under Assumption \ref{ass:ass13} along with the assumptions of Theorem \ref{thm:thm3}, as $N \rightarrow \infty$,$$\nabla_{\widehat{\boldsymbol\varphi_N}} g(\boldsymbol\varphi)^T J_N^{-1}(\widehat{\boldsymbol\varphi_N}) \nabla_{\widehat{\boldsymbol\varphi_N}} g(\boldsymbol\varphi)-Q_N\xrightarrow{p} 0\cdot$$
\end{theorem}
\noindent
The proof of Theorem~\ref{thm:thm4} is presented in Supplementary section S1.6. It follows that the sample counterpart of the sandwich covariance $\nabla_{\widehat{\boldsymbol\varphi_N}} g(\boldsymbol\varphi)^T J_N^{-1}(\widehat{\boldsymbol\varphi_N}) \nabla_{\widehat{\boldsymbol\varphi_N}} g(\boldsymbol\varphi)$ will be used to calculate standard errors for statistical inference.

\subsection{Estimated variances of the plug-in estimators of the six effects}
Applying Theorem \ref{thm:thm4}, leads to the closed-form expressions for the consistently estimated variance for each of the plug-in estimators of the six estimands. 

\begin{corollary}
    Let $j_N^{(a,b)}$ be the corresponding $(a,b)^{\text{th}}$ element of $J_N^{-1}(\widehat{\boldsymbol\varphi_N})$. The variance estimators of the six plug-in estimators are given as follows :
\begin{itemize}
    \item[(a)] $\widehat{\text{Var}}\{\widehat{\tau_1}(\boldsymbol d',  \boldsymbol d)\}=j_N^{(\beta_1,\beta_1)}(a-a')^2$;
    \item[(b)] $\widehat{\text{Var}}\{\widehat{\tau_2}(\boldsymbol d',  \boldsymbol d)\}=\{\widehat{\gamma_1}^2 j_N^{(\beta_3,\beta_3)}+\widehat{\beta_3}^2 j_N^{(\gamma_1,\gamma_1)}\} (a-a')^2$;
    \item[(c)] $\widehat{\text{Var}}\{\widehat{\tau_3}(\boldsymbol d',  \boldsymbol d)\}=\{\widehat{\gamma_2}^2 j_N^{(\beta_4,\beta_4)}+\widehat{\beta_4}^2j_N^{(\gamma_2,\gamma_2)}\} (a-a')^2\bar\Delta_{1N}^2$;
    \item[(d)] $\widehat{\text{Var}}\{\widehat{\tau_4}(\boldsymbol d',  \boldsymbol d)\}=j_N^{(\beta_2,\beta_2)} \bar\Delta_{2N}^2$;
    \item[(e)] $\widehat{\text{Var}}\{\widehat{\tau_5}(\boldsymbol d',  \boldsymbol d)\}=\{\widehat{\gamma_2}^2 j_N^{\beta_3\beta_3}+\widehat{\beta_3}^2 j_N^{\gamma_2\gamma_2}\} \bar\Delta_{2N}^2$;
    \item[(f)] $
         \widehat{\text{Var}}\{\widehat{\tau_6}(\boldsymbol d',  \boldsymbol d)\} = j_N^{(\beta_4,\beta_4)}\left\{\widehat{\gamma_1} \bar\Delta_{1N} +\widehat{\gamma_2} \bar\Delta_{2N}\right\}^2+\widehat{\beta_4}^2 \left[\bar\Delta_{2N}^2 +j_N^{(\gamma_2,\gamma_2)} \bar\Delta_{3N}^2 \right]+ 2\widehat{\beta_4}^2j_N^{(\gamma_1,\gamma_2)} \bar\Delta_{2N} \bar\Delta_{3N}\cdot$
 \end{itemize}    
\end{corollary}
\section{Simulation Experiments}\label{sec:simu} \subsection{Setup} \label{sec:setup}
Numerically we examine the behavior of the proposed methods across two distinct network structures, Network-1 and Network-2 described below, with network sizes $N\in\{100,200,800\}$. The setup details are given as follows.\\

\noindent
\textbf{Adjacency Matrix Specification:} (i) Network-1 of size $N$ is built upon a symmetric $1-$ adjacency matrix $E$ such that $\forall i \notin \{1,N\}$, $E_{ij}=1$ if $i=j \hspace{0.2cm} \text{or} \hspace{0.2cm}j=i+1,i-1$ or 0 otherwise. In addition, for the boundary nodes $i=1$, $E_{11}=E_{12}=E_{1N}=1$ and $i=N$, $E_{N1}=E_{N (N-1)}=E_{NN}=1$. See the left panel of Supplementary Figure S1 for an example of Network-1 with 10 units. (ii) Network-2 of size $N$ is constructed by a symmetric adjacency matrix $E$ using the Erdos Renyi Random Graph model, in which the $(i,j)^{\text{th}}$ connectivity is symmetric and generated by $E_{ij}=E_{ji}\sim\text{Bernoulli}(\pi)$ while $E_{ii}=1$, $\forall i,j\in\{1,2,\cdots, N\}$. The probability of node connection $\pi$, is calibrated according to network size ($N$) to ensure that the average connectivity degree (i.e., the average number of connections per node) equals to approximately 10. The right panel of Supplementary Figure S1 displays one resulting $E$ that defines a network of 10 units. Of note, the $E$ matrix is held fixed once it is generated.\\

\noindent
\textbf{LREN-SEM Model Specification:} For unit $i \in \{1,2,\cdots,N\}$, we simulate the binary exposure variables $A_i\overset{\text{i.i.d.}}{\sim}\text{Ber}(0.5)$ and a continuous confounder $C_i\overset{\text{i.i.d.}}{\sim}\mathcal{N}(0,1)$. Then we simulate continuous mediator and outcome, $M_i$ and $Y_i$ respectively, from the LREN-SEMs in \eqref{eq:eqM} and \eqref{eq:eqY} with $\boldsymbol\gamma=(-1,2,0.9,1.8,0.7)$ and $\boldsymbol \beta=(-2,1.5,0.8,1.2,0.4,2.1,1.3)$. The variance parameters for errors terms and random effects in Network-1 are set at: $(\sigma_m,\sigma_y,\sigma_{b^{\boldsymbol M}},\sigma_{b^{\boldsymbol Y}})=(1,1,1,1)$, while in Network-2, they are set as $(\sigma_m,\sigma_y,\sigma_{b^{\boldsymbol M}},\sigma_{b^{\boldsymbol Y}})=(1,1,0.5,0.5)$. Summary statistics of the performance are generated from $R=500$ rounds of simulations.\\

\noindent
\textbf{Metrics of Comparison:} Throughout simulations, evaluation criteria include estimation bias and Root Relative Mean Squared Error (RRMSE), namely $\text{Bias}(\widehat{\tau_k},\tau_k)=\frac{1}{R}\sum_{r=1}^R (\widehat{\tau_{k,r}}-\tau_{k})$ and $\text{RRMSE}(\widehat{\tau_k},\tau_k)=\sqrt{\frac{1}{R}\sum_{r=1}^R \left(\frac{\widehat{\tau_{k,r}}-\tau_{k}}{\tau_{k}}\right)^2}$, where $k=1,2,\cdots,6$ for the six estimands. In addition, we report average standard error (ASE), empirical standard error (ESE) as well as coverage probability to assess the appropriateness of the asymptotic variance estimation.   Proper coverage probability should be aligned with the nominal confidence level such as 0.95 for 95\% confidence intervals.

\subsection{Results of Network-1 Scenario} 
\textbf{Estimation accuracy:} Table \ref{tab:setup1} indicates that, across all considered sample sizes, the estimation biases of the six estimated estimands are all minimal. Notably, as the network size increases, the absolute bias diminishes, thereby corroborating Theorem \ref{thm:thm3} on the estimation consistency. The RRMSE exhibits a similar pattern of improvement with increasing network sizes; that is, the RRMSE values of all six $\widehat{\tau_k}$ are declining, the evidence supporting
the statistical reliability of the MLE method. The observed decrease in RRMSE, alongside the apparently smaller bias with larger network, reinforces the theoretical results in Theorem \ref{thm:thm3}.\\

\noindent
\textbf{Estimation efficiency:} Table \ref{tab:setup1} also indicates that as the network size increases, specifically for $N=200,800$, ESE and ASE get close to each other for all the six estimands.
This suggests that the asymptotic variance estimation is reliable. More importantly, it is worth noting  that as the network size increases, the coverage probability for every 
estimand appears to getting closer to the nominal level of 0.95. Such reliable coverage rates
are indicative of the validity of the 95\% confidence intervals constructed via the large sample variance estimation. 

\subsection{Results of Network-2 Scenario} 
\textbf{Estimation accuracy:} The findings in the aspects of both estimation bias and the Relative Root Mean Square Error (RRMSE) in the Network-2 setting, as reported in Supplementary Table S3, reiterate the evidence reported in the Network-1 scenario, and thus details of the discussion are omitted.\\

\noindent
\textbf{Estimation efficiency:} Supplementary Table S3 repeats the key evidence seen in Table \ref{tab:setup1} to confirm the theoretical results of Theorem \ref{thm:thm3}. Once again, ESE for all six variance estimators are close to ASE even for a small network $N=100$. The coverage probability of each estimand tends towards the nominal level of 0.95 with increasing network size. These stable performances again also reflect that the variance estimators maintain consistent accuracy and reliability as the network size increases. Such numerical stability and robustness is appealing in practical studies.
We include two Supplementary Tables S1 and S2 that report the detailed results of the LREN-SEMs model parameters $\boldsymbol \varphi$, demonstrating the accuracy of the MLE estimation under the two network structures used in the respective simulation scenarios with $N=100,200,800$.
\begin{table}[ht]
\centering
\resizebox{0.5\textwidth}{!}{
\begin{tabular}{|c|c|c|c|c|c|c|c|}
\hline
\textbf{Size(N)}          & \textbf{Effects} & \textbf{Actual} & \textbf{Bias} & \textbf{RRMSE} & \textbf{ESE} & \textbf{ASE}& \textbf{CP}\\ \hline
\multirow{7}{*}{\textbf{100}} & \textbf{$\tau_1$}   & 1.50            & -0.028                  & 0.341                                                                             & 0.443 & 0.426      & 0.950                                                                           \\ \cline{2-8} 
                              & \textbf{$\tau_2$}   & 2.40                         & -0.095                   & 0.259                                                                               & 0.458 & 0.83   & 0.960                                                                          \\ \cline{2-8} 
                              & \textbf{$\tau_3$}   & 0.18                          & -0.005                   & 0.741                                                                                & 0.132 & 0.134 & 0.940                                                                             \\ \cline{2-8} 
                              & \textbf{$\tau_4$}   & 0.80            & -0.003               & 0.790                            & 0.626                                                                     & 0.620   & 0.948                                                                             \\ \cline{2-8} 
                              & \textbf{$\tau_5$}   & 1.08            & -0.042               & 0.651                             & 0.653                                                                     & 0.674  & 0.950                                                                            \\ \cline{2-8} 
                              & \textbf{$\tau_6$}   & 0.80            & -0.026               & 0.433                             & 0.321                                                                    & 0.324   & 0.955                                                                           \\ \hline
\multirow{7}{*}{\textbf{200}} & \textbf{$\tau_1$}   & 1.50            & -0.008               & 0.202                            & 0.303                                                                     & 0.297       & 0.948                                                                     \\ \cline{2-8} 
                              & \textbf{$\tau_2$}   & 2.40            & 0.005               & 0.140                              & 0.340                                                                     & 0.341  & 0.966                                                                            \\ \cline{2-8} 
                              & \textbf{$\tau_3$}   & 0.18            & -0.008               & 0.493                          & 0.088                                                                     & 0.090    & 0.940                                                                               \\ \cline{2-8} 
                              & \textbf{$\tau_4$}   & 0.80            & 0.034              & 0.565                          & 0.451                                                                    & 0.437       & 0.942                                                                        \\ \cline{2-8} 
                              & \textbf{$\tau_5$}   & 1.08            & -0.017               & 0.425                             & 0.459                                                                     & 0.477    & 0.964                                                                            \\ \cline{2-8} 
                              & \textbf{$\tau_6$}   & 0.80            & -0.022             & 0.285                             & 0.227                                                                     & 0.224     & 0.934                                                                          \\ \hline
\multirow{7}{*}{\textbf{800}} & \textbf{$\tau_1$}   & 1.50            & 0.005              & 0.098                              & 0.148                                                                     & 0.148      & 0.942                                                                         \\ \cline{2-8} 
                              & \textbf{$\tau_2$}   & 2.40            & 0.008               & 0.071                              & 0.170                                                                     & 0.170   & 0.946                                                                            \\ \cline{2-8} 
                              & \textbf{$\tau_3$}   & 0.18            & 0.002               & 0.250                             & 0.046                                                                     & 0.046    & 0.958                                                                            \\ \cline{2-8} 
                              & \textbf{$\tau_4$}   & 0.80            & 0.003               & 0.275                            & 0.220                                                                     & 0.218     & 0.952                                                                          \\ \cline{2-8} 
                              & \textbf{$\tau_5$}   & 1.08            & 0.006               & 0.217                              & 0.234                                                                     & 0.239    & 0.958                                                                           \\ \cline{2-8} 
                              & \textbf{$\tau_6$}   & 0.80            & 0.007               & 0.136                              & 0.108                                                                     & 0.110   & 0.954                                                     \\\hline
\end{tabular}}
\vspace{0.5cm}
\caption{Summarized simulation results under the Network-1 scenario from 500 rounds of simulations. The Bias and Root Relative Mean Squared Error (RRMSE) for the six estimated effects $\hat\tau_1$ through $\hat\tau_6$, Empirical Standard Error (ESE), Average Standard Error (ASE) and Coverage probability (CP) are presented, respectively.}\label{tab:setup1}
\end{table}
\section{Analysis of the Gamer's Friendship Network}\label{sec:twitch}
We now illustrate our methodology through the social network data of Twitch users, which is sourced from the public API in Spring 2018, as documented by \citep{rozemberczki2021twitch}. The friendship network is structured with nodes representing Twitch users (or gamer) and edges denoting mutual follower relationships between them, so they are conveniently referred to as friends in the data analysis. The total dataset comprises 168,114 nodes or gamers, but a subset of 1,415 nodes used in our analysis due purely to the computational costs. As discussed in Section \ref{sec:effects}, our primary focus is to apply our proposed methodology to assess the impact of mature content—both from individual users and the average within their network neighbors—on the views received by a user’s account. This relationship is further analyzed for potential mediation by the account lifetime of both the individual user and their neighboring nodes (or friends). In the dataset analyzed, the average network degree is observed to be 67. Approximately 50\% of the gamers' accounts are flagged for having mature content. The average account lifetime within our sample is 2,035 days, with a standard deviation of 730 days, indicating a substantial variation in the duration that accounts have been active. The average number of views per user's account stands at approximately 573,003 with a standard deviation of 227,498, highlighting significant disparities in viewership across different accounts. This dataset provides a rich framework for investigating how content maturity and network characteristics influence user engagement on the Twitch platform.\\

\noindent
Our analysis results are summarized as follows. 
Supplementary Figure S2 presents the estimates of the six causal estimands along with 95\% confidence intervals (C.I.s). Below are several findings in terms of the influence of maturity on viewer engagement. 
\begin{itemize}
    \item \textbf{$\tau_1(A\rightarrow Y)$:} The direct effect of a user's account having matured content on the number of views is -0.051 [95\% C.I (-0.064, -0.039)] with a p-value $<0.05$. This implies that streams containing mature content tend to attract fewer views compared to non-mature streams. This could be due to a smaller audience base that either prefers or is permitted to view such content.
    \item \textbf{$\tau_2(A\rightarrow M\rightarrow Y)$:} The mediated effect of a user's account having mature content on the number of views through the user's account lifetime is -0.003 [95\% C.I. (-0.078, 0.002)], with a p-value $>0.05$. This finding suggests that 
    this mediation pathway is not statistically significant.
    \item \textbf{$\tau_3(A\rightarrow \boldsymbol M_{\mathcal{N}^{\dagger}}\rightarrow Y)$:} The mediated effect of a user's account having mature content on the number of views through the neighbors' lifetime is 0.0012 [95\% C.I. (0.0008, 0.0016)], with a p-value $<0.05$. This suggests that having mature content is correlated with a decrease in his or her friends' account lifetime, which leads to an increase in the number of views for the user. This phenomenon might occur because the reduced lifetime of friend's accounts lessens competition for viewers, thereby channeling more traffic to the user's account. 
    \item \textbf{$\tau_4(\boldsymbol A_{\mathcal{N}^{\dagger}}\rightarrow Y)$:} The spillover effect of the average number of friends having mature content on the number of views of a user is -0.031 [95\% C.I. (-0.041, -0.022)], with a p-value $<0.05$. This implies that being surrounded by channels that host mature content generally reduces a channel's viewership. This decrease in viewership can possibly be attributed to factors similar to the direct effects of mature content, such as limited audience reach. 
    \item \textbf{$\tau_5(\boldsymbol A_{\mathcal{N}^{\dagger}}\rightarrow M \rightarrow Y)$:} The spillover mediated effect of the average number of friends having mature content on the number of views through a user's lifetime is -0.006 [95\% C.I. (-0.008, -0.004)], with a p-value $<0.05$. This indicates that having friends with mature content decreases lifetime of the user's own account, which in turn leads to fewer views. 
    \item \textbf{$\tau_6(\boldsymbol A_{\mathcal{N}^{\dagger}}\rightarrow \boldsymbol M_{\mathcal{N}^{\dagger}} \rightarrow Y):$} The spillover mediated effect of the average number of friends having mature content on the number of views through neighbors' lifetimes is 0.005 [95\% C.I. (-0.005, 0.010)], with a p-value$ >0.05$. There is no data evidence to assert the mechanistic pathway; that is, changes in friends' content maturity and their account lifetimes do not unveil a reliable impact on a user’s viewership. 
\end{itemize}
In summary, the influence of mature content on Twitch viewership is multifaceted, with direct and neighborhood effects playing significant roles. The negative direct and spillover effects highlight challenges in attracting viewers to mature content, while the positive mediated effects through neighbors' account lifetimes offer insights into potential strategies for mitigating these challenges.

\section{Concluding Remarks}\label{sec:con}
In this work, we have developed a rigorous theoretical framework for causal mediation analysis within networked systems, extending classical methodology beyond the assumption of independent units. Our approach, grounded in a system of correlated Directed Acyclic Graphs (DAGs) and the Random Effects Network Structural Equation Model (REN-SEM)—incorporating extended exposure mappings that explicitly account for mediators—identifies six fundamental estimands. These estimands decompose causal effects into distinct direct, indirect, and spillover components. This decomposition is critical for dissecting complex mechanisms in fields such as social sciences, epidemiology, and environmental health, where the interplay among units is often as significant as the units themselves.\\

\noindent
A cornerstone of this contribution is the rigorous specification of the identifiability conditions and identification formula necessary to differentiate direct, mediation, spillover, and spillover-mediation pathways under a general REN-SEM framework. Building on this general theory, we provided a detailed operationalization via the Linear REN-SEM (LREN-SEM), establishing key regularity conditions for consistent estimation and deriving extended asymptotic normality results that explicitly account for network dependence. These theoretical findings are supported by comprehensive simulation studies, which demonstrate high robustness and reproducibility; notably, the empirical coverage probabilities consistently approximate the nominal level across the majority of settings. Furthermore, the application to the Twitch dataset has elucidated novel pathways in gamer behavior, specifically revealing how interactions on streaming platforms influence user engagement through distinct spillover mechanisms.\\

\noindent
Despite these advancements, several limitations warrant further investigation. First, as in classical mediation, our identification relies on the assumption of no unobserved confounding—a condition that is empirically unverifiable and remains a fundamental challenge in observational causal mediation analysis. Second, our framework currently employs an approximate interventional strategy to accommodate extended exposure mappings; future research could explore rigorous stochastic intervention definitions to better address the discreteness of neighborhood summary measures. Third, the complexity of the REN-SEM—characterized by dense network dependencies and latent heterogeneity—currently precludes the derivation of doubly robust estimators. Developing such estimators for network data is inherently difficult, particularly when distinguishing between mediated effects transmitted through a unit's own mediator versus those of their neighbors—a distinction that has not been deeply explored in prior literature. To achieve double robustness, future extensions could incorporate stronger conditional independence assumptions between mediators and outcomes—given the observed exposures and confounders. Alternatively, the framework of \citet{liu2025auto} might be generalized to accommodate the structural complexities of network mediation analysis. Nevertheless, the framework presented here constitutes a significant step toward a unified methodology for dissecting these complex mechanistic pathways.

\section*{Code and Data Availability}
The R-code is available in the Github Package \url{https://github.com/Ritoban1/NETMED}. Data is obtained from \url{https://snap.stanford.edu/data/twitch-social-networks.html}.
\section*{Acknowledgements and Funding}
The authors thank Drs. Bhramar Mukherjee and Seunggeun Lee for their suggestions on the manuscript. This research was supported in part by NSF DMS2113564 and NIH R01ES033565 grants.

\bibliographystyle{abbrvnat} 
\bibliography{references.bib}

\newpage
\pagenumbering{arabic}
\singlespacing
\section*{Supplementary Materials}
\setcounter{figure}{0}
\renewcommand{\thefigure}{S\arabic{figure}}
\setcounter{table}{0}
\renewcommand{\thetable}{S\arabic{table}}
\setcounter{equation}{0}
\renewcommand{\theequation}{S\arabic{equation}}
\setcounter{section}{0}
\renewcommand{\thesection}{S\arabic{section}}
\setcounter{theorem}{0}
\renewcommand{\thetheorem}{S\arabic{theorem}}
\section{Technical Proofs}
\subsection{Proof of Theorem 1}
\begin{proof}
    First we condition on the random effects to operate under the conditional independence assumptions. Using the double expectation property, we write
\begin{align*}
    &\Psi_i=\vec{\mathbb{E}}_i^{a \ddagger}\left\{\widetilde{Y_i}\left(a,\boldsymbol t_{1}^{a'},\widetilde{M_i}\left(a^*,\boldsymbol t_{1}^{a\dagger}\right),\widetilde{\boldsymbol T}^m_{1i}\left(a'',\boldsymbol T_{2i}^a,\boldsymbol T_{3i}^a\right)\right) \right\}\\
    &=\int_{\boldsymbol b^y_{\mathcal{N}_i}}\int_{\boldsymbol b^m_{\mathcal{N}_i}}\int_{\boldsymbol b^m_{\mathcal{N}^\ddagger_i}}\vec{\mathbb{E}}_i^{a \ddagger}\left\{\widetilde{Y_i}\left(a,\boldsymbol t_{1}^{a'},\widetilde{M_i}\left(a^*,\boldsymbol t_{1}^{a\dagger}\right),\widetilde{\boldsymbol T}^m_{1i}\left(a'',\boldsymbol T_{2i}^a,\boldsymbol T_{3i}^a\right)\right) |\boldsymbol b^y_{\mathcal{N}_i},\boldsymbol b^m_{\mathcal{N}_i},\boldsymbol b^m_{\mathcal{N}^\ddagger_i}\right\}dF(\boldsymbol b^m_{\mathcal{N}^\ddagger_i})dF(\boldsymbol b^m_{\mathcal{N}_i})dF(\boldsymbol b^y_{\mathcal{N}_i})
\end{align*}
\noindent
Thus we focus on identifying the term $\mathbb{E}(\Psi_i|\boldsymbol b^y_{\mathcal{N}_i},\boldsymbol b^m_{\mathcal{N}_i},\boldsymbol b^m_{\mathcal{N}^\ddagger_i})$.
\begin{align*}
    &\mathbb{E}(\Psi_i|\boldsymbol b^y_{\mathcal{N}_i},\boldsymbol b^m_{\mathcal{N}_i},\boldsymbol b^m_{\mathcal{N}^\ddagger_i})=\vec{\mathbb{E}}_i^{a \ddagger}\left\{\widetilde{Y_i}\left(a,\boldsymbol t_{1}^{a'},\widetilde{M_i}\left(a^*,\boldsymbol t_{1}^{a\dagger}\right),\widetilde{\boldsymbol T}^m_{1i}\left(a'',\boldsymbol T_{2i}^a,\boldsymbol T_{3i}^a\right)\right) |\boldsymbol b^y_{\mathcal{N}_i},\boldsymbol b^m_{\mathcal{N}_i},\boldsymbol b^m_{\mathcal{N}^\ddagger_i}\right\}=\\
    &\mathbb{E}_{\boldsymbol X_i}\left[\mathbb{E}_{\boldsymbol T^a_{3i}|\boldsymbol X_i}\left\{\mathbb{E}_{\boldsymbol T_{2i}^a|\boldsymbol T_{1i}^a=\boldsymbol t_{1}^{a \ddagger},\boldsymbol X_i}\mathbb{E}\left(\widetilde{Y_i}\left(a,\boldsymbol t_{1}^{a'},\widetilde{M_i}\left(a^*,\boldsymbol t_{1}^{a\dagger}\right),\widetilde{\boldsymbol T}^m_{1i}\left(a'',\boldsymbol t_{2i}^a,\boldsymbol t_{3i}^a\right)\right)|\boldsymbol X_i, \boldsymbol T_{2i}^a=\boldsymbol t_{2i}^a,\boldsymbol T_{3i}^a=\boldsymbol t_{3i}^a,\boldsymbol b^y_{\mathcal{N}_i},\boldsymbol b^m_{\mathcal{N}_i},\boldsymbol b^m_{\mathcal{N}^\ddagger_i}\right)\right\}\right]
\end{align*}
We simplify the innermost conditional expectation using Identifiability Conditions 1-5 of main text as shown below.
\begin{align*}
&\mathbb{E}\left(\widetilde{Y_i}\left(a,\boldsymbol t_{1}^{a'},\widetilde{M_i}\left(a^*,\boldsymbol t_{1}^{a\dagger}\right),\widetilde{\boldsymbol T}^m_{1i}\left(a'',\boldsymbol t_{2i}^a,\boldsymbol t_{3i}^a\right)\right)|\boldsymbol X_i, \boldsymbol T_{2i}^a=\boldsymbol t_{2i}^a,\boldsymbol T_{3i}^a=\boldsymbol t_{3i}^a,\boldsymbol b^y_{\mathcal{N}_i},\boldsymbol b^m_{\mathcal{N}_i},\boldsymbol b^m_{\mathcal{N}^\ddagger_i}\right)\\
&=\int_{\boldsymbol t_{1}^m}\int_{m}\mathbb{E}\left(\widetilde{Y_i}\left(a,\boldsymbol t_{1}^{a'},m,\boldsymbol t^m_1\right)|\boldsymbol X_i, \boldsymbol T_{2i}^a=\boldsymbol t_{2i}^a,\boldsymbol T_{3i}^a=\boldsymbol t_{3i}^a,\widetilde{M_i}\left(a^*,\boldsymbol t_{1}^{a\dagger}\right)=m,\widetilde{\boldsymbol T}^m_{1i}\left(a'',\boldsymbol t_{2i}^a,\boldsymbol t_{3i}^a\right)=\boldsymbol t^m_1,\boldsymbol b^y_{\mathcal{N}_i},\boldsymbol b^m_{\mathcal{N}_i},\boldsymbol b^m_{\mathcal{N}^\ddagger_i}\right)\\
&\hspace{4cm}dF(m,\boldsymbol t^m_1)_{\widetilde{M_i}\left(a^*,\boldsymbol t_{1}^{a\dagger}\right),\widetilde{\boldsymbol T}^m_{1i}\left(a'',\boldsymbol t_{2i}^a,\boldsymbol t_{3i}^a\right)|\boldsymbol X_i, \boldsymbol T_{2i}^a=\boldsymbol t_{2i}^a,\boldsymbol T_{3i}^a=\boldsymbol t_{3i}^a,\boldsymbol b^y_{\mathcal{N}_i},\boldsymbol b^m_{\mathcal{N}_i},\boldsymbol b^m_{\mathcal{N}^\ddagger_i}}
\end{align*}
Using Identifiability Conditions 1, 3 and 5 of main text, the above term reduces to
\begin{align*}
    & \int_{\boldsymbol t_{1}^m}\int_{m}\mathbb{E}\left(\widetilde{Y_i}\left(a,\boldsymbol t_{1}^{a'},m,\boldsymbol t^m_1\right)|\boldsymbol X_i,\boldsymbol b^y_{\mathcal{N}_i}\right)dF(m,\boldsymbol t^m_1)_{\widetilde{M_i}\left(a^*,\boldsymbol t_{1}^{a\dagger}\right),\widetilde{\boldsymbol T}^m_{1i}\left(a'',\boldsymbol t_{2i}^a,\boldsymbol t_{3i}^a\right)|\boldsymbol X_i,\boldsymbol b^m_{\mathcal{N}_i},\boldsymbol b^m_{\mathcal{N}^\ddagger_i}}
\end{align*}
It follows from Identifiability Condition 4 of main text that the above expectation can be written as
\begin{align*}
\int_{\boldsymbol t_{1}^m}\int_{m}\mathbb{E}\left(\widetilde{Y_i}\left(a,\boldsymbol t_{1}^{a'},m,\boldsymbol t^m_1\right)|\boldsymbol X_i,\boldsymbol b^y_{\mathcal{N}_i}\right)dF(m)_{\widetilde{M_i}\left(a^*,\boldsymbol t_{1}^{a\dagger}\right)|\boldsymbol X_i,\boldsymbol b^m_{\mathcal{N}_i}}dF(\boldsymbol t^m_1)_{\widetilde{\boldsymbol T}^m_{1i}\left(a'',\boldsymbol t_{2i}^a,\boldsymbol t_{3i}^a\right)|\boldsymbol X_i,\boldsymbol b^m_{\mathcal{N}_i},\boldsymbol b^m_{\mathcal{N}^\ddagger_i}}
\end{align*}
Moreover Identifiability Conditions 1 and 3 of main text reduces the above expectation to
\begin{align*}
&\int_{\boldsymbol t_{1}^m}\int_{m}\mathbb{E}\left(\widetilde{Y_i}\left(a,\boldsymbol t_{1}^{a'},m,\boldsymbol t^m_1\right)|A_i=a,\boldsymbol T^a_1=\boldsymbol t_{1}^{a'},\boldsymbol X_i,\boldsymbol b^y_{\mathcal{N}_i}\right)dF(m)_{\widetilde{M_i}\left(a^*,\boldsymbol t_{1}^{a\dagger}\right)|A_i=a^*,\boldsymbol T^a_1=\boldsymbol t_{1}^{a\dagger},\boldsymbol X_i,\boldsymbol b^m_{\mathcal{N}_i}}\\
& \hspace{4cm} dF(\boldsymbol t^m_1)_{\widetilde{\boldsymbol T}^m_{1i}\left(a'',\boldsymbol t_{2i}^a,\boldsymbol t_{3i}^a\right)|\boldsymbol X_i,A_i=a^{''},\boldsymbol T^a_1=\boldsymbol t_{1}^{a\ddagger},\boldsymbol T^a_2=\boldsymbol t_{2}^{a},\boldsymbol T^a_3=\boldsymbol t_{3}^{a},\boldsymbol b^m_{\mathcal{N}_i},\boldsymbol b^m_{\mathcal{N}^\ddagger_i}}
\end{align*}
Using Identifiability Condition 2 of main text, the above expectation becomes
\begin{align*}
&\int_{\boldsymbol t_{1}^m}\int_{m}\mathbb{E}\left(\widetilde{Y_i}\left(a,\boldsymbol t_{1}^{a'},m,\boldsymbol t^m_1\right)|A_i=a,\boldsymbol T^a_1=\boldsymbol t_{1}^{a'},M_i=m,\widetilde{\boldsymbol T}^m_{1i}=\boldsymbol t^m_1,\boldsymbol X_i,\boldsymbol b^y_{\mathcal{N}_i}\right)\\
& dF(m)_{\widetilde{M_i}\left(a^*,\boldsymbol t_{1}^{a\dagger}\right)|A_i=a^*,\boldsymbol T^a_1=\boldsymbol t_{1}^{a\dagger},\boldsymbol X_i,\boldsymbol b^m_{\mathcal{N}_i}} dF(\boldsymbol t^m_1)_{\widetilde{\boldsymbol T}^m_{1i}\left(a'',\boldsymbol t_{2i}^a,\boldsymbol t_{3i}^a\right)|\boldsymbol X_i,A_i=a^{''},\boldsymbol T^a_1=\boldsymbol t_{1}^{a\ddagger},\boldsymbol T^a_2=\boldsymbol t_{2}^{a},\boldsymbol T^a_3=\boldsymbol t_{3}^{a},\boldsymbol b^m_{\mathcal{N}_i},\boldsymbol b^m_{\mathcal{N}^\ddagger_i}}
\end{align*}
Finally using the consistency assumptions in Assumption 2 of the main text the above expectation becomes
\begin{align*}
&\int_{\boldsymbol t_{1}^m}\int_{m}\mathbb{E}\left(\left. Y_i \right|A_i=a,\boldsymbol T_{1i}^a=\boldsymbol t_{1}^{a'},M_i=m_i,\widetilde{\boldsymbol T}_{1i}^m=\boldsymbol t_{1}^m,\boldsymbol X_i,\boldsymbol b^y_{\mathcal{N}_i}\right)\\
& \hspace{1cm}  dF(m)_{\left.M_i\right|A_i=a^*,\boldsymbol T_{1i}^a=\boldsymbol t_{1}^{a \dagger},\boldsymbol X_i,\boldsymbol b^m_{\mathcal{N}_i}}\cdot dF(\boldsymbol t_{1}^m)_{\left.\widetilde{\boldsymbol  T}^m_{1i}\right|A_i=a'',\boldsymbol T_{1i}^a=\boldsymbol t_{1}^{a \ddagger},\boldsymbol T_{2i}^a=\boldsymbol t_{2}^{a},\boldsymbol T_{3i}^a=\boldsymbol t_{3}^{a},\boldsymbol X_i,\boldsymbol b^m_{\mathcal{N}_i},\boldsymbol b^m_{\mathcal{N}^\ddagger_i}}
\end{align*}
This proves Theorem 1 that
\begin{align*}
&\boldsymbol \Psi_i=\int_{\boldsymbol b^y_{\mathcal{N}_i}}\int_{\boldsymbol b^m_{\mathcal{N}_i}}\int_{\boldsymbol b^m_{\mathcal{N}^\ddagger_i}}\int_{\boldsymbol X_i}\int_{\boldsymbol t^a_{3}}\int_{\boldsymbol t_{2}^a}\int_{\boldsymbol t_{1}^m}\int_{m}\mathbb{E}\left(\left. Y_i \right|A_i=a,\boldsymbol T_{1i}^a=\boldsymbol t_{1}^{a'},M_i=m_i,\widetilde{\boldsymbol T}_{1i}^m=\boldsymbol t_{1}^m,\boldsymbol X_i,\boldsymbol b^y_{\mathcal{N}_i}\right)\\
& \hspace{1cm}  dF(m)_{\left.M_i\right|A_i=a^*,\boldsymbol T_{1i}^a=\boldsymbol t_{1}^{a \dagger},\boldsymbol X_i,\boldsymbol b^m_{\mathcal{N}_i}}\cdot dF(\boldsymbol t_{1}^m)_{\left.\widetilde{\boldsymbol  T}^m_{1i}\right|A_i=a'',\boldsymbol T_{1i}^a=\boldsymbol t_{1}^{a \ddagger},\boldsymbol T_{2i}^a=\boldsymbol t_{2}^{a},\boldsymbol T_{3i}^a=\boldsymbol t_{3}^{a},\boldsymbol X_i,\boldsymbol b^m_{\mathcal{N}_i},\boldsymbol b^m_{\mathcal{N}^\ddagger_i}}\\
& \hspace{1cm}dF(\boldsymbol t_{2}^a)_{\boldsymbol T_{2i}^a|\boldsymbol T_{1i}^a=\boldsymbol t_{1}^{a \ddagger},\boldsymbol X_i}\cdot dF(\boldsymbol t^a_{3})_{\boldsymbol T_{3i}^a|\boldsymbol X_i}\cdot dF(\boldsymbol b^m_{\mathcal{N}^\ddagger_i})\cdot dF(\boldsymbol b^m_{\mathcal{N}_i})\cdot dF(\boldsymbol b^y_{\mathcal{N}_i})\cdot
\end{align*}
\end{proof}

\subsection{Proof of Corollary 1}
\begin{proof}
In this proof, we would directly Theorem 1 of main text and simplify the expression of $\Psi_i$ under LREN-SEM. We obtain that
\begin{align*}
    \Psi_i&=\int_{\text{Supp(M)}}\int_{\text{Supp}(S_{1i}^m)}\mathbb{E}\left(\beta_0+\beta_1a+\beta_2 \widetilde{s_{1i}}^{a'}+\beta_3m_i +\beta_4s_{1i}^m+ \beta_5 \boldsymbol C_i+\beta_6 S_{5i} + \boldsymbol E_i^{T} \boldsymbol b^y\right)\\
& \hspace{3cm}  dF(m_i)_{\left.M_i\right|A_i=a^*,S_{1i}^a=\widetilde{s_{1i}}^{a \dagger},\boldsymbol X_i,\boldsymbol b^m_{\mathcal{N}_i}}\\
    & \hspace{3cm}  dF(s_{1i}^m)_{\left.S_{1i}^m\right|A_i=a'',S_{1i}^a=\widetilde{s_{1i}}^{a \ddagger},S_{2i}^a=s_{2i},S_{3i}^a=s_{3i},\boldsymbol X_i,\boldsymbol b^m_{\mathcal{N}_i},\boldsymbol b^m_{\mathcal{N}^\ddagger_i}}\\
& = \beta_0+\beta_1a+\beta_2 \widetilde{s_{1i}}^{a'}+\beta_3 \mathbb{E}\left(\left .M_i\right|A_i=a^*,S_{1i}^a=\widetilde{s_{1i}}^{a \dagger},\boldsymbol X_i,\boldsymbol b^m_{\mathcal{N}_i}\right) \\
& +\beta_4 \mathbb{E}\left(S_{1i}^m|A_i=a'',S_{1i}^a=\widetilde{s_{1i}}^{a \ddagger},S_{2i}^a=s_{2i},S_{3i}^a=s_{3i},\boldsymbol X_i,\boldsymbol b^m_{\mathcal{N}_i},\boldsymbol b^m_{\mathcal{N}^\ddagger_i}\right)+ \boldsymbol\beta_5^T \boldsymbol C_i+\boldsymbol \beta_6^T \boldsymbol S_{5i} + \boldsymbol E_i^{T} \boldsymbol b^y\\
& = \beta_0+\beta_1a+\beta_2 \widetilde{s_{1i}}^{a'}+\beta_3 \left(\gamma_0+\gamma_1a^*+\gamma_2 \widetilde{s_{1i}}^{a \dagger}+ \boldsymbol\gamma_3^T \boldsymbol C_i + \boldsymbol\gamma_4^T \boldsymbol S_{5i} + \boldsymbol E_i^{T}\boldsymbol b^m\right)\\
& +\beta_4\left(\gamma_0+\gamma_1\widetilde{s_{1i}}^{a \dagger}+\gamma_2 a'' \sum_{j\neq i}\frac{E_{ij}}{n_in_j}+ \gamma_2 S_{2i}^a + \gamma_2 S_{3i}^a+ \boldsymbol\gamma_3^T \boldsymbol S_{5i}+ \boldsymbol \gamma_4^T \boldsymbol S_{6i}\right)+ \boldsymbol\beta_5^T \boldsymbol C_i+\boldsymbol\beta_6^T \boldsymbol S_{5i} + \boldsymbol E_i^{T} \boldsymbol b^y \cdot
\end{align*}
Finally we obtain the expression for $\Psi_i$ using $\mathbb{E}(b^y_k)=\mathbb{E}(b^m_k)=0$ for all $k\in \{1,2,...,N\}$.
\begin{align*}
    &\Psi_i=\vec{\mathbb{E}}_i^{a \ddagger}\left\{\widetilde{Y_i}\left(a,\widetilde{s_{1i}}^{a'},\widetilde{M_i}\left(a^*,\widetilde{s_{1i}}^{a\dagger}\right),S^{m}_{1i}\left(a'',\widetilde{s_{1i}}^{a\ddagger},S_{2i}^a,S_{3i}^a\right)\right) \right\}\\
     &=\beta_0+\beta_1a+\beta_2 \widetilde{s_{1i}}^{a'}+\beta_3 \left(\gamma_0+\gamma_1a^*+\gamma_2 \widetilde{s_{1i}}^{a\dagger}\right)+\beta_4\left\{\gamma_0+\gamma_1\widetilde{s_{1i}}^{a\ddagger}+\gamma_2 a''\left( \sum_{j\neq i}\frac{E_{ij}}{n_i n_j}\right)\right.\\
& \left.+\gamma_2  \mathbb{E}_{\boldsymbol X_i}\mathbb{E}\left(S_{2i}^a|S_{1i}^a=\widetilde{s_{1i}}^{a\ddagger},\boldsymbol X_i\right)\right\}+\beta_4\gamma_2 \mathbb{E}_{\boldsymbol X_i} \mathbb{E}(S_{3i}^a|\boldsymbol X_i) + \mathbb{E}_{\boldsymbol X_i}\{(\boldsymbol\beta_5^T+\beta_3 \boldsymbol\gamma_3^T)\boldsymbol C_i+(\boldsymbol \beta_6^T+\beta_3\boldsymbol\gamma_4^T) \boldsymbol S_{5i}+\beta_4\boldsymbol\gamma_4^T \boldsymbol  S_{6i}\}\cdot
\end{align*}
\end{proof}

\subsection{Proof of (f) of Corollary 2}
\begin{proof}
Using Corollary 1 of main text, we obtain  
\begin{align*}
\tau_6(d',d)&=\beta_4\gamma_1\left\{\frac{1}{N}\sum_{i=1}^N (\widetilde{s_{1i}}^a-\widetilde{s_{1i}}^{a'})\right\}+ \beta_4\gamma_2\left[\frac{1}{N}\sum_{i=1}^N \mathbb{E}_{\boldsymbol X_i}\left\{\mathbb{E}\left(S_{2i}^a|S_{1i}^a=\widetilde{s_{1i}}^a,\boldsymbol X_i\right)-\mathbb{E}\left(S_{2i}^a|S_{1i}^a=\widetilde{s_{1i}}^{a'},\boldsymbol X_i\right)\right\}\right]
\end{align*}
Assumption 4 of main text implies that marginally $A_i\sim \text{Ber}(p)$. Using this assumption, the expectation
\begin{align*}
    & \mathbb{E}_{\boldsymbol X_i}\left[\mathbb{E}\left(S_{2i}^a|S_{1i}^a=\widetilde{s_{1i}}^a,\boldsymbol X_i\right)\right]=\mathbb{E}\left(S_{2i}^a|S_{1i}^a=\widetilde{s_{1i}}^a\right)\\
    & =\mathbb{E}\left(\sum_{j\neq i}\left.\frac{\sum_{k\neq j,k\in \mathcal{N}_i} E_{ij}E_{jk}A_k}{n_in_j}\right|S_{1i}^a=\widetilde{s_{1i}}^a\right)=\sum_{j\neq i}\frac{\sum_{k\neq j,k\in \mathcal{N}_i} E_{ij}E_{jk}\mathbb{E}(A_k|S_{1i}^a=\widetilde{s_{1i}}^a)}{n_i n_j}\\
    &=\sum_{j\neq i}\frac{\sum_{k\neq j,k\in \mathcal{N}_i} E_{ij}E_{jk}\mathbb{E}(A_k|\sum_{j\neq i} E_{ij}A_j=[s_1^an_i])}{n_i n_j}
\end{align*}
Since $A_k \in \mathcal{N}_i$ in the expectation term we obtain 
$$\mathbb{E}\left(A_k|\sum_{j\neq i} E_{ij}A_j=[s_1^an_i])\right)=\mathbb{E}\left(A_l|\sum_{j\neq i} E_{ij}A_j=[s_1^an_i]\right)$$
where $k\neq l$ and $k,l \in \mathcal{N}_i$. Hence we obtain $\forall k\in \mathcal{N}_i$, $\mathbb{E}\left(A_k|\sum_{j\neq i} E_{ij}A_j=[s_1^an_i]\right)=\widetilde{s_{1i}}^a$. Using this result
\begin{align*}
    \sum_{j\neq i}\frac{\sum_{k\neq j,k\in \mathcal{N}_i} E_{ij}E_{jk}\mathbb{E}(A_k|\sum_{j\neq i} E_{ij}A_j=[s_1^an_i])}{n_i n_j}=\sum_{j\neq i}\frac{\sum_{k\neq j,k\in \mathcal{N}_i} E_{ij}E_{jk}\cdot \widetilde{s_{1i}}^a}{n_i n_j}
\end{align*}
Therefore the final expression of $\tau_6(d',d)$ is 
\begin{align*}
& \tau_6(d',d)=\beta_4\gamma_1\left\{\frac{1}{N}\sum_{i=1}^N (\widetilde{s_{1i}}^a-\widetilde{s_{1i}}^{a'})\right\}+ \beta_4\gamma_2\left[\frac{1}{N}\sum_{i=1}^N \left\{\sum_{j\neq i}\frac{\sum_{k\neq j,k\in \mathcal{N}_i} E_{ij}E_{jk}\left(\widetilde{s_{1i}}^a-\widetilde{s_{1i}}^{a'}\right)}{n_i n_j }\right\}\right]\cdot
\end{align*}   
\end{proof}
\subsection{Hessian Calculation}\label{sec:hessian}

First the first order derivatives of the log likelihood are calculated.\\
\begin{align*}
    &\boldsymbol L_{\boldsymbol{\beta}}=\nabla_{\boldsymbol \beta}l(\boldsymbol \varphi)=-\boldsymbol X_{\boldsymbol y}^T(\Sigma_y)^{-1}\boldsymbol X_{\boldsymbol y}\boldsymbol \beta+\boldsymbol X_{\boldsymbol y}^T(\Sigma_y)^{-1}\boldsymbol Y\\
    &  \boldsymbol L_{\boldsymbol{\gamma}}=\nabla_{\boldsymbol{\gamma}}l(\boldsymbol \varphi)=-\boldsymbol X_{\boldsymbol m}^T(\Sigma_m)^{-1}\boldsymbol X_{\boldsymbol m}\boldsymbol \gamma+\boldsymbol X_{\boldsymbol m}^T(\Sigma_m)^{-1}\boldsymbol M\\
    & \boldsymbol L_{\boldsymbol{\alpha}}=\nabla_{\boldsymbol{\alpha}}l(\boldsymbol \varphi)=\sum_{i=1}^N \left(A_i-\frac{e^{\boldsymbol\alpha'\boldsymbol X_{ai}}}{1+e^{\boldsymbol\alpha'\boldsymbol X_{ai}}}\right)\boldsymbol X_{ai}  
\end{align*}
Let $\boldsymbol W_{\boldsymbol y}=\boldsymbol Y-\boldsymbol X_{\boldsymbol y}\boldsymbol \beta$ and $\boldsymbol W_{\boldsymbol m}=\boldsymbol M-\boldsymbol X_{\boldsymbol m}\boldsymbol \gamma$. Then we obtain 
\begin{align*}
   & \boldsymbol L_{\sigma^2_{\boldsymbol y}}=\nabla_{\sigma^2_{\boldsymbol y}}l(\boldsymbol \varphi)=-\frac{1}{2} \text{tr}(\Sigma_y^{-1})+\frac{1}{2} \boldsymbol W_{\boldsymbol y}^T\Sigma_y^{-2}\boldsymbol W_{\boldsymbol y}\\
    &  \boldsymbol L_{\sigma^2_{\boldsymbol b^{\boldsymbol Y}}}=\nabla_{\sigma^2_{\boldsymbol b^{\boldsymbol Y}}}l(\boldsymbol \varphi)=-\frac{1}{2} \text{tr}(\Sigma_y^{-1} EE^{T})+\frac{1}{2} \boldsymbol W_{\boldsymbol y}^T(\Sigma_y^{-1} EE^{T}\Sigma_y^{-1})\boldsymbol W_{\boldsymbol y}\\
    &  \boldsymbol L_{\sigma^2_{\boldsymbol m}}=\nabla_{\sigma^2_{\boldsymbol m}}l(\boldsymbol \varphi)=-\frac{1}{2} \text{tr}(\Sigma_m^{-1})+\frac{1}{2} \boldsymbol W_{\boldsymbol m}^T\Sigma_m^{-2}\boldsymbol W_{\boldsymbol m}\\
     &  \boldsymbol L_{\sigma^2_{\boldsymbol b^{\boldsymbol M}}}=\nabla_{\sigma^2_{\boldsymbol b^{\boldsymbol M}}}l(\boldsymbol \varphi)=-\frac{1}{2} \text{tr}(\Sigma_y^{-1} EE^{T})+\frac{1}{2} \boldsymbol W_{\boldsymbol m}^T(\Sigma_m^{-1} EE^{T}\Sigma_m^{-1})\boldsymbol W_{\boldsymbol m}
\end{align*}
The second order derivatives are given by
\begin{align*}
    & L_{\boldsymbol \beta \boldsymbol \beta}=\nabla^2_{\boldsymbol \beta \boldsymbol \beta}l(\boldsymbol \varphi)=-X_{\boldsymbol y}^T\Sigma_y^{-1}X_{\boldsymbol y}\\
    & L_{\boldsymbol \beta \sigma^2_{\boldsymbol y}}=\nabla^2_{\boldsymbol \beta \sigma^2_{\boldsymbol y}}l(\boldsymbol \varphi)=X_{\boldsymbol y}^T(\Sigma_y^{-1}\Sigma_y^{-1})X_{\boldsymbol y}\boldsymbol \beta-X_{\boldsymbol y}^T\Sigma_y^{-2}\boldsymbol Y\\
    & L_{\boldsymbol \beta \sigma^2_{\boldsymbol b^{\boldsymbol Y}}}=\nabla^2_{\boldsymbol \beta \sigma^2_{\boldsymbol b^{\boldsymbol Y}}}l(\boldsymbol \varphi)=X_{\boldsymbol y}^T(\Sigma_y^{-1} EE^{T}  \Sigma_y^{-1})X_{\boldsymbol y}\boldsymbol \beta-X_{\boldsymbol y}^T(\Sigma_y^{-1} EE^{T} \Sigma_y^{-1})\boldsymbol Y\\
     & L_{\sigma^2_{\boldsymbol y} \sigma^2_{\boldsymbol y}}=\nabla^2_{\sigma^2_{\boldsymbol y} \sigma^2_{\boldsymbol y}}l(\boldsymbol \varphi)=\frac{1}{2} \text{tr}(\Sigma_y^{-2})-\frac{1}{2} \boldsymbol W_{\boldsymbol y}^T(\Sigma_y)^{-3} \boldsymbol W_{\boldsymbol y}\\
     & L_{\sigma^2_{\boldsymbol b^{\boldsymbol Y}} \sigma^2_{\boldsymbol b^{\boldsymbol Y}}}=\nabla^2_{\sigma^2_{\boldsymbol b^{\boldsymbol Y}} \sigma^2_{\boldsymbol b^{\boldsymbol Y}}}l(\boldsymbol \varphi)=\frac{1}{2} \text{tr}(\Sigma_y^{-1} EE^{T}\Sigma_y^{-1} EE^{T})-\frac{1}{2} \boldsymbol W_{\boldsymbol y}^T(\Sigma_y^{-1} EE^{T}\Sigma_y^{-1} EE^{T} \Sigma_y^{-1}) \boldsymbol W_{\boldsymbol y}\\
     & L_{\sigma^2_{\boldsymbol y} \sigma^2_{\boldsymbol b^{\boldsymbol Y}}}=\nabla^2_{\sigma^2_{\boldsymbol y} \sigma^2_{\boldsymbol b^{\boldsymbol Y}}}l(\boldsymbol \varphi)=\frac{1}{2} \text{tr}(\Sigma_y^{-1} EE^{T}\Sigma_y^{-1})-\frac{1}{2} \boldsymbol W_{\boldsymbol y}^T(\Sigma_y^{-1}\Sigma_y^{-1} EE^{T}\Sigma_y^{-1}+\Sigma_y^{-1} EE^{T} \Sigma_y^{-1}\Sigma_y^{-1}) \boldsymbol W_{\boldsymbol y}
\end{align*}
\begin{align*}
       & L_{\boldsymbol \gamma \boldsymbol \gamma}=\nabla^2_{\boldsymbol \gamma \boldsymbol \gamma}l(\boldsymbol \varphi)=-X_{\boldsymbol m}^T\Sigma_m^{-1}X_{\boldsymbol m}\\
    & L_{\boldsymbol \gamma \sigma^2_{\boldsymbol m}}=\nabla^2_{\boldsymbol \gamma \sigma^2_{\boldsymbol m}}l(\boldsymbol \varphi)=X_{\boldsymbol m}^T(\Sigma_m^{-1}\Sigma_m^{-1})X_{\boldsymbol m}\boldsymbol \gamma-X_{\boldsymbol m}^T\Sigma_m^{-2}\boldsymbol M\\
    & L_{\boldsymbol \gamma \sigma^2_{\boldsymbol b^{\boldsymbol M}}}=\nabla^2_{\boldsymbol \gamma \sigma^2_{\boldsymbol b^{\boldsymbol M}}}l(\boldsymbol \varphi)=X_{\boldsymbol m}^T(\Sigma_m^{-1} EE^{T}  \Sigma_m^{-1})X_{\boldsymbol m}\boldsymbol \gamma-X_{\boldsymbol m}^T(\Sigma_m^{-1} EE^{T} \Sigma_m^{-1})\boldsymbol M\\
     & L_{\sigma^2_{\boldsymbol m} \sigma^2_{\boldsymbol m}}=\nabla^2_{\sigma^2_{\boldsymbol m} \sigma^2_{\boldsymbol m}}l(\boldsymbol \varphi)=\frac{1}{2} \text{tr}(\Sigma_m^{-2})-\frac{1}{2} \boldsymbol W_{\boldsymbol m}^T(\Sigma_m)^{-3} \boldsymbol W_{\boldsymbol m}\\
     & L_{\sigma^2_{\boldsymbol b^{\boldsymbol M}} \sigma^2_{\boldsymbol b^{\boldsymbol M}}}=\nabla^2_{\sigma^2_{\boldsymbol b^{\boldsymbol M}} \sigma^2_{\boldsymbol b^{\boldsymbol M}}}l(\boldsymbol \varphi)=\frac{1}{2} \text{tr}(\Sigma_m^{-1} EE^{T}\Sigma_m^{-1} EE^{T})-\frac{1}{2} \boldsymbol W_{\boldsymbol m}^T(\Sigma_m^{-1} EE^{T}\Sigma_m^{-1} EE^{T} \Sigma_m^{-1}) \boldsymbol W_{\boldsymbol m}\\
     & L_{\sigma^2_{\boldsymbol m} \sigma^2_{\boldsymbol b^{\boldsymbol M}}}=\nabla^2_{\sigma^2_{\boldsymbol m} \sigma^2_{\boldsymbol b^{\boldsymbol M}}}l(\boldsymbol \varphi)=\frac{1}{2} \text{tr}(\Sigma_y^{-1} EE^{T}\Sigma_m^{-1})-\frac{1}{2} \boldsymbol W_{\boldsymbol m}^T(\Sigma_m^{-1}\Sigma_m^{-1} EE^{T}\Sigma_m^{-1}+\Sigma_m^{-1} EE^{T} \Sigma_m^{-1}\Sigma_m^{-1}) \boldsymbol W_{\boldsymbol m}
\end{align*}
\begin{align*}
    L_{\boldsymbol \alpha \boldsymbol \alpha}=\nabla^2_{\boldsymbol \alpha \boldsymbol \alpha}l(\boldsymbol \varphi)=-\sum_{i=1}^N\frac{e^{\boldsymbol\alpha^T\boldsymbol X_{ai}}}{(1+e^{\boldsymbol\alpha^T\boldsymbol X_{ai}})^2} \boldsymbol X_{ai}\boldsymbol X_{ai}^T
\end{align*}
All the other terms in $L_N^{(2)}(\boldsymbol \varphi)$ are 0.

\subsection{Proof of Theorem 3}
Before proving this theorem, at first we prove the following Lemma.
\begin{lemma}\label{lem:Lemma1}
    For a given network of size $N$, $|\bar\Delta_{1N}|\leq 1, |\bar\Delta_{2N}|\leq 1$ and $|\bar\Delta_{3N}|\leq 1$.
\end{lemma}
\begin{proof}
Note that, $|\bar\Delta_{1N}|=\bar\Delta_{1N}=\frac{1}{N}\sum_{i=1}^N\sum_{j\neq i}\frac{E_{ij}}{n_in_j}$.
Since under LREN-SEM $\forall i \in \{1,2,\cdots,N\}$, $n_i\geq 1$. Using $\sum_{j \neq i}E_{ij}=n_i$, one can write,
\begin{align*}
    |\bar\Delta_{1N}|\leq \frac{1}{N}\sum_{i=1}^N\sum_{j\neq i}\frac{E_{ij}}{n_i}=1 \cdot
\end{align*}
\noindent
By the triangle inequality one can write, 
$$|\bar\Delta_{2N}|=\left|\frac{1}{N}\sum_{i=1}^N (\widetilde{s_{1i}}^a-\widetilde{s_{1i}}^{a'})\right| \leq \frac{1}{N}\sum_{i=1}^N |\widetilde{s_{1i}}^a-\widetilde{s_{1i}}^{a'}|$$
Since $\forall i \in \{1,2,\cdots,N\}$, $s_{1i}^a,s_{1i}^{a'} \in [0,1]$, $|s_{1i}^a-s_{1i}^{a'}|\leq 1$. This implies that, $|[s_{1i}^an_i]-[s_{1i}^{a'} n_i]|=n_i\left|\frac{[s_{1i}^an_i]}{n_i}-\frac{[s_{1i}^{a'} n_i]}{n_i}\right|=n_i|\widetilde{s_{1i}}^a-\widetilde{s_{1i}}^{a'}|\leq n_i$ and hence we obtain, $|\bar\Delta_{2N}|\leq 1$.\\

\noindent
Finally, $\bar\Delta_{3N} = N^{-1}\sum_{i=1}^N \sum_{j\neq i} (n_in_j)^{-1} \sum_{k\neq j,k\in \mathcal{N}^{\dagger}_i} E_{ij}E_{jk} \left(\widetilde{s_{1i}}^a-\widetilde{s_{1i}}^{a'}\right)$. Hence again using the result that $\forall i \in \{1,2,\cdots,N\}$, $|\widetilde{s_{1i}}^a-\widetilde{s_{1i}}^{a'}|\leq 1$ and triangle inequality,
\begin{align*}
    |\bar\Delta_{3N}|&=N^{-1}\sum_{i=1}^N \left|\sum_{j\neq i} (n_in_j)^{-1} \sum_{k\neq j,k\in \mathcal{N}^{\dagger}_i} E_{ij}E_{jk} \left(\widetilde{s_{1i}}^a-\widetilde{s_{1i}}^{a'}\right)\right|\\
    & \leq N^{-1}\sum_{i=1}^N \sum_{j\neq i} (n_in_j)^{-1} \sum_{k\neq j,k\in \mathcal{N}^{\dagger}_i} E_{ij}E_{jk} \left|\widetilde{s_{1i}}^a-\widetilde{s_{1i}}^{a'}\right|\\
    &\leq N^{-1}\sum_{i=1}^N \left\{\sum_{j\neq i} (n_in_j)^{-1} \sum_{k\neq j,k\in \mathcal{N}^{\dagger}_i} E_{ij}E_{jk}\right\}=N^{-1}\sum_{i=1}^N (n_i)^{-1}\left\{\sum_{j\neq i}  E_{ij}\right\}\\
    & \leq N^{-1}\sum_{i=1}^N \left\{\sum_{j\neq i} (n_in_j)^{-1} E_{ij}\sum_{k\neq j}E_{jk}\right\}=1 \cdot
\end{align*}
\end{proof}
\noindent
Next we prove Theorem 3 of main text.
\begin{proof}
Since $g(.)$ might depend on the network size $N$, we cannot apply Continuous Mapping Theorem to prove the consistency of $g(\widehat{\boldsymbol \varphi_N})$.
By the Multivariate Taylor Theorem of $g(\widehat{\boldsymbol \varphi_N})$ around $\boldsymbol \varphi_0$, we obtain 
\begin{equation}
    g(\widehat{\boldsymbol \varphi_N})=g(\boldsymbol\varphi_0)+\nabla_{\boldsymbol \varphi_0} g(\boldsymbol\varphi)^T (\widehat{\boldsymbol \varphi_N}-\boldsymbol \varphi_0)+\frac{1}{2}(\widehat{\boldsymbol \varphi_N}-\boldsymbol \varphi_0)^T\nabla_{\boldsymbol \varphi'}^2 g(\boldsymbol\varphi) (\widehat{\boldsymbol \varphi_N}-\boldsymbol \varphi_0)\cdot \label{eq:taylor}
\end{equation}
where $\boldsymbol\varphi'$ lies in a $\epsilon-$ neighborhood of $\boldsymbol\varphi_0$. By Theorem 2, we obtain $(\widehat{\boldsymbol \varphi_N}-\boldsymbol \varphi_0)\xrightarrow{p}\boldsymbol 0$. Using Lemma 1 and compactness of the parameter space $\Theta$, one can easily observe that this Assumption 8(i)  holds for all the six estimands. Hence $\nabla_{\boldsymbol \varphi_0} g(\boldsymbol\varphi)^T (\widehat{\boldsymbol \varphi_N}-\boldsymbol \varphi_0)=o_p(1)$. Similarly we obtain $(\widehat{\boldsymbol \varphi_N}-\boldsymbol \varphi_0)^T\nabla_{\boldsymbol \varphi'}^2 g(\boldsymbol\varphi) (\widehat{\boldsymbol \varphi_N}-\boldsymbol \varphi_0)=o_p(1)$. 
Hence, $g(\widehat{\boldsymbol \varphi_N})-g(\boldsymbol\varphi_0)=o_p(1)$, which implies $g(\widehat{\boldsymbol \varphi_N})-g(\boldsymbol \varphi_0) \xrightarrow{p} 0$. This completes the proof of consistency.\\

\noindent
Next we prove the asymptotic normality. Multiplying equation \eqref{eq:taylor} by $Q_N^{-\frac{1}{2}}$ we obtain,
\begin{align}
    & Q_N^{-\frac{1}{2}}\{g(\widehat{\boldsymbol \varphi_N})-g(\boldsymbol\varphi_0)\}=Q_N^{-\frac{1}{2}}\nabla_{\boldsymbol\varphi_0} g(\boldsymbol\varphi)^T(\widehat{\boldsymbol \varphi_N}-\boldsymbol \varphi_0)+ \nonumber\\
    & \hspace{5cm}\frac{1}{2}Q_N^{-\frac{1}{2}}(\widehat{\boldsymbol \varphi_N}-\boldsymbol \varphi_0)^T\nabla_{\boldsymbol \varphi'}^2 g(\boldsymbol\varphi) (\widehat{\boldsymbol \varphi_N}-\boldsymbol \varphi_0)\cdot \label{eq:taylor2}
\end{align}
By the spectral decomposition theorem for a positive definite symmetric matrix, $B_N^{-1}(\boldsymbol\varphi_0)$, one can write $B_N^{-1}(\boldsymbol\varphi_0)=\sum_{k=1}^K\lambda_{N,k}\boldsymbol v_{N,k} \boldsymbol v_{N,k}^T$,
where $\boldsymbol v_{N,1},\boldsymbol v_{N,2},\cdots,\boldsymbol v_{N,K}$ are the orthonormal eigenvectors corresponding to the eigenvalues,
$\lambda_{N,1}\geq\lambda_{N,2}\geq\cdots \geq\lambda_{N,K}$ of $B_N^{-1}(\boldsymbol\varphi_0)$.
Since $\{\boldsymbol v_{N,1},\boldsymbol v_{N,2},\cdots,\boldsymbol v_{N,K}\}$ spans $\mathbb{R}^K$, one can write $\nabla_{\boldsymbol\varphi_0} g(\boldsymbol\varphi)=\sum_{k=1}^K z_{N,k}\boldsymbol v_{N,k}$.
Hence,
\begin{align*}
    Q_N=\nabla_{\boldsymbol\varphi_0} g(\boldsymbol\varphi)^T B_N^{-1}(\boldsymbol\varphi_0)\nabla_{\boldsymbol\varphi_0} g(\boldsymbol\varphi)=\left(\sum_{k=1}^K z_{N,k}\boldsymbol v_{N,k}\right)^T\left(\sum_{k=1}^K\lambda_{N,k}\boldsymbol v_{N,k} \boldsymbol v_{N,k}^T\right)\left(\sum_{k=1}^K z_{N,k}\boldsymbol v_{N,k}\right)
\end{align*}
Since $\{\boldsymbol v_{N,1},\boldsymbol v_{N,2},\cdots,\boldsymbol v_{N,K}\}$ are orthonormal eigenvectors, we obtain $Q_N=\sum_{k=1}^K\lambda_{N,k}(z_{N,k})^2$.
Using Assumption 8(ii), we obtain $ Q_N=O(N^{-\delta_1})$.\\

\noindent
The first term in the R.H.S of equation \eqref{eq:taylor2}, can be written as
\begin{align*}
    Q_N^{-\frac{1}{2}}\nabla_{\boldsymbol\varphi_0} g(\boldsymbol\varphi)^T(\widehat{\boldsymbol \varphi_N}-\boldsymbol \varphi_0)=Q_N^{-\frac{1}{2}}\nabla_{\boldsymbol\varphi_0} g(\boldsymbol\varphi)^TB_N^{-\frac{1}{2}}(\boldsymbol\varphi_0)B_N^{\frac{1}{2}}(\boldsymbol\varphi_0)(\widehat{\boldsymbol \varphi_N}-\boldsymbol \varphi_0)
\end{align*}
where the term, $Q_N^{-\frac{1}{2}}\nabla_{\boldsymbol\varphi_0} g(\boldsymbol\varphi)^TB_N^{-\frac{1}{2}}(\boldsymbol\varphi_0)$ is assessed as follows. Noting that $B_N^{-\frac{1}{2}}(\boldsymbol\varphi_0)=\sum_{k=1}^K\sqrt{\lambda_{N,k}}\boldsymbol v_{N,k} \boldsymbol v_{N,k}^T$, we obtain
\begin{align*}
   \nabla_{\boldsymbol\varphi_0} g(\boldsymbol\varphi)^TB_N^{-\frac{1}{2}}(\boldsymbol\varphi_0)&=\left(\sum_{k=1}^K z_{N,k}\boldsymbol v_{N,k}\right)^T\left(\sum_{k=1}^K\sqrt{\lambda_{N,k}}\boldsymbol v_{N,k} \boldsymbol v_{N,k}^T\right)=\sum_{k=1}^K z_{N,k}\sqrt{\lambda_{N,k}}\boldsymbol v_{N,k}^T
\end{align*}
One can write $Q_N^{-\frac{1}{2}}\nabla_{\boldsymbol\varphi_0} g(\boldsymbol\varphi)^TB_N^{-\frac{1}{2}}(\boldsymbol\varphi_0)=\left(\sum_{k=1}^K Q_N^{-\frac{1}{2}}\sqrt{\lambda_{N,k}} z_{N,k}\boldsymbol v_{N,k}^T\right)$. Under Assumption 8(ii) we have $z_{N,k}\boldsymbol v_{N,k}^T=\boldsymbol v_{N,k}^T O(1)$ $\forall k \in \{1,2,\cdots,K\}$. Hence for all $r<k\leq K$, we have,
\begin{align*}
      Q_N^{-\frac{1}{2}}\sqrt{\lambda_{N,k}} z_{N,k}\boldsymbol v_{N,k}^T&=O(N^{\frac{1}{2}\delta_1})O(\lambda_{N,k}^{\frac{1}{2}})\boldsymbol v_{N,k}^T=O(N^{\frac{1}{2}\delta_1})O(N^{-\frac{1}{2}\delta_{r+1}})\boldsymbol v_{N,k}^T\\
      &=O(N^{(-\frac{1}{2}(\delta_{r+1}-\delta_1))})\boldsymbol v_{N,k}^T=o(1)\boldsymbol v_{N,k}^T
\end{align*}
Hence for all $r<k\leq K$, we have $\lim_{N\rightarrow \infty}Q_N^{-\frac{1}{2}}z_{N,k}\sqrt{\lambda_{N,k}}\boldsymbol v_{N,k}^T=\boldsymbol 0^T$.\\

\noindent
Next for all $1 \leq k\leq r$, we have
\begin{align*}
    Q_N^{-\frac{1}{2}}\sqrt{\lambda_{N,k}} z_{N,k}\boldsymbol v_{N,k}^T& =\boldsymbol u_{N,k}^T=\frac{\sqrt{\lambda_{N,k}}z_{N,k}\boldsymbol v_{N,k}^T}{\sqrt{\sum_{k=1}^K\lambda_{N,k}z_{N,k}^2}}
\end{align*}
We observe that $\boldsymbol u_{N,k}^T\boldsymbol u_{N,k}=\frac{\lambda_{N,k}z^2_{N,k}}{\sum_{k=1}^K\lambda_{N,k}z_{N,k}^2}$. Hence $||u_{N,k}||^2_2=\boldsymbol u_{N,k}^T\boldsymbol u_{N,k}=\frac{\lambda_{N,k}z^2_{N,k}}{\sum_{k=1}^K\lambda_{N,k}z_{N,k}^2}$ is bounded between 0 and 1 for $\forall N\in \mathbb{N}$ and for all $1 \leq k\leq r$. Hence by the Bolzano–Weierstrass theorem for finite-dimensional bounded norm vector sequence, we obtain there exist a sub-sequence of $\{\boldsymbol u_{N_t,k}^T\}_{t=1}^{\infty}$, of $\{\boldsymbol u_{N,k}^T\}$, such that $\lim_{t\rightarrow \infty}\boldsymbol u_{N_t,k}^T=\boldsymbol w_k^T$, $\forall k\in\{1,2,\cdots,r\}$, where $\boldsymbol w_1^T,\boldsymbol w_2^T,\cdots,\boldsymbol w_r^T\in \mathbb{R}^K /\ \{0\}$. Therefore by Assumption 8(iii), we have for all $1 \leq k\leq r$, $\lim_{N\rightarrow \infty}Q_N^{-\frac{1}{2}}z_{N,k}\sqrt{\lambda_{N,k}}\boldsymbol v_{N,k}^T=\boldsymbol w_k^T$. Hence we obtain that 
\begin{align}
    \lim_{N\rightarrow \infty}Q_N^{-\frac{1}{2}}\nabla_{\boldsymbol\varphi_0} g(\boldsymbol\varphi)^TB_N^{-\frac{1}{2}}(\boldsymbol\varphi_0)=\boldsymbol w^{*T}:=\sum_{k=1}^r\boldsymbol w_k^T \label{eq:slutsky}
\end{align}
Since for all $N\geq 1$, using the expression of $Q_N=\{\nabla_{\boldsymbol\varphi_0} g(\boldsymbol\varphi)\}^T B_N^{-1}(\boldsymbol\varphi_0)\{\nabla_{\boldsymbol\varphi_0} g(\boldsymbol\varphi)\}$,
$$Q_N^{-\frac{1}{2}}\nabla_{\boldsymbol\varphi_0} g(\boldsymbol\varphi)^TB_N^{-\frac{1}{2}}(\boldsymbol\varphi_0)B_N^{-\frac{1}{2}}(\boldsymbol\varphi_0)\nabla_{\boldsymbol\varphi_0} g(\boldsymbol\varphi)Q_N^{-\frac{1}{2}}=Q_N^{-1}\nabla_{\boldsymbol\varphi_0} g(\boldsymbol\varphi)^TB_N^{-\frac{1}{2}}(\boldsymbol\varphi_0)B_N^{-\frac{1}{2}}(\boldsymbol\varphi_0)\nabla_{\boldsymbol\varphi_0} g(\boldsymbol\varphi)=1,$$ therefore $\boldsymbol w^{*T}\boldsymbol w^*=1$. Next we deal with the second term in the RHS of equation \eqref{eq:taylor2}. Using assumption 8(i), there exists a matrix $M$ such that, $\sup_{N,\boldsymbol \varphi'}||\nabla^2_{\boldsymbol\varphi'} g(\boldsymbol\varphi)||_F<||M||_F$.
\begin{align*}
   &||Q_N^{-\frac{1}{2}}(\widehat{\boldsymbol \varphi_N}-\boldsymbol \varphi_0)^T\nabla_{\boldsymbol \varphi'}^2 g(\boldsymbol\varphi) (\widehat{\boldsymbol \varphi_N}-\boldsymbol \varphi_0)||_F\\
   &=||Q_N^{-\frac{1}{2}}(\widehat{\boldsymbol \varphi_N}-\boldsymbol \varphi_0)^TB_N^{\frac{1}{2}}(\boldsymbol\varphi_0)B_N^{-\frac{1}{2}}(\boldsymbol\varphi_0)\nabla_{\boldsymbol \varphi'}^2 g(\boldsymbol\varphi) B_N^{-\frac{1}{2}}(\boldsymbol\varphi_0)B_N^{\frac{1}{2}}(\boldsymbol\varphi_0)(\widehat{\boldsymbol \varphi_N}-\boldsymbol \varphi_0)||_F\\
   &=||Q_N^{-\frac{1}{2}}\left\{B_N^{\frac{1}{2}}(\boldsymbol\varphi_0)(\widehat{\boldsymbol \varphi_N}-\boldsymbol \varphi_0)\right\}^TB_N^{-\frac{1}{2}}(\boldsymbol\varphi_0)\nabla_{\boldsymbol \varphi'}^2 g(\boldsymbol\varphi) B_N^{-\frac{1}{2}}(\boldsymbol\varphi_0)\left\{B_N^{\frac{1}{2}}(\boldsymbol\varphi_0)(\widehat{\boldsymbol \varphi_N}-\boldsymbol \varphi_0)\right\}||_F\\
   & \leq |Q_N^{-\frac{1}{2}}|(||B_N^{\frac{1}{2}}(\boldsymbol \varphi_0)(\widehat{\boldsymbol \varphi_N}-\boldsymbol \varphi_0)||_F)^2(||B_N^{-\frac{1}{2}}(\boldsymbol \varphi_0)||_F)^2||M||_F
\end{align*}
Using Theorem 2, we have  $B_N^{\frac{1}{2}}(\boldsymbol \varphi_0)(\widehat{\boldsymbol \varphi_N}-\boldsymbol \varphi_0)\xrightarrow{d}\mathcal{N}(\boldsymbol 0,\mathcal{I}_{K\times K})$. Hence we obtain $||B_N^{\frac{1}{2}}(\boldsymbol \varphi_0)(\widehat{\boldsymbol \varphi_N}-\boldsymbol \varphi_0)||_F=O_p(1)$. Moreover, $|Q_N^{-\frac{1}{2}}|=O(N^{\frac{\delta_1}{2}})$. One can write 
\begin{align*}
    (||B_N^{-\frac{1}{2}}(\boldsymbol \varphi_0)||_F)^2=\text{tr}(B_N^{-1}(\boldsymbol \varphi_0))=\sum_{k=1}^K \lambda_{N,k}=O(N^{-\delta_1})
\end{align*}
Hence we obtain $||Q_N^{-\frac{1}{2}}(\widehat{\boldsymbol \varphi_N}-\boldsymbol \varphi_0)^T\nabla_{\boldsymbol \varphi'}^2 g(\boldsymbol\varphi) (\widehat{\boldsymbol \varphi_N}-\boldsymbol \varphi_0)||_F=O_p(N^{-\frac{\delta_1}{2}})=o_p(1)$. Using this result, one can write in equation \eqref{eq:taylor2},
\begin{align}
    & Q_N^{-\frac{1}{2}}\{g(\widehat{\boldsymbol \varphi_N})-g(\boldsymbol\varphi_0)\}=Q_N^{-\frac{1}{2}}\nabla_{\boldsymbol\varphi_0} g(\boldsymbol\varphi)^TB_N^{-\frac{1}{2}}(\boldsymbol\varphi_0)B_N^{\frac{1}{2}}(\boldsymbol\varphi_0)(\widehat{\boldsymbol \varphi_N}-\boldsymbol \varphi_0)+o_p(1)\label{eq:taylor3}
\end{align}
Since $B_N^{\frac{1}{2}}(\boldsymbol \varphi_0)(\widehat{\boldsymbol \varphi_N}-\boldsymbol \varphi_0)\xrightarrow{d}\mathcal{N}(\boldsymbol 0,\mathcal{I}_{K\times K})$ from Theorem 2, applying the Slutsky's Theorem and equations \eqref{eq:slutsky} and \eqref{eq:taylor3}, that
\begin{align*}
      & Q_N^{-\frac{1}{2}}\{g(\widehat{\boldsymbol \varphi_N})-g(\boldsymbol \varphi_0)\}\xrightarrow{d}\mathcal{N}(0,1)\cdot 
\end{align*}
\end{proof}

\subsection{Proof of Theorem 4}
\begin{proof}
Theorem 3 shows that the asymptotic variance of $g(\widehat{\boldsymbol \varphi_N})$ is\\ $\nabla_{\boldsymbol \varphi_0} g(\boldsymbol\varphi)^T B_N^{-\frac{1}{2}}(\boldsymbol \varphi_0) \nabla_{\boldsymbol\varphi_0} g(\boldsymbol \varphi)$. 
Since $
    \mathbb{E}\{||B_N^{-\frac{1}{2}}(\boldsymbol \varphi_0) J_N(\boldsymbol \varphi_0)  B_N^{-\frac{1}{2}}(\boldsymbol \varphi_0)-\mathcal{I}_{K\times K}||_F^2\}\rightarrow 0
$ in Assumption 6, we have $B_N^{-\frac{1}{2}}(\boldsymbol \varphi_0)J_N(\boldsymbol \varphi_0) B_N^{-\frac{1}{2}}(\boldsymbol \varphi_0)\xrightarrow{p}\mathcal{I}$, as $N\rightarrow \infty$. Noting the fact that matrix inverse $g(H)=H^{-1}$ is an element-wise continuous function of $H$, where $H$ is a square invertible matrix, applying the continuous mapping theorem at individual elements, we obtain weak consistency for the inverse sandwich covariance matrix; that is, for any $(i,j)$ element, 
\begin{align*}
    [\{B_N^{-\frac{1}{2}}(\boldsymbol \varphi_0)J_N(\boldsymbol \varphi_0) B_N^{-\frac{1}{2}}(\boldsymbol \varphi_0)\}^{-1}]^{(i,j)}\xrightarrow{p}\mathcal{I}_{K\times K}^{(i,j)}, \mbox{as $N\rightarrow \infty$.}
\end{align*}
This implies $B_N^{\frac{1}{2}}(\boldsymbol \varphi_0)J_N^{-1}(\boldsymbol \varphi_0) B_N^{\frac{1}{2}}(\boldsymbol \varphi_0) \xrightarrow{p}\mathcal{I}_{K\times K}$, as $N\rightarrow \infty$, where $J_N^{-1}(\boldsymbol \varphi_0)$ is well defined under Assumption 8(i). Let $\Omega_p(1)$ be a $K\times K$ matrix where each element is $o_p(1)$. Therefore we obtain 
\begin{align}
   B_N^{\frac{1}{2}}(\boldsymbol \varphi_0)J_N^{-1}(\boldsymbol \varphi_0) B_N^{\frac{1}{2}}(\boldsymbol \varphi_0) - \mathcal{I}_{K\times K}=\Omega_p(1) \label{eq:eq4.3.1}
\end{align}
Multiplying equation \eqref{eq:eq4.3.1} by $B_N^{-1}(\boldsymbol \varphi_0)$, we obtain 
$$B_N^{-\frac{1}{2}}(\boldsymbol \varphi_0)B_N^{\frac{1}{2}}(\boldsymbol \varphi_0) J_N^{-1}(\boldsymbol \varphi_0)  B_N^{\frac{1}{2}}(\boldsymbol \varphi_0) B_N^{-\frac{1}{2}}(\boldsymbol \varphi_0)-B_N^{-1}(\boldsymbol \varphi_0)=\Omega_p(1)B_N^{-1}(\boldsymbol \varphi_0).$$
Under Assumption 5, we have $\lim_{N\rightarrow \infty}||B_N^{-1}(\boldsymbol \varphi_0)||_F=0$, and thus $\Omega_p(1)B_N^{-1}(\boldsymbol \varphi_0)=\Omega_p(1)$. Furthermore, we obtain as $N\rightarrow \infty$, $ J_N^{-1}(\boldsymbol \varphi_0)-B_N^{-1}(\boldsymbol \varphi_0)\xrightarrow{p}\mathcal{O}_{K\times K}$. By Assumption 8(i) of $\sup_N\nabla_{\boldsymbol \varphi_0} g(\boldsymbol \varphi)^T \nabla_{\boldsymbol \varphi_0} g(\boldsymbol \varphi)<\infty$, we have $\nabla_{\boldsymbol \varphi_0} g(\boldsymbol \varphi)^T \Omega_p(1)\nabla_{\boldsymbol \varphi_0} g(\boldsymbol \varphi)=o_p(1)$. Therefore, we obtain,
\begin{align*}
    \nabla_{\boldsymbol\varphi_0} g(\boldsymbol \varphi)^T  J_N^{-1}(\boldsymbol\varphi_0)  \nabla_{\boldsymbol\varphi_0}g(\boldsymbol \varphi)-\nabla_{\boldsymbol\varphi_0} g(\boldsymbol \varphi)^T B_N^{-1}(\boldsymbol\varphi_0) \nabla_{\boldsymbol\varphi_0} g(\boldsymbol \varphi)\xrightarrow{p} 0 \cdot
\end{align*}
Using the expression $J_N(\boldsymbol \varphi)$ from the Hessian Calculation in Supplementary Section S1.4 and compactness of the parameter space $\Theta$ and under Assumptions 9(i) and 9(ii) we obtain that
$$\sup_N||\nabla_{\boldsymbol\varphi_0} \{ \nabla_{\boldsymbol\varphi} g(\boldsymbol \varphi)^T  J_N^{-1}(\boldsymbol\varphi)  \nabla_{\boldsymbol\varphi}g(\boldsymbol \varphi)-\nabla_{\boldsymbol\varphi} g(\boldsymbol \varphi)^T B_N^{-1}(\boldsymbol\varphi) \nabla_{\boldsymbol\varphi} g(\boldsymbol \varphi)\}||_F<\infty$$
$$\sup_N||\nabla^2_{\boldsymbol\varphi'} \{ \nabla_{\boldsymbol\varphi} g(\boldsymbol \varphi)^T  J_N^{-1}(\boldsymbol\varphi)  \nabla_{\boldsymbol\varphi}g(\boldsymbol \varphi)-\nabla_{\boldsymbol\varphi} g(\boldsymbol \varphi)^T B_N^{-1}(\boldsymbol\varphi) \nabla_{\boldsymbol\varphi} g(\boldsymbol \varphi)\}||_F<\infty$$
where $\boldsymbol\varphi'$ lies in an $\epsilon$ neighborhood of $\boldsymbol\varphi_0$. Moreover given that $\widehat{\boldsymbol\varphi_N}\xrightarrow{p} \boldsymbol \varphi_0$, using the same lines of arguments used to prove the consistency of $g(\widehat{\boldsymbol\varphi_N})$ in Theorem 3 , we can establish
\begin{align}
    \nabla_{\widehat{\boldsymbol\varphi_N}} g(\boldsymbol \varphi)^T J_N^{-1}(\widehat{\boldsymbol\varphi_N})  \nabla_{\widehat{\boldsymbol\varphi_N}} g(\boldsymbol \varphi)-\nabla_{\widehat{\boldsymbol\varphi_N}} g(\boldsymbol \varphi)^T B_N^{-1}(\widehat{\boldsymbol\varphi_N}) \nabla_{\widehat{\boldsymbol\varphi_N}} g(\boldsymbol \varphi)\xrightarrow{p}0. \label{eq:step1}
\end{align}
Similarly, we obtain 
 $$\sup_N||\nabla_{\boldsymbol\varphi_0}\{\nabla_{\boldsymbol \varphi} g(\boldsymbol \varphi)^T  B_N^{-1}(\boldsymbol \varphi) \nabla_{\boldsymbol \varphi} g(\boldsymbol \varphi)\}||_F <\infty,\hspace{0.2cm} \text{and}$$
$$\sup_N||\nabla^2_{\boldsymbol\varphi'}\{\nabla_{\boldsymbol \varphi} g(\boldsymbol \varphi)^T  B_N^{-1}(\boldsymbol \varphi) \nabla_{\boldsymbol \varphi} g(\boldsymbol \varphi)\}||_F <\infty$$ 
where $\boldsymbol\varphi'$ lies in an $\epsilon$ neighborhood of $\boldsymbol\varphi_0$. Using similar arguments leads to
\begin{align}
    \nabla_{\widehat{\boldsymbol\varphi_N}} g(\boldsymbol \varphi)^T B_N^{-1}(\widehat{\boldsymbol\varphi_N}) \nabla_{\widehat{\boldsymbol\varphi_N}} g(\boldsymbol \varphi)-\nabla_{\boldsymbol \varphi_0} g(\boldsymbol \varphi)^T  B_N^{-1}(\boldsymbol \varphi_0) \nabla_{\boldsymbol \varphi_0} g(\boldsymbol \varphi)\xrightarrow{p} 0. \label{eq:step2}
\end{align}
Adding equations \eqref{eq:step1} and \eqref{eq:step2}, we obtain 
\begin{align*}
     \nabla_{\widehat{\boldsymbol\varphi_N}} g(\boldsymbol \varphi)^T J_N^{-1}(\widehat{\boldsymbol\varphi_N})  \nabla_{\widehat{\boldsymbol\varphi_N}} g(\boldsymbol \varphi)-\nabla_{\boldsymbol \varphi_0} g(\boldsymbol \varphi)^T  B_N^{-1}(\boldsymbol \varphi_0) \nabla_{\boldsymbol \varphi_0} g(\boldsymbol \varphi)\xrightarrow{p}0 \cdot
\end{align*}
\end{proof}

\section{Supplementary Tables and Figures}
\begin{table}[H]
\centering
\begin{tabular}{|c|c|c|c|c|}
\hline
\textbf{Parameter}                  & \textbf{Actual} & \textbf{\begin{tabular}[c]{@{}c@{}}Estimated\\ ($N=100$)\end{tabular}} & \textbf{\begin{tabular}[c]{@{}c@{}}Estimated\\ ($N=200$)\end{tabular}} & \textbf{\begin{tabular}[c]{@{}c@{}}Estimated\\ ($N=800$)\end{tabular}} \\ \hline
\textbf{$\beta_0$}                  & -2.00           & -1.93                                                                  & -2.02                                                                  & -2.00                                                                  \\ \hline
\textbf{$\beta_1$}                  & 1.50            & 1.47                                                                   & 1.49                                                                   & 1.51                                                                   \\ \hline
\textbf{$\beta_2$}                  & 0.80            & 0.80                                                                   & 0.83                                                                   & 0.80                                                                   \\ \hline
\textbf{$\beta_3$}                  & 1.20            & 1.16                                                                   & 1.20                                                                   & 1.20                                                                   \\ \hline
\textbf{$\beta_4$}                  & 0.40            & 0.40                                                                   & 0.39                                                                   & 0.40                                                                   \\ \hline
\textbf{$\beta_5$}                  & 2.10            & 2.06                                                                   & 2.09                                                                   & 2.10                                                                   \\ \hline
\textbf{$\beta_6$}                  & 1.30            & 1.29                                                                   & 1.31                                                                   & 1.30                                                                   \\ \hline
\textbf{$\gamma_0$}                 & -1.00           & -0.97                                                                  & -1.01                                                                  & -1.01                                                                  \\ \hline
\textbf{$\gamma_1$}                 & 2.00           & 1.93                                                                   & 2.00                                                                   & 2.01                                                                   \\ \hline
\textbf{$\gamma_2$}                 & 0.90            & 0.87                                                                   & 0.89                                                                   & 0.90                                                                   \\ \hline
\textbf{$\gamma_3$}                 & 1.80            & 1.73                                                                   & 1.79                                                                   & 1.80                                                                   \\ \hline
\textbf{$\gamma_4$}                 & 0.70            & 0.67                                                                   & 0.68                                                                   & 0.70                                                                   \\ \hline
\textbf{$\sigma_y$}                 & 1.00            & 0.95                                                                   & 0.98                                                                   & 1.00                                                                   \\ \hline
\textbf{$\sigma_m$}                 & 1.00            & 0.93                                                                   & 0.98                                                                   & 1.00                                                                   \\ \hline
\textbf{$\sigma_{\boldsymbol b^y}$} & 1.00            & 0.90                                                                   & 0.98                                                                   & 1.00                                                                   \\ \hline
\textbf{$\sigma_{\boldsymbol b^m}$} & 1.00            & 0.93                                                                   & 0.98                                                                   & 1.00                                                                   \\ \hline
\end{tabular}
\vspace{0.5cm}
\caption{Average estimates of the REN-SEM model parameters over 500 replications under Simulation-1 with Network sizes, $N=100,200,800$.}
\end{table}

\begin{table}[H]
\centering
\begin{tabular}{|c|c|c|c|c|}
\hline
\textbf{Parameter}                  & \textbf{Actual} & \textbf{\begin{tabular}[c]{@{}c@{}}Estimated\\ ($N=100$)\end{tabular}} & \textbf{\begin{tabular}[c]{@{}c@{}}Estimated\\ ($N=200$)\end{tabular}} & \textbf{\begin{tabular}[c]{@{}c@{}}Estimated\\ ($N=800$)\end{tabular}} \\ \hline
\textbf{$\beta_0$}                  & -2.00           & -2.05                                                                  & -1.99                                                                  & -1.98                                                                  \\ \hline
\textbf{$\beta_1$}                  & 1.50            & 1.51                                                                   & 1.51                                                                   & 1.49                                                                   \\ \hline
\textbf{$\beta_2$}                  & 0.80            & 0.89                                                                   & 0.77                                                                   & 0.78                                                                   \\ \hline
\textbf{$\beta_3$}                  & 1.20            & 1.19                                                                   & 1.19                                                                   & 1.19                                                                   \\ \hline
\textbf{$\beta_4$}                  & 0.40            & 0.39                                                                   & 0.41                                                                   & 0.40                                                                   \\ \hline
\textbf{$\beta_5$}                  & 2.10            & 2.10                                                                   & 2.11                                                                   & 2.08                                                                   \\ \hline
\textbf{$\beta_6$}                  & 1.30            & 1.29                                                                   & 1.30                                                                   & 1.27                                                                   \\ \hline
\textbf{$\gamma_0$}                 & -1.00           & -0.99                                                                  & -1.02                                                                  & -0.99                                                                  \\ \hline
\textbf{$\gamma_1$}                 & 2.00            & 2.00                                                                   & 1.98                                                                   & 1.98                                                                   \\ \hline
\textbf{$\gamma_2$}                 & 0.90            & 0.90                                                                   & 0.98                                                                   & 0.99                                                                   \\ \hline
\textbf{$\gamma_3$}                 & 1.80            & 1.78                                                                   & 1.79                                                                   & 1.79                                                                   \\ \hline
\textbf{$\gamma_4$}                 & 0.70            & 0.69                                                                   & 0.70                                                                   & 0.71                                                                   \\ \hline
\textbf{$\sigma_y$}                 & 1.00            & 0.95                                                                   & 0.97                                                                   & 0.99                                                                   \\ \hline
\textbf{$\sigma_m$}                 & 1.00            & 0.96                                                                   & 0.97                                                                   & 0.99                                                                   \\ \hline
\textbf{$\sigma_{\boldsymbol b^y}$} & 0.50            & 0.48                                                                   & 0.49                                                                   & 0.49                                                                   \\ \hline
\textbf{$\sigma_{\boldsymbol b^m}$} & 0.50            & 0.48                                                                   & 0.49                                                                   & 0.49                                                                   \\ \hline
\end{tabular}
\vspace{0.5cm}
\caption{Average estimates of the REN-SEM model parameters over 500 replications under Simulation-2 with Network sizes, $N=100,200,800$.}
\end{table}

\begin{table}{H}
\centering
\resizebox{0.9\textwidth}{!}{
\begin{tabular}{|c|c|c|c|c|c|c|c|}
\hline
\textbf{Size(N)}          & \textbf{Effects} & \textbf{Actual} & \textbf{Bias} & \textbf{RRMSE}& \textbf{ESE} & \textbf{ASE} & \textbf{CP}\\ \hline
\multirow{7}{*}{\textbf{100}} & \textbf{$\tau_1$}             & 1.50            & 0.008               & 0.250                          & 0.367                                                                    & 0.391      & 0.954                                                                             \\ \cline{2-8} 
                              & $\tau_2$            & 2.40            & -0.012               & 0.175                              & 0.405                                                                    & 0.428    & 0.947                                                                           \\ \cline{2-8} 
                              & $\tau_3$            & 0.04            & -0.001               & 1.778                              & 0.066                                                                    & 0.072   & 0.938                                                                            \\ \cline{2-8} 
                              & $\tau_4$            & 0.80            & 0.086               & 1.725                              & 1.375                                                                    & 1.361   & 0.930                                                                          \\ \cline{2-8} 
                              & $\tau_5$            & 1.08            & -0.002               & 1.305                              & 1.408                                                                    & 1.381    & 0.954                                                                          \\ \cline{2-8} 
                              & $\tau_6$            & 0.83            & -0.029               & 0.798                            & 0.655                                                                    & 0.715    & 0.960                                                                            \\ \hline
\multirow{6}{*}{\textbf{200}} & $\tau_1$            & 1.50            & 0.012               & 0.182                             & 0.256                                                                    & 0.263  & 0.951                                                                             \\ \cline{2-8} 
                              & $\tau_2$            & 2.40            & -0.030               & 0.135                             & 0.287                                                                    & 0.298   & 0.952                                                                             \\ \cline{2-8} 
                              & $\tau_3$            & 0.04            & 0.004               & 1.231                              & 0.044                                                                    & 0.046     & 0.936                                                                          \\ \cline{2-8} 
                              & $\tau_4$            & 0.80            & -0.029               & 1.134                            & 0.908                                                                    & 0.948   & 0.964                                                                             \\ \cline{2-8} 
                              & $\tau_5$            & 1.08            & 0.086               & 0.966                             & 1.040                                                                    & 1.020     & 0.942                                                                          \\ \cline{2-8} 
                              & $\tau_6$            & 0.81            & 0.022               & 0.522                              & 0.423                                                                    & 0.437 & 0.968                                                                               \\ \hline
\multirow{6}{*}{\textbf{800}} & $\tau_1$            & 1.50            & -0.011              & 0.132                            & 0.131                                                                    & 0.129  & 0.946                                                                              \\ \cline{2-8} 
                              & $\tau_2$            & 2.40            & -0.027               & 0.117                             & 0.146                                                                    & 0.147  & 0.956                                                                              \\ \cline{2-8} 
                              & $\tau_3$            & 0.04            & -0.001               & 0.532                              & 0.020                                                                    & 0.020     & 0.937                                                                         \\ \cline{2-8} 
                              & $\tau_4$            & 0.80            & -0.016               & 0.564                             & 0.447                                                                   & 0.422   & 0.939                                                                            \\ \cline{2-8} 
                              & $\tau_5$            & 1.08            & -0.020               & 0.446                             & 0.472                                                                    & 0.460   & 0.952                                                                            \\ \cline{2-8} 
                              & $\tau_6$            & 0.80            & -0.002               & 0.261                              & 0.195                                                                    & 0.189    & 0.949                                                                          \\ \hline
\end{tabular}}
\vspace{0.5cm}
\caption{Summarized simulation results under the Network-2 scenario from 500 rounds of simulations. The Bias and Root Relative Mean Squared Error (RRMSE) for the six estimated effects $\hat\tau_1$ through $\hat\tau_6$, Empirical Standard Error (ESE), Average Standard Error (ASE) and Coverage probability (CP) are presented, respectively.}\label{tab:setup2}
\end{table}

\begin{figure}
\centering
 \includegraphics[scale=0.42]{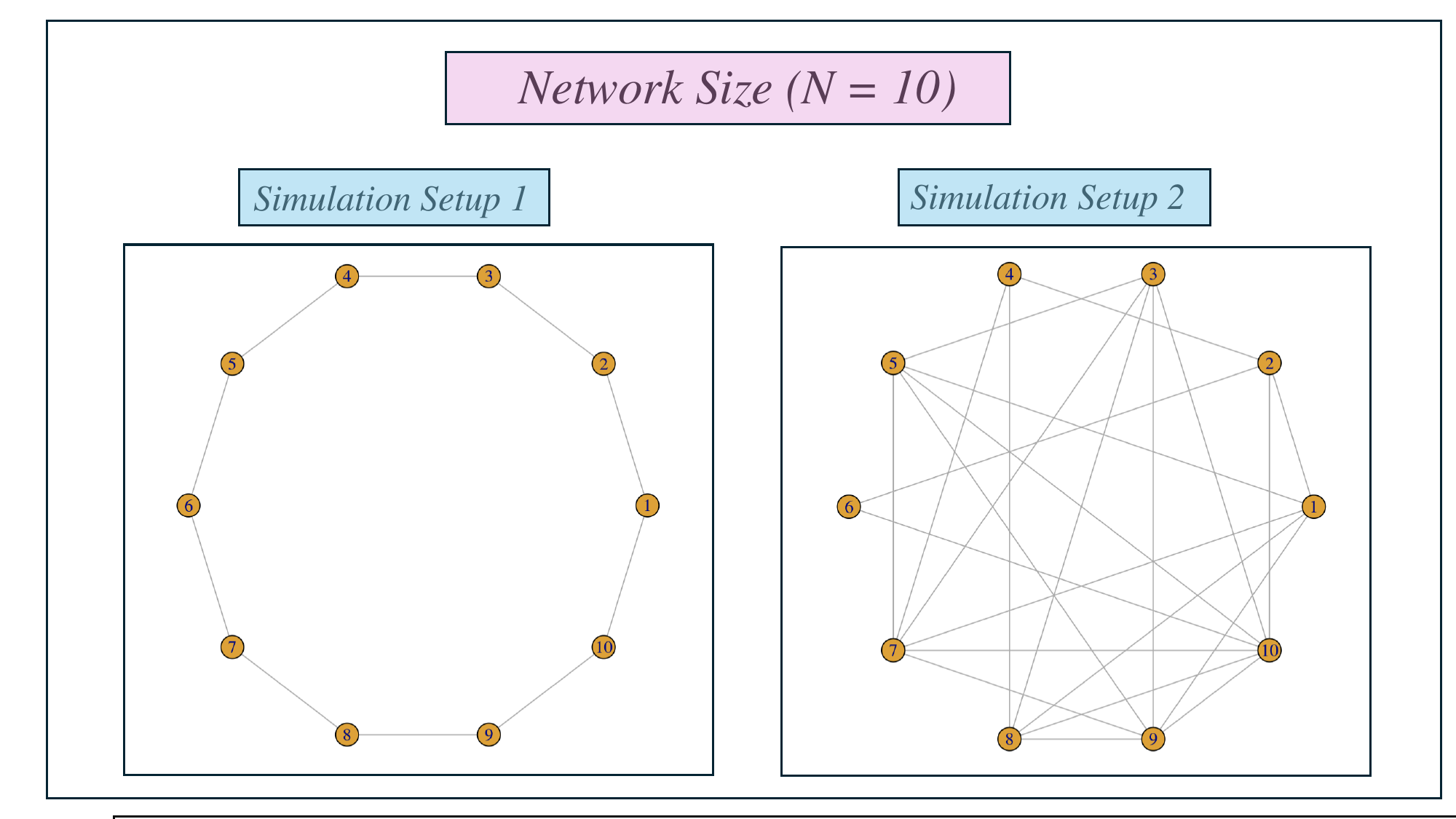}
 \caption{Network Examples under the two simulation setups generated using sample size, $N=10$ for the purpose of illustration. The nodes are present on the circumference of the circle, where $i^{\text{th}}$ and $j^{\text{th}}$ nodes are connected if $E_{ij}=1$ and the self loops are omitted. For simulation setup 2, we used $p=0.4$ to generate the network under Erdos-Renyi Random Model.}
 \label{fig:simugraphs}
\end{figure}

\begin{figure}
\centering
 \includegraphics[scale=0.4]{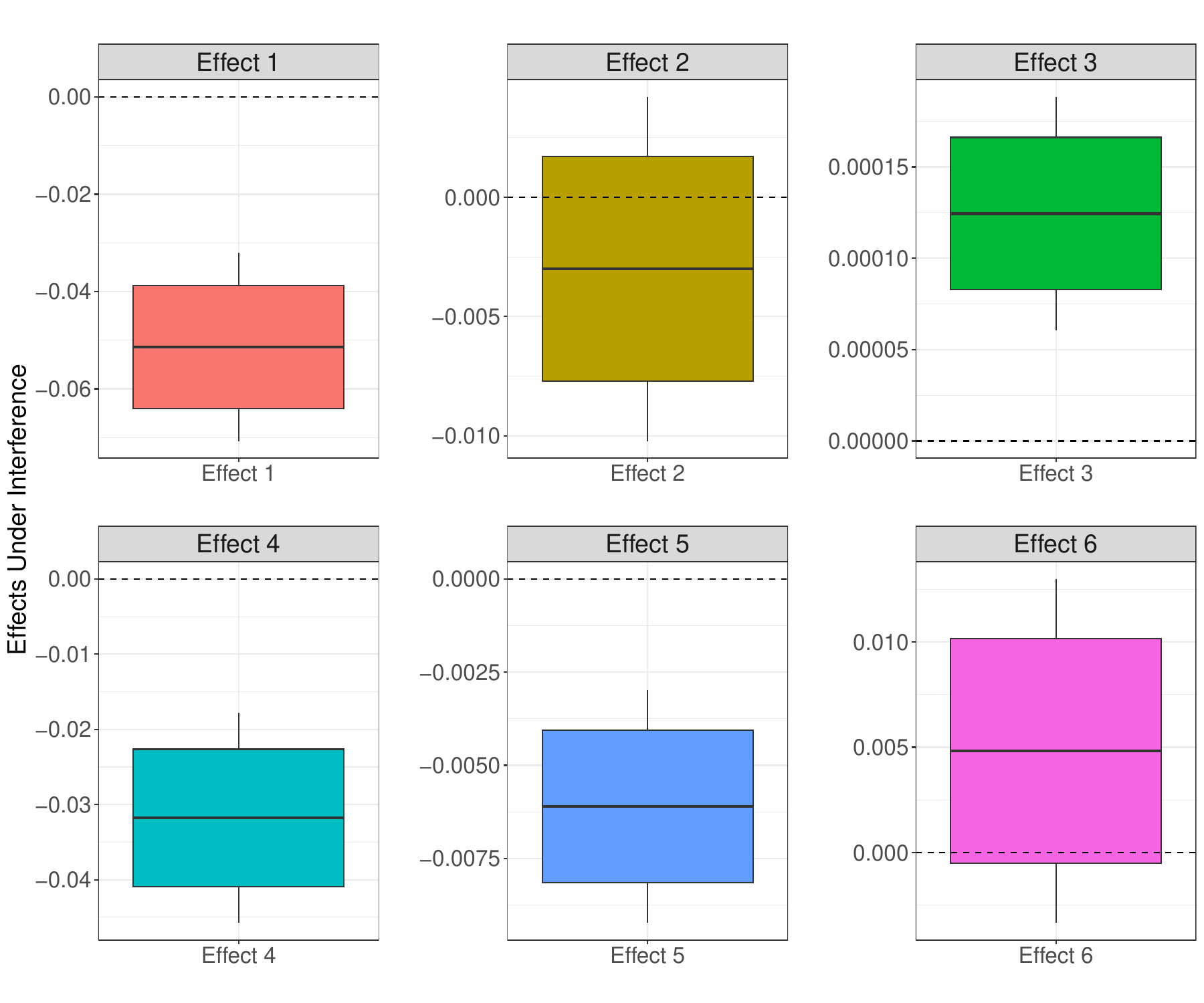}
 \caption{Mean Estimates of each of the six causal estimands along with 95\% C.I applied to the Twitch Network Dataset. In this analysis we study the impact of mature content—both from individual users and the average within their network neighbors—on the views received by a user’s account mediated via the lifetimes of both the individual and neighbors' accounts.}
   \label{fig:twitch}
\end{figure}
\end{document}